\documentclass[12pt]{article}

\usepackage{epsfig,latexsym,amsfonts,amsmath,amsthm,amssymb,amsbsy,multirow,slashed,wasysym,textcomp,dsfont,comment,mathtools,cancel,cite,diagbox,datetime,appendix,BOONDOX-calo}
\usepackage{tocloft}
\usepackage{bbold}
\usepackage{sidecap}
\usepackage{adjustbox}
\usepackage{pgfplots}
\usepackage{oplotsymbl}
\usepackage{tikzpagenodes}
\usepackage{adforn}
\usepackage{tikz}
\usetikzlibrary{shapes,arrows,cd,chains,decorations.markings,decorations.pathmorphing,calc}
\tikzset{
->-/.style args={#1rotate#2}{decoration={markings, mark=at position #1 with {\arrow[scale=1.5,rotate = #2 ]{stealth}}}, postaction={decorate}}
}
\tikzset{curve/.style={settings={#1},to path={(\tikztostart)
    .. controls ($(\tikztostart)!\pv{pos}!(\tikztotarget)!\pv{height}!270:(\tikztotarget)$)
    and ($(\tikztostart)!1-\pv{pos}!(\tikztotarget)!\pv{height}!270:(\tikztotarget)$)
    .. (\tikztotarget)\tikztonodes}},
    settings/.code={\tikzset{quiver/.cd,#1}
        \def\pv##1{\pgfkeysvalueof{/tikz/quiver/##1}}},
    quiver/.cd,pos/.initial=0.35,height/.initial=0}
\tikzset{tail reversed/.code={\pgfsetarrowsstart{tikzcd to}}}
\tikzset{2tail/.code={\pgfsetarrowsstart{Implies[reversed]}}}
\tikzset{2tail reversed/.code={\pgfsetarrowsstart{Implies}}}

\usetikzlibrary{shapes}
\tikzstyle{bag} = [align=center]
\usetikzlibrary{decorations.pathmorphing}
\usepackage[normalem]{ulem}
\usepackage[mathscr]{eucal}
\usepackage[version=3]{mhchem}
\usepackage{physics}
\usepackage{hyperref}
\usepackage{subcaption}

\usepackage{empheq}
\usepackage[most]{tcolorbox}
\newtcbox{\mymath}[1][]{%
    nobeforeafter, math upper, tcbox raise base,
    enhanced, colframe=blue!30!black,
    colback=blue!30, boxrule=1pt,
    #1}

 \newcommand{\badat}{\begin{alignedat}}
 \newcommand{\eadat}{\end{alignedat}}
  
\def\be{\begin{equation}}
\def\ee{\end{equation}}

\def\p{\partial}

\usepackage{color}

\newcommand{\pink}[1]{\textcolor{\pink}{#1}}

\definecolor{dblue}{rgb}{0.2,0.50,0.80}

\def\A{\mathcal{A}}

\def\I{\mathcal{I}}

\def\O{\mathcal{O}}

\def\S{\mathcal{S}}

\def\bz{{\bar z}}
\def\bw{{\bar w}}

\def\tP{\widetilde{P}}

\def\bz{{\bar z}}
\def\bw{{\bar w}}

\def\scri{\mathcal I}
\def\AdS{\mathrm{AdS}}
\def\CFT{\mathrm{CFT}}

\def\CarCFT{\mathrm{CarCFT}}
\def\Mink{\mathrm{Mink}}
\def\calC{\mathcal{C}}
\def\calI{\mathcal{I}}

\def\bbR{\mathbb{R}}
\def\bbZ{\mathbb{Z}}
\def\bbC{\mathbb{C}}

\def\calO{\mathcal{O}}

\def\calS{\mathcal{S}}

\def\calT{\mathcal{T}}
\def\calB{\mathcal{B}}
\def\calH{\mathcal{H}}

\def\ins{\!{}_{\mathrm{in}}}
\def\lquad{\!\!\!\!}

\tikzset{snake it/.style={decorate, decoration=snake}}

\numberwithin{equation}{section} 
\pgfplotsset{compat=1.17} 

\begin{document}

\begin{titlepage}
\thispagestyle{empty}
\begin{flushright}
\end{flushright}
\bigskip

\begin{center}

\baselineskip=13pt {\LARGE \scshape{
Multiparticle States \\ [.5em]
for the Flat Hologram
    }
}

\vskip1cm 

\centerline{
    {Justin Kulp}${}^{1,2}$ and 
    {Sabrina Pasterski}${}^3$
}

\bigskip\bigskip
\bigskip\bigskip

\centerline{
    ${}^1$
    Simons Center for Geometry and Physics, Stony Brook University, Stony Brook, NY 11794, USA
}
\vspace{1em}
\centerline{
    ${}^2$
    C. N. Yang Institute for Theoretical Physics, Stony Brook University, Stony Brook, NY 11794, USA
}
\vspace{1em}
\centerline{
    ${}^3$ 
    Perimeter Institute for Theoretical Physics, Waterloo, ON N2L 2Y5, Canada
}

\bigskip\bigskip
\end{center}

\begin{abstract}
We use the extrapolate dictionary to revisit the spectrum of operators in Celestial CFT. Under the Celestial CFT map, each state in the 4D Hilbert space should map to one in the 2D Hilbert space. This implies that, beyond the familiar single particle states/operators, there should be multiparticle operators appearing in the celestial OPE. We extend the existing flat-space dictionary by constructing composite primaries from both Carrollian and Celestial perspectives. In the process, we demonstrate some subtleties in deriving the Poincar\'e primary condition from the Carrollian limit, clarify the compatibility of principal series representations with highest weight representations and unitarity in Celestial CFT, and discuss how the celestial OPE block expansion emerges from a 2D CFT standpoint.
\end{abstract}

\bigskip \bigskip \bigskip \bigskip 
 
\end{titlepage}

\setcounter{tocdepth}{2}
\tableofcontents

\section{Introduction}\label{sec:Intro}
Flat space holography studies quantum gravity and aspects of quantum field theory in ``bulk'' asymptotically flat spacetimes by encoding observables in a lower dimensional dual theory supported ``at the boundary of spacetime.'' One observable of interest in asymptotically flat space is the bulk $\mathcal{S}$-matrix, which characterizes the response of the theory to sources in the asymptotic past and future. Consequently, any holographic dual encodes information about scattering experiments, with the $\calS$-matrix represented by collections of boundary correlation functions.

The pursuit of a dual dynamical boundary theory is both conceptually and practically motivated. Conceptually, the $\calS$-matrix of QFT only depends on observables supported at asymptotically early/late times, so many details of bulk physics are largely irrelevant. However, most approaches to computing $\calS$-matrix elements still rely on bulk correlators, which are then ``pushed to the boundary'' by some suitable prescription (reviewed in Section \ref{sec:SMatrix}). The desire for a purely boundary-based description is underscored when gravity is involved, as all particles couple universally to gravity, and bulk local operators are not gauge-invariant. As a result, the only observables in a gravitational theory are boundary or global observables, as emphasized by the holographic principle \cite{tHooft:1993Dmi, Susskind:1994vu}. Of course, one could work with gauge non-invariant intermediate quantities, so long as final results are well-defined, but we should never be forced to work with ill-defined quantities, even in principle, suggesting a purely boundary-based description must exist. 

An interesting role is played by so-called ``large gauge transformations,'' which have non-compact support and extend to the conformal boundary of spacetime. This has led to a number of practical successes for flat-space holography. The ``infrared triangle'' connects soft-theorems, the memory effect, and asymptotic symmetries of spacetime, as different realizations of the same underlying concept \cite{Strominger:2014pwa, Pasterski:2015tva, Pasterski:2015zua, Strominger:2017zoo}. The Ward identities for the bulk asymptotic symmetries then have a natural holographic interpretation as currents in a codimension-\textit{two} CFT. Pursuing these ideas led to the change of bases (described in Section \ref{sec:ExtrapolateDictionary}) whereby $\cal S$-matrices are recast as correlators in a Celestial CFT. Celestial holography techniques have further revealed exceptional symmetry structures, like $w_{1+\infty}$ algebras (and related deformations and quantizations) in scattering amplitudes \cite{Strominger:2021mtt,Guevara:2021abz, Himwich:2021dau, Mago:2021wje, Bu:2022iak, Drozdov:2023qoy,Mason:2023mti}. These techniques have also been used to ``bootstrap'' tree and loop-level scattering amplitudes, for gauge theories with and without matter or supersymmetry, via the recently developed ``chiral algebra bootstrap'' \cite{Costello:2022upu, Costello:2022wso, Costello:2023vyy} (see also \cite{Zeng:2023qqp, Garner:2023izn, Fernandez:2024tue, Garner:2024tis,Fernandez:2024qnu} for examples of the chiral algebra bootstrap).

Thus far, the two dominant approaches to flat space holography are the Celestial Conformal Field Theory (CCFT) and the Carrollian Conformal Field Theory (CarCFT) approaches. In CarCFT, flat space scattering amplitudes are encoded in conformally covariant correlation functions of a \textit{null} CFT living on the codimension-one lightlike boundary of spacetime $\calI^{\pm}$ \cite{Ciambelli:2018wre, Ciambelli:2018ojf, Bagchi:2019xfx,Bagchi:2019clu,Gupta:2020dtl,Freidel:2022bai, Ciambelli:2019lap, Ciambelli:2023xqk}. Simply put, the CarCFT approach is a large radius/ultra-relativistic limit of the usual AdS/CFT correspondence \cite{Maldacena:1997re, Witten:1998qj,Susskind:1998vk, Giddings:1999jq, Polchinski:1999ry, Penedones:2010ue, Paulos:2016fap, Hijano:2019qmi, Komatsu:2020sag, Compere:2020lrt,Li:2021snj, PipolodeGioia:2022exe, Bagchi:2023fbj, deGioia:2023cbd}. We will review the construction of CarCFT and discuss this limit in Sections \ref{sec:ExtrapolateDictionary} and \ref{sec:Carrollian} respectively. On the other hand, as introduced above, CCFT amounts to examining the $\calS$-matrix in a basis of boost eigenstates, so that Lorentz covariance manifests as conformal symmetry of a Euclidean CFT in codimension-two, with scattering states mapped to conformal (quasi)-primaries on the celestial sphere \cite{Pasterski:2016qvg, Raclariu:2021zjz, Pasterski:2021rjz, Donnay:2023mrd}. 

The respective approaches each have their own intrinsic strengths. For example, the CarCFT is most simply related to AdS/CFT (but see also \cite{Bittleston:2024rqe}), and is naturally related to detector operators \cite{Hofman:2008ar,Hu:2022txx,Hu:2023geb}. Meanwhile, the CCFT is defined on a standard Euclidean spacetime, most naturally supports both massive and massless particles, and has both bottom-up as well as top-down constructions \cite{Costello:2022jpg, Costello:2023hmi, Bittleston:2024efo}. However, the two approaches turn out to be compatible and provably equivalent (see e.g. \cite{Donnay:2022aba,Bagchi:2022emh, Donnay:2022wvx}, the review of \cite{Ciambelli:2022vot}, and references within for comparisons between the two approaches).

One of the key ingredients to the practical successes above in flat space holography, and especially in understanding the structure of scattering amplitudes, is the celestial OPE \cite{Pate:2019lpp, Banerjee:2020kaa, Himwich:2021dau, Jiang:2021csc, Adamo:2021zpw, Bittleston:2022jeq, Costello:2022upu}. However, a precise understanding of the operator spectrum and OPE data of the dual CFTs is missing in general. Understanding this data is necessary if we hope to use CFT techniques to describe bulk scattering, e.g. using conformal block decompositions of $\cal S$-matrix elements as an alternate way of organizing on-shell amplitudes to avoid complicated Feynman diagrammatics \cite{Nandan:2019jas, Fan:2021isc, Atanasov:2021cje, Fan:2021pbp, Guevara:2021tvr, Hu:2022syq, De:2022gjn, Fan:2022vbz, Fan:2022kpp, Garcia-Sepulveda:2022lga, Jorge-Diaz:2022dmy}.

At the level of the spectrum, we can already learn something missing from earlier work. In \cite{Pasterski:2017kqt}, only the CCFT dictionary for single particle states was examined. This is also the case for holographic CarCFTs as we will see in Section \ref{sec:ExtrapolateDictionary}. However, operators that create ``single particle states'' when acting on the vacuum are only a subset of operators in the theory. In the absence of IR divergences, the theory should look approximately free at the boundary with full Fock spaces of ``in'' and ``out'' states. Then, so long as we have a 2D state operator correspondence~\cite{Crawley:2021ivb}, we should be able to deduce the full spectrum of celestial operators. Indeed, in tree-level \cite{Lam:2017ofc} and all-loop resummed large-$N$ amplitudes \cite{Garcia-Sepulveda:2022lga}, it was shown that the conformal block expansion of CCFT four-point correlators includes operator exchanges that \textit{do not} correspond to bulk one-particle operators, with scaling dimensions instead resembling ``double trace primaries.'' It was also shown that accounting for these exchanges was necessary for bulk unitarity. Moreover, omissions of operators in the OPE are expected to spoil associativity of the OPE \cite{Costello:2022upu} (see also \cite{Ren:2022sws, Bhardwaj:2022anh, Ball:2022bgg, Bittleston:2022jeq}). A better accounting of the multiparticle operators will also be useful for organizing perturbation theory in the bulk, where higher derivative terms appear indirectly as loop order corrections to the OPE coefficients for single particle operators. 

In this paper, we explore the spectrum of the boundary CFTs much further and define and construct the analog of ``multiparticle states'' or ``double trace operators'' in the boundary CFTs of interest. Our approach is particularly inspired by the conglomeration/projector formalism of \cite{Fitzpatrick:2010zm, Fitzpatrick:2011dm, SimmonsDuffin:2012uy} for describing multiparticle states in AdS/CFT. We examine the representations that appear in the flat limit of the conglomerated operators and point out subtleties of handling massive states prepared by composites of massless operators in the Carrollian picture. In particular the Poincar\'e primary conditions are seen to arise from the usual flat limit of 3D primaries plus the $m=0$ constraint. On the other hand, Celestial CFT is more amenable to treating massive and massless states on a similar footing. We then take a representation theoretic approach that lands us on a nice clarification of the single particle representations as principal series, {\it not} highest weight before revisiting the ``celestial OPE block expansion'' \cite{Czech:2016xec, Gadde:2017sjg, Guevara:2021tvr} in this language. Finally, we prepare a setup analogous to~\cite{Pasterski:2017kqt} using bulk to boundary propagators for the multiparticle operators.

\paragraph{Outline of the Rest of the Paper} The rest of this paper proceeds as follows:
\begin{enumerate}
    \item[{\hyperref[sec:Intro]{$\S$.1.}}] In the remainder of the Introduction we discuss some further motivation and background for the rest of the paper. In Section \ref{sec:SMatrix}, we discuss basic assumptions of the $\calS$-matrix and motivate the need for a theory of flat space holography. With this we move to Section \ref{sec:ExtrapolateDictionary}, where we review the flat space extrapolate dictionary and definition of the boundary operators creating ``one-particle states.'' Then, in Section \ref{sec:Conglom}, we discuss the need for ``multiparticle states'' in flat space holography and the techniques for describing multiparticle operators in AdS/CFT, which will guide us through the remainder of the text.
    \item[{\hyperref[sec:Carrollian]{$\S$.2.}}] In Section \ref{sec:Carrollian}, we turn to the construction of Carrollian CFT. In Section \ref{sec:carrollianPrimary} we define Carrollian primaries and study the algebraic conditions to which they correspond. Then, in Section \ref{sec:Casimirs}, we compare this to the primary conditions implied by taking the large radius limit of AdS/CFT and show how the Poincar\'e primary conditions follow from the usual CFT primary conditions and massless constraint. In Section \ref{sec:MultiCarroll} we consider the flat limits of the first few composite operators and find that more general CarCFT representations are necessary, and point to some subtleties about using primaries at null infinity to prepare massive states.  Finally, in Section \ref{sec:CarToCCFT} we show that the Carrollian to Celestial map extends beyond the single particle sector \ref{sec:Carrollian}.
    \item[{\hyperref[sec:Celestial]{$\S$.3.}}] In Section \ref{sec:Celestial} we give an extensive representation-theoretic study of celestial CFT, mimicking traditional constructions in QFT. In Section \ref{sec:HilbertSpaces} we set up the machinery we will be using in later sections, reviewing the Wigner's momentum space constructions and describing the procedure of decomposing representations of the Poincar\'e group into representations of the Lorentz group. The notion of ``channel'' introduced here will be important for our discussion of the celestial OPE. In Section~\ref{sec:SingleParticle} we use this representation theory machinery to revisit some subtleties of single particle states. In particular, we comment on the representations appearing to be highest weight rather than principal series and examine the action of translations on celestial states. Section~\ref{sec:multOPE} then attacks the question of multiparticle states in CCFT from an OPE block perspective using the machinery of decomposing Poincar\'e irreps into $SL(2,\mathbb{C})$ irreps. In particular, in Section \ref{sec:unitarity} we discuss the unitary of the boundary celestial CFT, arguing that it is equivalent to bulk unitarity, and introduce the OPE block expansion. Then, in Section \ref{sec:MultiCelestial} we discuss and extend the celestial OPE block with the ``channels'' introduced from the previous sections.  We 
    comment on the relation to the usual celestial OPE and previous studies of conformal block and partial wave expansions. Finally we show how to co-opt momentum space results to prepare celestial primaries in Section~\ref{sec:Weinberg}. This gives us two different approaches to multiparticle states in CCFT.
    \item[{\hyperref[sec:FlatAlg]{$\A$.A.}}] In Appendix \ref{sec:FlatAlg} we discuss the symmetry algebras involved in AdS/CFT kinematics and carefully describe the In\"on\"u-Wigner contraction from AdS/CFT to flat space in different bases. This Appendix underpins the investigations in Section \ref{sec:Carrollian}. 
    \item[{\hyperref[sec:SL2CReps]{$\A$.B.}}] In Appendix \ref{sec:SL2CReps} we review the representation theory of $SL(2,\bbC)$ and discuss the $z$-basis, where unitary irreducible representations are realized on $L^2(\bbC)$. The discussions in this Appendix are essential for the investigations in Section \ref{sec:Celestial}.
\end{enumerate}

Some precedents for composite operators in celestial CFT appear in analyses of the conformally soft sector. Namely, the Sugawara-like construction of a 2D celestial stress tensor \cite{McLoughlin:2016uwa,Nande:2017dba,Fan:2020xjj} from the spin-1 currents for large gauge transformations and recent studies of the loop-level consistency of the celestial $w_{1+\infty}$~\cite{Strominger:2021mtt} in the twisted holography constructions of \cite{Costello:2022wso, Costello:2022upu, Costello:2022jpg, Bittleston:2022jeq}. Meanwhile, multi-collinear factorization has been studied from the amplitudes side in \cite{Ebert:2020nqf} and used for examining multiparticle contributions to the celestial OPE by the authors in~\cite{Ball:2023sdz,Guevara:2024ixn} in parallel to this work. In the final stages of this project, some overlapping results appeared which we should point out for the reader. In particular, \cite{deGioia:2024yne,Alday:2024yyj} has overlap with Section~\ref{sec:Casimirs} in that the flat limit of single particle representations in the bulk point configuration is discussed. In addition, \cite{Iacobacci:2024laa} has some overlap with Section~\ref{sec:SingleParticle}. 
However, our results are compatible but complementary, focusing heavily on distributional aspects. As a result, we find their work to be an excellent complement having completed the omitted cases in our own work.

\subsection{Scattering and Flat Space Holography}\label{sec:SMatrix}
In sufficiently well-behaved quantum field theories, like those with a mass gap, local operators create finite energy excitations whose correlation functions decay rapidly with separation distance. As a result, one can essentially ignore complicated bulk interactions in describing scattering states and we are left with well-defined Fock spaces of asymptotic ``free particle'' in/out scattering states:
\begin{alignat}{2}\label{eq:in}
	&\ket{p_{m},\dots,p_{1}}_{\mathrm{in}} 
	&&:= a_{p_m}^\dagger \cdots a_{p_1}^\dagger \ket{0}\,,\\
	&\ket{p_{m},\dots,p_{1}}_{\mathrm{out}}\label{eq:out}
	&&:= b_{p_m}^\dagger \cdots b_{p_1}^\dagger \ket{0}\,.
\end{alignat}
By definition, the $\calS$-matrix is the unitary operator giving the amplitude for a state describing well-defined in-particles at some early time to transition to a state describing well-defined out-particles at some late time, i.e.
\begin{equation}
	b_i = \calS^\dagger a_i \calS\,,\quad
	b_i^\dagger = \calS^\dagger a_i^\dagger \calS\,,
\end{equation}
see \cite{Prabhu:2022zcr, Caron-Huot:2023vxl} for additional axioms. This intuition provides the physical basis for various rigorous algebraic theories of scattering \cite{Haag:1958vt, Araki:1962zhd, Collins:2019ozc, bogolubov1990haag}. E.g. in Haag-Ruelle scattering theory, one argues that (for sufficiently nice theories) there exist unitarily equivalent (isometric embeddings of) Fock spaces of asymptotic ``free particles'' in the Hilbert space that are created by bulk fields at infinite times \cite{bogolubov1990haag}.

A typical procedure for computing $\calS$-matrix elements in theories which admit a standard Fock space representation is the LSZ prescription. The LSZ prescription computes the $\calS$-matrix elements from the time-ordered vev of local fields acting on the (interacting) vacuum by Fourier transforming bulk expectation values to momentum space and extracting the leading singularities of the external particles at the poles $p_i^2 = -m_{i,\mathrm{phys}}^2$, putting the external particles on-shell \cite{Collins:2019ozc}. At least one conceptual downside of the LSZ procedure is that it involves bulk expectation values as an intermediary, despite the fact that the $\cal S$-matrix should only be a function of the boundary data where the in and out states are prepared. 

One step towards a holographic description of the $\cal S$-matrix comes from an analogy to the ``extrapolate dictionary'' description of AdS/CFT, where boundary local operators are created by pushing suitably rescaled bulk fields to the boundary of spacetime \cite{Banks:1998dd, Polchinski:1999ry, Harlow:2011ke}. Indeed, suppose that we have a weakly-interacting bulk scalar (for simplicity) field theory, which looks free on/near $\scri^{\pm}$. In this case, the classical bulk field $\phi(X)$ can be expanded near $\scri^{\pm}$ as
\begin{equation}\label{freedat}
	\phi(X) 
	\overset{\mathcal{I}^-}{=} \frac{1}{r}\Phi^{\mathrm{in}}(v,z,\bz)+O(r^{-2}\log r),~~~  
        \phi(X)\overset{\mathcal{I}^+}{=} \frac{1}{r}\Phi^{\mathrm{out}}(u,z,\bz)+O(r^{-2}\log r)\,.
\end{equation}
$\Phi(X)^{\mathrm{in}/\mathrm{out}}$ are free fields which are canonically quantized to give the in/out operators $\{a_{p},a_{p}^\dagger\}$ and $\{b_{p},b_{p}^\dagger\}$ above (see also \cite{Donnay:2022wvx}) and, in our holographic analogue, will correspond to boundary primary operators.\footnote{Section 4.3 of~\cite{He:2020ifr} and the recent works~\cite{Kim:2023qbl, Jain:2023fxc, Kraus:2024gso} flesh out the relation between LSZ, extrapolate, and path integral formulations in insightful detail, and recommend \cite{Susskind:1998vk, Giddings:1999jq, Polchinski:1999ry, Paulos:2016fap, vanRees:2022itk} for a complementary discussion on the LSZ formalism and the relationship of the $\calS$-matrix to flat space limits of QFT in AdS.}

\subsection{The Flat Space Extrapolate Dictionary}\label{sec:ExtrapolateDictionary}
Let us now describe the bulk-to-boundary map constructing boundary local operators from bulk fields. By assumption, our theory will look free at the boundary of spacetime, so we start with a free theory for ease of manipulation below, with the understanding that expressions hold for interacting fields near the boundary of spacetime.

Consider a free massless scalar\footnote{The following discussion has a straightforward generalization to arbitrary spin-$s$ fields, but the scalar case is sufficient to see the structure of the argument without adding complications, see \cite{Donnay:2022sdg}.} in (3+1)D Minkowski spacetime:
\be
S = -\frac{1}{2}\int d^4 x \, \p_\mu\phi \p^\mu \phi\,.
\ee
An on-shell field can be expanded in a basis of plane waves
\be
	\phi(x)=\int\frac{d^3p}{(2\pi)^3}\frac{1}{2p^0}[a_p e^{ip\cdot x}+a_p^\dagger e^{-ip\cdot x}]\,, \label{eq:scalarModeExp}
\ee
where $a_p^\dagger$ acts on the vacuum to create the one-particle plane wave state $\ket{p}$. While the plane wave basis may be most familiar, we can use an alternative choice of expansion for the field $\phi$, with different wavefunctions and creation operators. In particular, given: a bulk field $\phi(x)$; a wavefunction $\varphi(x;y)$ satisfying the Klein-Gordon equation; and an inner product $(\,\cdot\,,\,\cdot\,)_\Sigma$ on the space of single particle wavefunctions on the Cauchy surface $\Sigma$, we can define the operator $\O^\epsilon_{\Sigma}$ on $\Sigma$ by
\begin{equation}\label{eq:3Dmu}
	\O^{\epsilon}_\Sigma(y) := (\phi(x),\varphi(x_{-\epsilon};y)^*)_\Sigma \,.
\end{equation}
The positive and negative frequency mode operators are selected by the $i\epsilon$ prescription $x^0_\pm=x^0\mp i\epsilon$. The key to this construction and producing alternatives to the plane-wave basis is in understanding the alternative spaces of wavefunctions. 

To see \eqref{eq:3Dmu} in action, let's recover the standard momentum space basis. For the case of a bulk scalar operator our inner product is just the Klein-Gordon inner product. Using coordinates $X^i$ for the surface $\Sigma$ and $X^0$ for the normal direction, the Klein-Gordon inner product takes the form:
\begin{equation}
	\left( \phi_1, \phi_2 \right)_\Sigma = -i \int_{\Sigma} d^3X^i \, [\phi_1(X) \partial_{X^0} \phi_2^*(X) - \partial_{X^0} \phi_1(X) \phi_2^*(X)]\,.
\end{equation}
A complete set of wavefunctions satisfying the free scalar equation of motion are plane waves of definite momentum $\varphi(x;p)= e^{i x\cdot p}$. The Klein-Gordon inner product between plane waves is:
\begin{equation}
	(e^{i x\cdot p}, e^{i x \cdot p'})_{\Sigma} =  (2\pi)^{3} 2E_p \delta^{3}(p-p^\prime)\,,
\end{equation}
and thus we recover $a_p^\dagger$ as
\begin{equation}
	(\phi(x), e^{i x\cdot p})_{\Sigma} = a^{\dagger}_p\,.
\end{equation}

\paragraph{Pushing Bulk Fields to the Boundary} An operator on the null boundary is defined by taking $\Sigma=\cal{I}^\epsilon$ and should act on the vacuum to create in/out states, even in the weakly interacting theory. Let us focus on the out states created at $\scri^+$ for concreteness. To describe  $\calI^+$, it is convenient to switch to coordinates
\be
	x^\mu=\frac{1}{2}\Big(u+r(1+z\bz), r(z+\bz), -ir(z-\bz), -u+r(1-z\bz)\Big)\,,
\label{equ:Xmu-Bondi}
\ee
in which case the flat Minkowski metric becomes
\be
ds^2=-\,dudr+r^2\, dzd\bz\,.
\label{equ:Minkmetric}
\ee
Here $z,\bar{z}\in\bbC$ parametrizes the celestial sphere (minus a point at infinity). In these coordinates, the surface $\Sigma = \calI^+$ is obtained by taking $r\to\infty$.  

We can parametrize an on-shell massless four momentum by
\be\label{qi}
k^\mu= \omega q^\mu(w,\bw)\,,\quad 
q^\mu=(1+w\bw,w+\bw,i(\bw-w),1-w\bw)\,.
\ee
In taking the limit of \eqref{eq:scalarModeExp} as $r\to\infty$, a saddle point approximation reduces the three dimensional integral to one that is tangent to the null boundary.\footnote{But see also \cite{Jorstad:2024yzm}.} The limit also identifies the momentum $(w,\bw)$ and position $(z,\bz)$ celestial spheres. In the end, we obtain the boundary field operator
\begin{align}
	\Phi^{\mathrm{out}}(u,z,\bar{z}) 
	&=        \lim_{r\rightarrow\infty}r\,\phi(u,r,z,\bz)\label{eq:hatA}\\
	&\hphantom{:}=-\frac{i}{(2\pi)^2} \int_0^\infty d\omega \left[b(\omega,z,\bz)e^{-i\omega (1+z\bz) u}-b^\dagger(\omega,z,\bz)e^{i\omega (1+z\bz) u}\right]\,,
\end{align}
described in \eqref{freedat}. Note the power of $r$ is consistent with the expansion in~\eqref{freedat} above and the modes capture the free radiative data of the massless scalar field $\phi$. As discussed in~\cite{Jorstad:2023ajr}, this power is more generally related to the bulk scaling dimension $r^{\Delta_{4D}-s}$, which matches the radiative falloffs corresponding to the free theory when the bulk is weakly coupled. For in states, one can repeat the above discussion with the advanced time coordinate $v=t+r$ in place of $u$. These boundary operators $\Phi^{\mathrm{in}/\mathrm{out}}$, defined by pushing bulk fields to infinity and rescaling, will define Carrollian primaries in Section \ref{sec:Carrollian}, and most naturally mimic the relationship between fields and boundary operators in the AdS/CFT correspondence.

\begin{figure}
	\centering
	\begin{minipage}[c]{.45\textwidth}
		\centering
		\begin{tikzpicture}[scale=3, baseline=(current bounding box.center)]
			\draw[thick](0,0) -- (1,1) node[right] {$i^0$} --(0,2)node[above] {$i^+$} --(-1,1) --(0,0)  node[below] {$i^-$} ;
			\draw[ultra thick,blue] (0+.0,0+.0) -- (-.9+.0,1+.05);
			\draw[ultra thick,red] (0-.0,2-.0) -- (1-.1,1-0.05);
			\node at (1/2+.2,3/2+.2) {$\cal{I}^+$};
			\node at (1/2+.2,1/2-.2) {$\cal{I}^-$};
			\draw[ultra thick, red] (0-.0,2-.0) -- (.5-.0,.8+0.1);
			\draw[ultra thick, red] (0-.0,2-.0) -- (.4-.0,.9);
			\draw[ultra thick, blue] (0+.0,0+.0) -- (-.5+.0,.8+.11);
			\draw (0,1) ellipse (1cm and .11cm);
		\end{tikzpicture}
	\end{minipage}
	\hfill
	\begin{minipage}[c]{.45\textwidth}
		\centering
		\begin{tikzpicture}[scale=1, baseline=(current bounding box.center)]
			\draw[thick] (0,0) circle (2cm);
			\draw[fill,blue] (-.3,1) circle (.2em);
			\draw[fill,blue] (-1-.2,0) circle (.2em);
			\draw[fill,red] (0.2,1) circle (.2em);
			\draw[fill,red] (1,-.5+.1) circle (.15em);
			\draw[fill,red] (1-.1,-.6) circle (.15em);
			\draw[red,dashed] (1-.05,-.5) circle (.65em);
		\end{tikzpicture}
	\end{minipage}
	\caption{Massless momentum and celestial scattering states are prepared with operators smeared along generators of null infinity with the appropriate kernel for the $u$ (resp. $v$)-integral. Here we will be concerned with the collinear limit of operators smeared along generators of $\mathcal{I}^+$ (shown in red) viewed from the bulk and (dimensionally reduced) boundary perspective.
		\label{in_out_states}}
\end{figure}
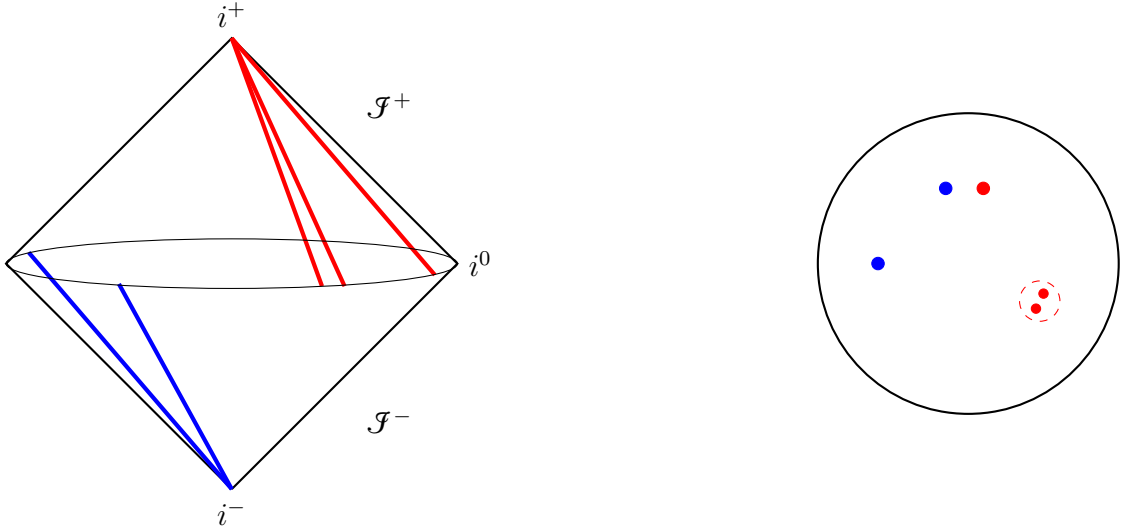

The field $\Phi^{\mathrm{out}}$ encodes the asymptotic behaviour of $\phi$ at $\mathcal{I}^+$ and creates a superposition of single particle out-states smeared in energy $\omega$, all with momenta pointing towards the $(z,\bar{z})$ direction on the celestial sphere. Phrasing it in the reverse, massless momentum eigenstates with definite energy $\omega$ and fixed direction $(z,\bz)$ on the celestial sphere are prepared by operators that are local on the celestial sphere, but smeared along the generator $u$ of null infinity
\begin{equation}
    \ket{\omega,z,\bz}_{\mathrm{out}} = \int_{-\infty}^{\infty} du\, e^{-iu\omega} \,\Phi^{\mathrm{out}}(u,z,\bz)\ket{0}\,.
\end{equation}
This is illustrated in Figure \ref{in_out_states}.\footnote{We might question the validity of taking the Cauchy surface $\Sigma$ to null infinity in \eqref{eq:3Dmu}. At least classically, the $\phi$ phase space consists of points $(\phi, n^\mu \nabla_\mu \phi)$ on a spacelike initial data Cauchy surface $\Sigma$. For massless theories in Minkowski spacetime (but not general asymptotically flat spacetimes, see Footnote 19 of \cite{Prabhu:2022zcr} and also \cite{Geroch:1978us, Tjoa:2022mly}), $\I^{-}$ is also a good initial data Cauchy surface, with initial data specified by $(\Phi, \p_u\Phi)$. In this case, the algebra of boundary observables obtained from quantizing classical observables on $\calI^-$ is ``as good'' for scattering as the algebra of observables obtained from quantizing the classical bulk observables for describing physics at infinity \cite{Ashtekar:1987tt, Ashtekar:1981sf, Prabhu:2022zcr}. Note that the derivative normal to the spacelike Cauchy surface becomes a derivative along the null-direction of $\I^-$, which will appear again in discussing null-derivatives of Carrollian primaries in \eqref{carprim2}.}

\paragraph{Celestial Primaries} An alternative to Carrollian primaries are Celestial Conformal Field Theory (CCFT) primaries. To define a CCFT primary from the bulk scalar field, we must use a scalar conformal primary wavefunction $\varphi_{\Delta, J}(x;z,\bar{z})$. The wavefunction is called ``scalar'' because it satisfies the free Klein-Gordon equation for a bulk field in the $(0,0)$-rep of the Lorentz group; while the $R = (\Delta, J)$ indices indicate that it transforms as an $R$-primary under $SL(2,\bbC)$. For a $J=0$ CCFT primary the wavefunction is
\begin{equation}
	\varphi_{\Delta}(x;z,\bar{z})=\frac{1}{(-q(z,\bar{z})\cdot x)^\Delta}\,.
\end{equation}

The Lorentz invariance of the Klein-Gordon inner product guarantees that $\calO^\epsilon_R$ in \eqref{eq:3Dmu} transforms appropriately under the specified boundary representation of $SL(2,\bbC)$. In particular, the natural outgoing scalar boost eigenstates take the form
\begin{align}
	{\cal O}^+_\Delta(z,\bz)
	&=(\phi(x),\varphi_\Delta(x_{-};z,\bar{z}))_{\mathcal{I}^+}\\
	&=\Gamma[\Delta]\lim\limits_{r\rightarrow\infty}\int du (u-i\varepsilon)^{-\Delta}  \phi(u,r,z,\bz)\,.
	\label{uDelta}
\end{align}
Above we use standard CCFT conventions and write $\Delta$ as a subscript on the CCFT primary even though it can be thought of as a coordinate in this discussion. Analogous operators $\cal{O}_\Delta^-$ defined on $\mathcal{I}^-$ prepare incoming celestial states. We can also write the celestial basis in terms of the momentum space modes
\begin{align}\label{eq:CCFTOp1}
	\calO_{\Delta}^{+}(z,\bar{z}) 
	&= \int_0^\infty d\omega\,\omega^{\Delta-1} a_p^{\dagger}\,,\\
	\calO_{\Delta}^{-}(z,\bar{z}) 
	&= \int_0^\infty d\omega\,\omega^{\Delta-1} b_p\,.\label{eq:CCFTOp2}
\end{align}
We will more carefully examine the construction of these one-particle states and conformal primary wavefunctions in Section \ref{sec:Celestial}.

\paragraph{Trading Bases} By construction, all of these different descriptions in terms of momentum modes $b^\dagger(\omega,z,\bz)$, Carrollian operators $\Phi^{\mathrm{out}}(u,z,\bz)$, and Celestial operators $\calO_{\Delta}(z,\bar{z})$, are just creation operators for different bases of ($\delta$-normalizable) in/out states in the Hilbert space, and are all related by integral transforms. The three choices can be thought of as being parametrized by $\bbR_b \times \mathbb{CP}^1$, corresponding to the momentum $b = \omega$, position $b = u$, and conformal bases $b = \Delta$ respectively, see \cite{Donnay:2022sdg} for more details. For example, the fact that we can trade between the momentum and position bases with a 1d Fourier transform implicitly underpins the ``Soft Theorem = Ward Identity'' results of \cite{He:2014laa, Kapec:2014opa}. 

\subsection{Multiparticle States and Conglomeration}\label{sec:Conglom}
In the previous sections, we related the usual in/out creation and annihilation operators to ``primary'' operators in a Carrollian Conformal Field Theory (CarCFT) and a Celestial Conformal Field Theory (CCFT) by the extrapolate dictionary.\footnote{We will give the algebraic definition of CarCFT and CCFT primaries in the subsequent sections.} For example, a massless momentum eigenstate can be written as
\begin{equation}
    \ket{\omega,z,\bz}\ins = \int du \, e^{-i\omega u}\, \Phi^{\mathrm{in}}(u,z,\bz) \ket{0} = \int_{1-i\infty}^{1+i\infty} \frac{d\Delta}{2\pi i} \, \omega^{-\Delta} \calO_{\Delta}^+(z,\bz) \ket{0}\,,
\end{equation}
with similar integral transforms for the other creation and annihilation operators.

However, the Carrollian $\Phi(u,z,\bz)$ and Celestial $\calO_{\Delta}(z,\bz)$ operators defined above only create weighted linear-combinations of \textit{one-particle} states when acting on the vacuum. Of course, the theory has more than just one-particle states, it also has multiparticle states $\ket{p_m,\dots, p_1}\ins := a_{p_m}^\dagger \cdots a_{p_1}^\dagger \ket{0}$ (we focus on two-particle states for simplicity). A straightforward description of two-particle states follows immediately from the integral transforms, and a usual momentum-basis two-particle state is created by a product of $\Phi$ or $\calO$ acting on the vacuum. For example, in the celestial case
\begin{equation}\label{p12delta}
    \ket{p_1, p_2}\ins = a_{p_2}^\dagger a_{p_1}^\dagger \ket{0} = \int_{\mathrm{PCS}} \frac{d\Delta_2}{2\pi i} \frac{d\Delta_1}{2\pi i} \, \omega_1^{-\Delta_1} \omega_2^{-\Delta_2}\,\calO_{\Delta_2}^+(z_2,\bz_2) \calO_{\Delta_1}^+(z_1,\bz_1) \ket{0}\,, 
\end{equation}
so multiple $\Phi$ or $\calO$ acting on the vacuum should create multiparticle states, but for particles with well-defined Carrollian or Celestial weights rather than momentum, e.g.
\begin{equation}
    \ket{\Delta_1,z_1,\bz_1, \Delta_2, z_2, \bz_2}\ins := \calO^+_{\Delta_1}(z_1,\bz_1) \calO^+_{\Delta_2}(z_2,\bz_2) \ket{0}\,.
\end{equation}

Understanding an analog of these states in the Carrollian and Celestial descriptions is important for completeness, but also because it naturally introduces massive particles into the discussion. Two massless particles have an effective mass in the coupled basis
\begin{equation}\label{eq:PP}
    P^2\ket{p_1,p_2} = 2 p_1 \cdot p_2 \ket{p_1,p_2} = -4\omega_1\omega_2 \abs{z_1-z_2}^2\ket{p_1,p_2}\,,
\end{equation}
with a larger mass based on their separation on the celestial sphere. i.e. only in the collinear or soft limit is the resulting product state massless. In the previous section, our discussion of the extrapolate dictionary focused mainly on radiative data reaching null-infinity $\scri^{\pm}$, so it is instructive to push this issue to better understand holographic descriptions of massive scattering. We discuss these points in more depth Sections \ref{sec:MultiCarroll} and \ref{sec:multOPE}-\ref{sec:Weinberg}.

In order to push these issues, let us recall what happens in the usual AdS/CFT picture (see e.g. \cite{Heemskerk:2009pn, Penedones:2010ue, Fitzpatrick:2011jn, Fitzpatrick:2011hu, Fitzpatrick:2011dm}). In AdS/CFT, when the bulk theory is free, the boundary theory is Mean Field Theory (MFT). Suppose that $\calO$ is the dual primary operator to the bulk free field $\phi$, so that $\calO$ acting on the vacuum creates some (generically spatially delocalized) excitation of the bulk field $\phi$. One can construct families of ``double trace'' primaries $[\calO\calO]_{n,\ell}$ which create two-particle states in the AdS bulk and have dimension $2\Delta+2n+\ell$ and spin $\ell$. In MFT, these operators are just normal ordered products of operators with their derivatives, and in bulk language, these primaries are dual to composites built from two $\phi$'s and derivatives. In theories which are approximately free in the bulk, one can still define composite operators and these double trace operators, although the operators collect anomalous dimensions.\footnote{Indeed, one of the most important results of the conformal bootstrap in $d>2$ dimensions is that limits of this classical intuition survive in any CFT: Given operators with twist $\tau_1 = \Delta_1-\ell_1$ and $\tau_2 = \Delta_2-\ell_2$, then for each fixed $n = 0,1,2,\dots$, there exists an infinite family of primaries $[\calO_1\calO_2]_{n,\ell}$ with increasing spin $\ell$ and twists approaching $\tau_1+\tau_2+2n$ as $\ell \to \infty$.} The bulk $\calS$-matrix can then be viewed as the matrix of anomalous dimensions (at least to first order in perturbation theory)~\cite{Fitzpatrick:2011dm}. 

One way to think about these double-trace primaries in the MFT (which loosens as one leaves MFT) is that they are the primaries that appear on the right hand side of the OPE
\begin{equation}\label{eq:CFTOPE}
    \calO(x_1) \calO(x_2)\ket{0} = \sum_k C(x_{12},\partial_{2})[\calO\calO]_k(x_2) \ket{0}\,.
\end{equation}
In this case, it is completely obvious that $[\calO\calO]_k$ exist and create two-particle states when acting on the vacuum. To ``extract'' these two-particle operators from the OPE, one integrates them against certain integral kernels in a process called ``conglomeration.'' Let us use this to \textit{define} what we mean by two-particle operators in our (boundary) CFTs of interest, even slightly away from MFT.

Turning back to our bulk flat space theories, we can ask if there are analogs of the multiparticle operators $[\Phi\Phi]$ or $[\calO\calO]$ in the Carrollian and Celestial theories. Based on our expression above, we are asking if there is a notion of OPE in Carrollian and Celestial theories mimicking \eqref{eq:CFTOPE}, with the property that (primary) operators appearing on the right hand side of this expansion act on the vacuum to create two particle states. We explore the Carrollian limit of these composite operators in Section~\ref{sec:MultiCarroll}. Conglomeration is intimately connected to the ``OPE block'' and partial wave decomposition of CFT correlators \cite{SimmonsDuffin:2012uy, Karateev:2018oml}, and as we will see in Section \ref{sec:MultiCelestial}, the OPE block is extremely natural for 2D CCFTs \cite{Czech:2016xec, Gadde:2017sjg, Guevara:2021tvr, Garcia-Sepulveda:2022lga, Chang:2023ttm} and suggest an alternative to the usual celestial OPE. Another route to the celestial multiparticle primaries starting from momentum space partial waves is explored in Section~\ref{sec:Weinberg}.

\section{Carrollian CFT and the Flat Space Limit}\label{sec:Carrollian}
Results from AdS/CFT can be imported to flat space holography in the Carrollian picture by a suitable flat space limit. We start in Section \ref{sec:carrollianPrimary} by defining Carrollian primaries, then track the fate of AdS/CFT primaries in the flat space limit in Section \ref{sec:Casimirs}. We use this to discuss some consequences for Carrollian multiparticle states in Section \ref{sec:MultiCarroll}.  
To prepare us for our discussion of celestial primaries in Section~\ref{sec:Celestial}, we briefly show how the ``dimensional reduction of null infinity'' extends from the single particle case in~\eqref{eq:CCFTOp2} to multiparticle Carrollian primaries in Section \ref{sec:CarToCCFT}. We provide some mathematical background and discuss the (by now standard) ``algebraic'' version of this limit in Appendix \ref{sec:FlatAlg}.

\subsection{Carrollian Primaries}\label{sec:carrollianPrimary}
In order to recast the $\cal S$-matrix in terms of correlators of operators in a  Carrollian theory at null infinity, we would like to identify what the relevant symmetries are and how they should act on these objects. To do so, we start with the conformal killing equation and conformal primary conditions.

\paragraph{Conformal Isometries of Carrollian Spacetime}
Formally, we are interested in a CFT living on a null hypersurface with metric
\be
    ds^2=-0du^2+dzd\bz\,.
\ee
Since the metric is degenerate, we should take care whenever the inverse metric would appear in more familiar contexts. For example, when the metric is invertible, one automatically has
\be\label{eq:liedelta}
\mathcal{L}_\xi (g^{\alpha\mu}g_{\mu\nu})=(\mathcal{L}_\xi g^{\alpha\mu})g_{\mu\nu}+ g^{\alpha\mu}(\mathcal{L}_\xi g_{\mu\nu}) =0
\ee
for any killing vector $\xi$. Since this expression requires both the Killing vector equation for $g_{\mu\nu}$ as well as its inverse $g^{\mu\nu}$, it is not well-defined for degenerate metrics. Instead we consider our null-metric as the $c\rightarrow0$ limit of the metric
\be\label{cmet}
    ds^2=-c^2 du^2+dzd\bz \,.
\ee
In this case, lowering indices gives
\be
\xi^a=(\xi^u,\xi^z,\xi^\bz)~\mapsto \xi_b=g_{ba}\xi^a=(-c^2 \xi^u,\frac{1}{2}\xi^\bz,\frac{1}{2}\xi^z)\,.
\ee

Conformal isometries are diffeomorphisms that leave the metric invariant up to a Weyl rescaling, i.e. ${\cal L}_\xi g=\alpha g$. The Carrollian limit of the usual conformal killing equation is still just the usual conformal killing equation
\be\label{ckv}
	\lim_{c\rightarrow 0}\, [\nabla_\mu\xi_\nu+\nabla_\nu\xi_\mu]=\alpha g_{\mu\nu}\,.
\ee
Meanwhile, \eqref{eq:liedelta} has a leading $c\to 0$ singular term which is extracted by
\be\label{ckvinv}
    \lim_{c\rightarrow 0}\,c^2 \,[\nabla^\mu\xi^\nu+\nabla^\nu\xi^\mu]=-\alpha g^{\mu\nu}\,.
\ee
This can be thought of as encoding the conformal killing equation condition for the ``inverse'' of the degenerate metric. Together, \eqref{ckv} and \eqref{ckvinv} give the following equations:
\be
    \p_u\xi^u=\alpha\,,
    ~~\frac{1}{2}\p_u\xi^z=0\,,
    ~~\frac{1}{2}\p_\bz \xi^z=0\,,
    ~~\p_z \xi^z+\p_\bz \xi^\bz=\alpha\,,
\ee
which are solved by 
\be\label{eq:xiyf}
    \xi^z=Y(z),~~\xi^\bz=\bar Y(\bz),~~~\xi^u=f(z,\bz)+u\alpha\,,
\ee
where $\alpha =\frac{1}{2}(\p Y+\bar \p \bar Y)$ \cite{Donnay:2022aba, Bagchi:2022emh}. This is none other than the BMS algebra with supertranslations $f(z,\bz)$ and superrotations $Y(z)$ and $\bar{Y}(\bz)$.\footnote{By contrast, solving only~\eqref{ckv} would allow for $f(u,z,\bz)$, the so called Carrollian diffeomorphisms of~\cite{Ciambelli:2018wre}. The combination of  \eqref{ckv} and \eqref{ckvinv} can be rephrased in terms of the intrinsic Carrollian structure of a degenerate metric and null normal $(q_{ab},n^a)$ with $n^aq_{ab}=0$. By contrast solving~\eqref{ckv} to all orders $c^2$ before taking the limit is just the global Poincar\'e group \eqref{eq:tr}-\eqref{lorentz}. See Appendix B of~\cite{Bagchi:2022emh} for a compact discussion of these Carroll structures. See also~\cite{deGioia:2023cbd}, where this procedure is phrased in terms of a flat limit of the bulk, for which $c\sim\frac{1}{R}$ is the radius of curvature. We thank Luca Ciambelli and Daniel Weiss for discussions on this point.} 

In (3+1)D, the \textit{global} translations are generated by 
\be\label{eq:tr}
    f_0=1+z\bz\,,\quad 
    f_1=-(z+\bz)\,,\quad
    f_2=-i(\bz-z)\,,\quad
    f_3=-1+z\bz\,,
\ee
and the global Lorentz transformations are generated by 
\be\begin{array}{lll}\label{lorentz}
    Y^z_{12}=iz\,,\quad
        & Y^z_{13}=-\frac{1}{2}(1+z^2)\,,\quad
        & Y^z_{23}=-\frac{i}{2}(1-z^2)\,,\\
   Y^z_{03}=z\,,\quad
        & Y^z_{02}=-\frac{i}{2}(1+z^2)\,,\quad
        & Y^z_{01}=-\frac{1}{2}(1-z^2)\,.
\end{array}
\ee
As we will see in the next section, just as the global $SO(2,3)$ isometries of $\AdS_4$ act as conformal isometries of the boundary $\CFT_3$, the Poincar\'e group is the global part of the conformal isometries of a 3D Carrollian CFT living at null infinity $\mathcal{I}^+$ (see also Appendix \ref{app:BMS}).

\paragraph{Carrollian Primaries} Following~\cite{Donnay:2022aba,Donnay:2022wvx}, we define a Carrollian primary of weight $(k,\bar k)$ living at $\mathcal{I}^+$ to transform under the action of \eqref{eq:xiyf} at $\mathcal{I}^+$ as:
\be\label{carprim}
    \delta_\xi\Phi_{(k,\bar k)}(u,z,\bz)
        := [(f+u\alpha)\p_u+Y\p+k\p Y +\bar Y{\bar\p}+\bar k\bar \p \bar Y ]\Phi_{(k,\bar k)}(u,z,\bz)\,.
\ee
In ``holographic'' setups, where a Carrollian CFT lives at $\mathcal{I}^+$, bulk fields pushed to the boundary by the extrapolate dictionary construction give primary fields in the Carrollian CFT. 

 For example, the scalar extrapolate dictionary operator $\Phi(u,z,\bz)$ from \eqref{eq:hatA} is a Carrollian primary of weight $(\frac{1}{2},\frac{1}{2})$.\footnote{See also (2.51) of \cite{Donnay:2022wvx}, as well  as~\cite{Salzer:2023jqv,Nguyen:2023miw} for more on the extrapolate perspective.} This follows immediately from the global transformation properties of the bulk scalar field. In particular, the bulk vector field
\be
    \badat{3}\label{eq:xiyfbulk}
    \xi(Y,f) 
        &=
        \left(1+\frac{u}{2r}\right)Y\p-\frac{u}{2r}\p^2 Y\bar \p -\frac{1}{2}(u+r)\p Y\p_r +\frac{u}{2}\p Y\p_u+c.c.\\
        &\quad
        +f\p_u-\frac{1}{r}(\bar \p f \p+\p  f \bar \p)+\p\bar\p f\p_r
    \eadat
\ee
acts as an isometry of Minkowski space for \eqref{eq:tr}--\eqref{lorentz} (or an
asymptotic symmetry of asymptotically flat spacetimes for generic $f(z,\bz)$, $Y(z)$, $\bar Y(\bz)$). Meanwhile, the scalar field has leading radiative falloff:
\be\label{eq:phi}
\phi(u,r,z,\bz)=\frac{1}{r}\Phi(u,z,\bz)+ \dots \,.
\ee
Taking the Lie derivative of~\eqref{eq:phi} along~\eqref{eq:xiyfbulk} gives~\eqref{carprim} with Carrollian weight $(k,\bar k)=(\frac{1}{2},\frac{1}{2})$. More generally, the field corresponding to a spin $J$ radiative degree of freedom will have
\be\label{kkbar}
k=\frac{1}{2}(1+J)\,,~~~\bar k=\frac{1}{2}(1-J)\,.
\ee

One special property of these conformal primaries on a null surface is that the descendants along the null generators are also primaries. Namely, \eqref{carprim} implies
\be\badat{3}\label{carprim2}
    \p_u \delta_\xi \Phi_{(k,\bar k)}
        &=[(f+u\alpha)\p_u+Y\p+k\p Y +\bar Y\bar \p+\bar k \bar \p \bar Y ]\p_u\Phi_{(k,\bar k)}+\alpha \p_u \Phi_{(k,\bar k)} \\
        &=[(f+u\alpha)\p_u+Y\p+(k+\tfrac{1}{2})\p Y +\bar Y\bar\p+(\bar k+\tfrac{1}{2})\bar \p \bar Y ]\p_u\Phi_{(k,\bar k)} \,,\\
\eadat\ee
which matches the transformation of a Carrollian primary with shifted weights 
\be 
    k'=k+\frac{1}{2}\,,\quad \bar k' = \bar k+\frac{1}{2}\,.
\ee
We will return to this feature when relating Carrollian and Celestial primaries in Section \ref{sec:CarToCCFT}.

While we will be focusing on the global symmetries, it is natural to make a brief comment about the BMS group here. The Lorentz group preserves the $u=0$ slice of null infinity, and the Carrollian primaries at $u=0$ are just $SL(2,\mathbb{C})$ primaries. Just as the superrotations enhance the Lorentz group to a Virasoro symmetry (see Appendix \ref{app:BMS}), the subleading soft theorem implies that these Carrollian primaries are promoted to Virasoro primaries, with descendants generated by subleading soft graviton insertions.  The rest of the BMS algebra can be generated from the global translations and the superrotations. 
The enhancement from Poincar\'e to BMS extends the transformation law~\eqref{carprim} to arbitrary $f(z,\bz)$ and $Y(z)$, and our single particle states are now BMS primaries.
While the above analysis has been performed for $\mathcal{I}^+$, the upshot of the ``Soft Theorem = Ward Identity'' results of \cite{He:2014laa, Kapec:2014opa} is that there is a diagonal subgroup of $\mathrm{BMS}^+\times \mathrm{BMS}^-$ that is a symmetry of the full $\calS$-matrix. We will focus on the global part in what follows.

\paragraph{Poincar\'e Primary Conditions} Looking at~\eqref{eq:phi} above, we see that by virtue of the falloff in $r$, radial translations will annihilate the operator $\Phi$. Let's phrase this in terms of a Carrollian primary at $u=z=\bz=0$. In this case, it is natural to break the Poincar\'e generators into holomorphic and anti-holomorphic Lorentz generators:
\begin{equation}
\begin{alignedat}{3}
\label{embedding_generators_pm}
	 L_0
	 	&=\frac{1}{2}(M^{12}+i M^{+-}) \, ,~~ 
	 & L_{-1}
	 	&=\frac{1}{2}(-M^{2+}-i M^{1+}) \, ,~~
    & L_1
    	&=\frac{1}{2} (-M^{2-}+i M^{1-})\, ,~~\\
 	\bar L_0
 		&=\frac{1}{2} (-M^{12}+i M^{+-})\, ,~~
 	&\bar L_{-1}
 		&=\frac{1}{2} (M^{2+}-i M^{1+})\, ,~~
   & \bar L_1 
   		&=\frac{1}{2} (M^{2-}+i M^{1-})\, ,~~
\end{alignedat}
\end{equation}
and label the translations as
\be\label{otherP}
	P_{\frac{1}{2},\frac{1}{2}}=P^+\,,~~~
	P_{-\frac{1}{2},-\frac{1}{2}}=P^-\,,~~~
	P_{\frac{1}{2},-\frac{1}{2}}=P^1- iP^2\,,~~~
	P_{-\frac{1}{2},\frac{1}{2}}=P^1+ iP^2\,,
\ee
where we have switched to lightcone coordinates $x^{\pm}=\frac{1}{2}(x^0\pm x^3)$. With this redefinition, the Poincar\'e generators obey the following algebra
\begin{equation}\label{eq:BMSAlg}
\begin{alignedat}{3}
    [L_m,L_n] &=(m-n)L_{m+n}\,,\quad
    &[\bar{L}_m,\bar{L}_n] &=(m-n)\bar{L}_{m+n}\,,\\
    [L_{n},P_{k,l}] &=\left(\tfrac{n}{2}-k\right)P_{k+n,l}\,,\quad
    &[\bar{L}_{n},P_{k,l}] &=\left(\tfrac{n}{2}-l\right)P_{k,l+n}\,.
\end{alignedat}
\end{equation}
Note that these complex linear combinations have the feature that $L_n^\dagger =-\bar L_{n}$ and $P_{a,b}^\dagger=P_{b,a}$. The former makes applying the standard CFT$_2$ radial quantization treatment to CCFT more subtle (though see~\cite{Crawley:2021ivb} for a redefined inner product meant to remedy this).  While taking complex linear combinations of the Poincar\'e generators may seem like an unnecessary detour, rewriting them in terms of the $L_n$ and $P_{k,l}$  makes it easy to extend the global isometries (for $n\in\{-1,0,1\}$ and $k,l\in \{-\frac{1}{2},+\frac{1}{2}\}$) to the full BMS algebra.

With these redefinitions, the Carrollian primary condition \eqref{carprim} and \eqref{eq:tr}--\eqref{lorentz} implies the Carrollian primary at $u=z=\bz=0$ is annihilated by:
\be
\label{prim}
    L_1\,,\,\,
    \bar{L}_1\,,\,\,
    P_{\frac{1}{2},\frac{1}{2}}\,,\,\,
    P_{\frac{1}{2},-\frac{1}{2}}\,,\,\,
    P_{-\frac{1}{2},\frac{1}{2}}\,,
\ee
while
\be\label{l0}
    [L_0,\Phi(0,0,0)] =k\,\Phi(0,0,0)\,,\quad
    [\bar L_0,\Phi(0,0,0)] =\bar k\, \Phi(0,0,0)\,.
\ee
Descendants are generated by $\{ L_{-1},\bar L_{-1}, P_{-\frac{1}{2},-\frac{1}{2}}\}$.\footnote{For the holographic Carrollian operators defined in the previous section, $\Phi(0,0,0)$ acts on $\ket{0}$ and creates a state $\propto \int_0^\infty d\omega \ket{\omega,0,0}$. Naively, this is just some state in a unitary irreducible representation of the Poincar\'e group. Now we realize a slight tension, the global BMS algebra is the Poincar\'e algebra, but the Poincar\'e group has no unitary irreducible highest weight representations. We will revisit these subtleties in Section \ref{sec:SingleParticle}, where the distributional nature of states plays an important role in discussing these as-if highest weight representations.}  Let us now see how this arises from the flat limit of AdS.

\subsection{The Flat Limit of CFT Primary Conditions}\label{sec:Casimirs}
 As studied recently in \cite{PipolodeGioia:2022exe, deGioia:2023cbd, deGioia:2024yne, Alday:2024yyj}, the flat space limit of AdS/CFT can be naturally identified with a Minkowski/CarCFT pair. We can try to use such flat limits to better understand holographic CarCFTs. Specifically, we may ask about the fate of boundary CFT operators in the limiting CarCFT, e.g. what happens to primaries in the limit. We work in (3+1)D for simplicity, but identical manipulations can be performed in general $d$ using the algebra in Appendix \ref{sec:FlatAlg}.

\paragraph{Flat Space from AdS}
Consider the global AdS$_{4}$ metric\footnote{It will be slightly easier to use the round sphere metric in this section. This will not change our identification of the generators annihilating a primary at the north pole. See Footnote 13 of~\cite{Dumitrescu:2015fej} for a discussion of mapping from round to flat celestial spheres in Minkowski space.} 
\be
    ds^2=-f dt^2+\frac{1}{f}dr^2+r^2 d\Omega_{2}^2\,,\quad 
    f=1+\frac{r^2}{l^2}\,. 
\ee This can be written in a Bondi-like gauge
\be\label{roundAdS}
    ds^2=-fdu^2-2dudr+r^2 d\Omega_{2}^2 \,,
\ee
by substituting
\be
    t=u+l \arctan{\frac{r}{l}} \,.
\ee
We see that, at large $r$ and fixed $t$, sending $l\rightarrow\infty$ is a Carrollian limit of the boundary metric where $c=l^{-1}$. Meanwhile, in terms of our $u$ coordinate, the metric near $t=\frac{\pi}{2}l$ neatly matches to null infinity in flat space.

We can lift this story to embedding space $\mathbb{R}^{2,3}$ with metric $\mathrm{diag}(-1,+1,+1,+1,-1)$ and coordinates $X^A$ where $A = 0, 1, \dots, 4=d+1$, via
\be
X^A=l\,(\cosh\rho\sin(t/l),\sinh\rho\, \hat{n},\cosh\rho \cos(t/l))\,.
\ee
Here, $t$ is defined as above, we have introduced $\rho$ via
\be
r=l\sinh\rho\,, 
\ee
and we use stereographic coordinates $z,\bz $ on the celestial sphere so that
\be
\hat n=\left ( \frac{z+\bar{z}}{1+z \bar{z}}, i \frac{\bar{z}-z}{1+z \bar{z}},  \frac{1-z \bar{z}}{1+z \bar{z}}\right)\,.
\ee
Now an operator at $u=z=\bz=0$ is at the point 
\be\label{XA}
X^A=(r, 0,0,r,l)\,. 
\ee
To go to the bulk point configuration where the operator is on the boundary at $t=\frac{\pi}{2}l$, we take $r\rightarrow \infty$. To go to the flat limit, we take $l\rightarrow\infty$ in all of the expressions above. 

\paragraph{Picking a Conformal Frame}
The $SO(2,3)$ generators take a simple form in embedding space (see Appendix \ref{sec:FlatAlg})
\be
J_{AB}=i(X_A\p_B-X_B\p_A)\,.
\ee
We want to identify these generators with the boundary CFT generators $\{\mathbf{D}, \mathbf{K}_i, \mathbf{P}_i, \mathbf{M}_{ij}\}$. 
The point on the conformal boundary corresponding to $u=z=\bz=0$, namely the  $r\rightarrow\infty$ limit of~\eqref{XA}, can be thought of as a ray on the lightcone of embedding space
\be
X^1=X^2=X^-=X^{-1}=0\,.
\ee
The ray is stabilized by the generators 
\be
 J_{+,d+1}\,,\,
 J_{+1}\,,\, 
 J_{+2}\,,\, 
 J_{+-}\,,\, 
 J_{12}\,,\,
 J_{1,d+1}\,,\,
 J_{2,d+1}\,,
\ee
which span the special conformal generators $\mathbf{K}_i$ annihilating the operator, dilatation $\mathbf{D}$, and $SO(1,2)$ boost/rotation generators in the Lorentz subgroup of the boundary CFT$_{3}$. 
Indeed we have suggestively ordered things so that
\begin{equation}\badat{3}\label{eq:kdm}
    \{\mathbf K_0, \mathbf K_1,\mathbf{K}_2\} &\sim
    \{J_{+,d+1},J_{+1},J_{+2}\}\, \\
    \mathbf D &\sim 
   J_{+-}\,\\
    \{\mathbf M_{12},\mathbf M_{10},\mathbf M_{20}\} & \sim
    \{J_{12},J_{1,d+1},J_{2,d+1}\} \,.
    \eadat
\end{equation}
By $\sim$ we mean that the generators are linear combinations of the other set. Meanwhile
\begin{equation}\badat{3}
    \{\mathbf P_0, \mathbf P_1,\mathbf{P}_2\} \sim 
    \{J_{-,d+1},J_{-1},J_{-2}\}\,.
    \eadat
\end{equation}

\paragraph{Taking the Flat Limit}
Now, for the flat limit we want to identify
\be
    M_{\mu\nu} := J_{\mu\nu}\,,\qquad
    P_{\mu} := \frac{1}{l} J_{\mu,d+1}\,,
\ee
so that $M_{\mu\nu}$ and $P_\mu$ will become the appropriate Lorentz transformations and translations in flat space. In the flat limit, the special conformal transformations, defining the CFT primary condition, become:
\begin{equation}
    \{\mathbf K_0, \mathbf K_1,\mathbf{K}_2\} \rightsquigarrow 
    \{\mathbf{K}^{\mathrm{car}},\mathbf{K}_1^{\mathrm{car}},\mathbf{K}_2^{\mathrm{car}}\} \sim \{P_{\frac{1}{2},\frac{1}{2}},L_1,\bar{L}_1\}\,,
\end{equation}
where $\rightsquigarrow$ means the generators on the left become the generators on the right (up to linear combinations) under the \.In\"on\"u-Wigner contraction. 

In this limit there are two interesting $ISO(2)$s that appear. First, the $ISO(2)$ subgroup of the Lorentz group which stabilizes a null-momentum, ie the little group of the four momentum $\omega(1,0,0,1)$, is spanned by
\be
\{J_{+1},J_{+2}, J_{12}\} ~\rightsquigarrow~ \{L_1,\bar L_1, L_0-\bar L_0\}\,.
\ee
We thus see that the statement that there are no continuous spin representations is equivalent to the 2D primary condition, while the (3+1)D helicity is just the 2D CFT spin. This fact, and the fact that the momentum eigenstate is aligned along the reference direction, are enough to imply \eqref{prim} annihilate the state.
Second, the boundary spin $SO(1,2)$ \.In\"on\"u-Wigner contracts to another $ISO(2)$, spanned by 
\be\label{siso2}
\{M_{12}, P_1, P_2 \}~\rightsquigarrow~ \{L_0-\bar L_0, P_{\frac{1}{2},-\frac{1}{2}},P_{-\frac{1}{2},\frac{1}{2}}\}\,.
\ee
In CFT language, this involves the Carrollian boosts 
\begin{equation}
    \{\mathbf M_{01}, \mathbf M_{02}\} \rightsquigarrow 
    \{\mathbf{B}_1^{\mathrm{car}},\mathbf{B}_2^{\mathrm{car}}\} \sim \{P_{\frac{1}{2},-\frac{1}{2}},P_{-\frac{1}{2},\frac{1}{2}}\}\,. 
\end{equation}
Due to $P_{\frac{1}{2},\frac{1}{2}}$ annihilating the state, we see that the Carrollian boosts control the value of $P^2$ (see \eqref{eq:P2} below).
We also note that the CFT time translation generator becomes the Carrollian time translation generator, which is the $\partial_u$ null weight-shifting operator discussed in \eqref{carprim2}:
\begin{equation}\label{eq:timeLimit}
    \mathbf{P}_0 \rightsquigarrow \mathbf{H}^{\mathrm{car}} = P_{-\frac{1}{2},-\frac{1}{2}}\,,
\end{equation}
while the spacelike derivatives become
\begin{equation}
    \{\mathbf{P}_1,\mathbf{P}_2\} \rightsquigarrow \{L_{-1},\bar{L}_{-1}\}\,.
\end{equation}

\paragraph{Deforming Representations} To date, a number of references have thoroughly discussed the intricacies of the group contractions involved in the flat space limit of AdS/CFT, usually at the level of vector fields or the algebra, as in Appendix \ref{sec:FlatAlg}. We would also like to better understand what happens to representations under this contraction. With the identification of generators in the flat limit, we can start looking at Casimirs.

Casimir elements can essentially be used to classify unitary irreducible representations.\footnote{By definition, Casimirs are elements of the center of the universal enveloping algebra, which has dimension $\mathrm{rank}(\mathfrak{g})$ for $\mathfrak{g}$ semisimple. For (split) orthogonal groups they always take the form of a trace $\Tr(J^{2k})$ for $k = 1, \dots, r$, where the trace $\Tr$ is over the Lie algebra indices.} The quadratic Casimir of $\mathfrak{so}(2,d)$ can be written in the various bases considered in Appendix \ref{sec:FlatAlg}:
\begin{align}\label{cas2}
    \calC_2(\mathfrak{so}(2,d))
        &= \frac{1}{2}J_{AB} J^{BA} \\
        &= -\frac{1}{2}M_{\mu\nu}M^{\mu\nu} + l^2 P_\mu P^{\mu}\label{cas22}\\
        &= \mathbf{D}^2+\frac{1}{2}(\mathbf{K}_a \mathbf{P}^a + \mathbf{P}_a \mathbf{K}^a) -\frac{1}{2}\mathbf{M}_{ab}\mathbf{M}^{ab}\,.\label{cas23}
\end{align}
We suppress the quartic Casimir $\calC_4 = J_{AB} J^{BC} J_{CD} J^{DA}$ as it is long and not particularly enlightening. Under the contraction $\mathfrak{so}(2,d) \rightsquigarrow \mathfrak{iso}(1,d)$ the quadratic Casimir deforms to
\begin{align}
    \calC_2(\mathfrak{iso}(1,d))
        &= l^2 P_\mu P^{\mu}\\
        &= -\mathbf{H}\mathbf{K}+\mathbf{B}_i \mathbf{B}^i\,.
\end{align}
Similarly, the quartic Casimir deforms to the quartic Casimir of the Poincar\'e algebra (see e.g. \cite{Bekaert:2006py} for details and relations), which is the square of the Pauli-Lubanski pseudo-vector $W^\mu = \epsilon^{\mu\nu\rho\sigma} P_\nu M_{\rho\sigma}$ in (3+1)D.

The action of the $\mathfrak{so}(2,d)$ CFT generators on a conformal primary $\mathcal{O}_I$ are given in \eqref{eq:Conf1}--\eqref{eq:Conf4}. The quadratic conformal Casimir acts as 
\begin{equation}
    \calC_2(\mathfrak{so}(2,d)) \calO_I = -(\Delta(\Delta-d)+\ell(\ell+d-2))\calO_I \,.
\end{equation}
 Up to normalization of the generators, we see that \eqref{cas23} is consistent with the expected relation
\be\label{deltadis}
\Delta(\Delta-d) = m^2 l^2
\ee
for an extrapolated scalar operator. Note that while the spin/Lorentz part $\mathfrak{so}(1,d) \subset \mathfrak{so}(2,d)$, spanned by $M_{\mu\nu}$, is not deformed in the flat space limit, the subgroup preserving the origin is not the same as the one fixing an operator at $u=z=\bz=0$ on the boundary, spanned by ${\mathbf{M}_{ab}}$. Still, we can use our identification of generators above to relate $\ell$ to the helicity in the massless case.

The references \cite{PipolodeGioia:2022exe, deGioia:2023cbd, deGioia:2024yne, Alday:2024yyj} cover aspects of this flat limit relevant to preparing massless single particle states at null infinity from the bulk point configuration in AdS. A point that arose in those investigations is that identifying the correct scaling dimensions is delicate in the massless limit. From our identification of generators in~\eqref{eq:kdm} we see that by acting on a spinless primary at $t=\frac{\pi}{2}l$ on the boundary of AdS, the $l^2P^2$ term will only contribute an $O(l^0)$ term from commuting $P_{\frac{1}{2},\frac{1}{2}}$ through $P_{-\frac{1}{2},-\frac{1}{2}}$ (equivalently ${\mathbf K}_0$ through ${\mathbf P}_0$). Meanwhile, the Carrollian weight measured by $M_{+-}$ or $\mathbf{D}$ is nonzero, and these two terms combine to give the left hand side of~\eqref{deltadis}.
Looking at that equation, one might naively want to force $\Delta=\{0,d\}$ for a massless field. However, as pointed out in~\cite{deGioia:2024yne} and~\cite{Alday:2024yyj} ,we can still have a field that is massless in the flat limit for any $O(l^0)$ dimension $\Delta$. Indeed, from \eqref{cas23}, the expected Carrollian weight $\Delta=1$ for a massless scalar corresponds to an AdS mass that scales like $1/l$, with $m^2$ negative but above the BF bound. 

It would be interesting to connect this physics-motivated discussion of flat space limits with earlier work in representation theory. The problem of understanding infinitesimal Unitary Irreducible Representations (UIRs) under algebra contractions was already studied in the original work of \.In\"on\"u and Wigner \cite{inonu1953contraction}. The process is generally subtle and can be ambiguous, e.g. depending on how elements are scaled in the contraction. At the level of group representations, the process is partially understood \cite{mickelsson1972contractions, dooley1983contractions, dooley1985contractions, cishahayo1994inonu}. For the case relevant to (3+1)D physics, some partial understanding was obtained in \cite{de1992quantum, el1993phase} and especially \cite{ElGradechi:1992te}. The upshot of \cite{ElGradechi:1992te} is UIRs of AdS with non-zero mass and spin $(m,s)$ should deform to UIRs of Poincar\'e with the same mass and spin $(m,s)$, while the massless states are more complicated. It would be extremely interesting to extend this analysis to massless particles and compare this representation-theoretic treatment to the physics understanding of the flat space limit, but we leave this for future work.

\subsection{Multiparticle States in Carrollian CFT}\label{sec:MultiCarroll}
With our understanding of the primary conditions for operators in the bulk point configuration, we can also track the conglomerated operators of \cite{Fitzpatrick:2011dm} in this limit. We will use the two-particle case as a concrete example. As introduced in Section~\ref{sec:Conglom} above, two particle operators in the MFT are captured by ``double trace'' primaries $[\calO\calO]_{n,\ell}$ with dimension $2\Delta+2n+\ell$ and spin $\ell$. These are constructed by acting with the appropriate differential operators on the constituent single particle primaries so that the result is a 3D primary. 

In \cite{Fitzpatrick:2010zm}, it was shown that states with large $n$, created by acting with $[\calO_1\calO_2]_{n,\ell,J}$ on the vacuum, can be expressed as
\be
|n,\ell,J\rangle=\frac{|2p|^{\frac{d-2}{2}}}{(2\pi)^d\sqrt{2RE}}\int d\hat{p}\,Y_{\ell,J}(\hat p)\int d^dq \, f(q)|q+p\rangle|q-p\rangle\,,
\ee
where $f$ is a Gaussian wavepacket with width $\sqrt{E/R}$ and $n=ER$. i.e. the double trace operators are picking out particular partial waves of the two particle states. We will encounter similar expressions in Section~\ref{sec:Weinberg} when we compose bulk-to-boundary propagators with the momentum space partial waves to construct multiparticle celestial primaries.

Our goal here is to build up local two-particle Carrollian primaries. We will focus on explicit forms of the first few low-lying $n=0,1$ double trace primaries. These will illustrate some subtleties of conglomeration in this limit, as well as the appearance of Carrollian reps that are more general than the representations~\eqref{carprim} with
\be 
k=\frac{1}{2}(1+J)\,,~~~\bar k=\frac{1}{2}(1-J)\,,
\ee
from~\cite{Donnay:2022aba,Donnay:2022wvx}, which were relevant for the single particle massless states.

\paragraph{$\mathbf{n=0}$, $\mathbf{\ell=0}$} The simplest CFT conglomerate operator is just
\be
\left[\mathcal{O}_1 \mathcal{O}_2\right]_{0,0}=\mathcal{O}_1 \mathcal{O}_2\,.
\ee
The Carrollian limit of this composite operator is straightforward, especially in the case of scalar $\calO_1$ and $\calO_2$, since then the rescaling of the time coordinate does not affect the representations. Alternatively, starting directly with spinning or non-spinning Carrollian primaries, the collinear limit defines a composite Carrollian primary  
\be
:\!\Phi_{(k_1,\bar k_1)}\Phi_{(k_2,\bar k_2)}\!: \,=\, {\bf \Phi}_{(k_1+k_2,\bar k_1+\bar k_2)}\,.
\ee
In this case, we already see the spectrum of Carrollian primaries is more general than the representations of the form \eqref{kkbar}. For example the normal ordered product of two operators obeying~\eqref{kkbar} with $k_i+\bar k_i=1$  will have weights $(k',\bar k')$ that add up to $k+\bar k'=2$. Due to the feature~\eqref{carprim2}, additional $u$-derivatives on the operators will also give us composite Carrollian primaries.

As we will show in more detail in Section~\ref{sec:CarToCCFT}, we can build celestial primaries from composite primaries on null infinity before we perform a dimensional reduction to the celestial sphere. Similar products in the celestial basis were examined in~\cite{Ball:2023sdz,Guevara:2024ixn}, where the authors considered multi-collinear limits of amplitudes to extract celestial OPEs to subleading order.

\paragraph{$\mathbf{n=0}$, $\mathbf{\ell=1}$}  Next we turn to composite operators constructed from two scalars and one derivative. In MFT, we find that the combination
\be\label{eq:cong01}
\left[\mathcal{O}_1 \mathcal{O}_2\right]_{0,1; \mu}=\Delta_2\left(\partial_\mu \mathcal{O}_1\right) \mathcal{O}_2-\Delta_1 \mathcal{O}_1\left(\partial_\mu \mathcal{O}_2\right)
\ee
is a primary. Now consider what happens when we contract the boundary metric~\eqref{cmet} to the $c\rightarrow0$ Carrollian limit. While we never need to raise and lower indices, it is natural to examine the time and spacelike components separately. Below, we assume that $\calO_1$ and $\calO_2$ are Poincar\'e primaries, and thus annihilated by the operators
\be
    L_1\,,\,\,
    \bar{L}_1\,,\,\,
    P_{\frac{1}{2},\frac{1}{2}}\,,\,\,
    P_{\frac{1}{2},-\frac{1}{2}}\,,\,\,
    P_{-\frac{1}{2},\frac{1}{2}}\,.
\ee

We start by considering the time component of the conglomerate CFT primary (recall \eqref{eq:timeLimit}). The analog of \eqref{eq:cong01} is
\begin{equation}
    [\calO_1\calO_2]_{0,1;u} := \Delta_2(P_{-\frac{1}{2},-\frac{1}{2}} \mathcal{O}_1) \mathcal{O}_2-\Delta_1 \mathcal{O}_1(P_{-\frac{1}{2},-\frac{1}{2}} \mathcal{O}_2)\,.
\end{equation}
Then, using the algebra~\eqref{eq:BMSAlg}, we find:
\begin{equation}
    \begin{alignedat}{3}
        P_{\frac{1}{2},\frac{1}{2}}[\calO_1\calO_2]_{0,1;u}
            &=0\,,\\
        L_{1}[\calO_1\calO_2]_{0,1;u}
            &=\Delta_2([L_{1},P_{-\frac{1}{2},-\frac{1}{2}}] \mathcal{O}_1) \mathcal{O}_2-\Delta_1 \mathcal{O}_1([L_{1},P_{-\frac{1}{2},-\frac{1}{2}}]\mathcal{O}_2)
                &&=0\,,\\
        {\bar L}_{1}[\calO_1\calO_2]_{0,1;u}
            &=\Delta_2([{\bar L}_{1},P_{-\frac{1}{2},-\frac{1}{2}}] \mathcal{O}_1) \mathcal{O}_2-\Delta_1 \mathcal{O}_1([{\bar L}_{1},P_{-\frac{1}{2},-\frac{1}{2}}]\mathcal{O}_2)
                &&=0\,,
    \end{alignedat}
\end{equation}
so $[\calO_1\calO_2]_{0,1;u}$ satisfies the Carrollian limit of the 3D CFT primary condition. Moreover
\begin{equation}
        P_{-\frac{1}{2},\frac{1}{2}}[\calO_1\calO_2]_{0,1;u} = 0\qquad \text{and}\qquad
        P_{\frac{1}{2},-\frac{1}{2}}[\calO_1\calO_2]_{0,1;u} = 0\,,
\end{equation}
so $[\calO_1\calO_2]_{0,1;u}$ is also a Poincar\'e primary. Under closer inspection, it is not surprising that the $[\calO_1\calO_2]_{0,1;u}$ operator is a Poincar\'e primary because the individual components
\begin{equation}
    (P_{-\frac{1}{2},-\frac{1}{2}}\calO_1)\calO_2\,,\quad
    \calO_1(P_{-\frac{1}{2},-\frac{1}{2}}\calO_2)\,,
\end{equation}
are just $n=0$, $\ell=0$ conglomerate primaries on their own since $P_{-\frac{1}{2},-\frac{1}{2}} = \partial_u$ descendants are primaries.

Now we consider the spacelike components. The 3D CFT operators $\mathbf{P}_1$ and $\mathbf{P}_2$ become $L_{-1}$ and $\bar{L}_{-1}$ in the flat limit (via suitable linear recombinations). As a result, we define
\begin{align}
    [\calO_1\calO_2]_{0,1;z} &:= k_2 (L_{-1}\calO_1)\calO_2 - k_1 \calO_1(L_{-1}\calO_2)\,,\\
    [\calO_1\calO_2]_{0,1;\bz} &:= \bar{k}_2 (\bar{L}_{-1}\calO_1)\calO_2 - \bar{k}_1 \calO_1(\bar{L}_{-1}\calO_2)\,.
\end{align}
In the strict flat limit of~\eqref{eq:cong01} above, we want $k_1=\bar k_1=\frac{\Delta_1}{2}$ and $k_2=\bar k_2=\frac{\Delta_2}{2}$ since we started with an expression for scalar primaries. However, it is instructive to leave the left and right weights independent in the manipulations below, to see what holds without this constraint. Following the ``holomorphic'' conglomerate, we find that
\begin{align}
    P_{\frac{1}{2},\frac{1}{2}}[\calO_1\calO_2]_{0,1;z}
        &=k_2([P_{\frac{1}{2},\frac{1}{2}},L_{-1}] \mathcal{O}_1) \mathcal{O}_2-k_1 \mathcal{O}_1([P_{\frac{1}{2},\frac{1}{2}},L_{-1}]\mathcal{O}_2)
        = 0\,,\\
    L_1[\calO_1\calO_2]_{0,1;z}
        &=k_2([L_1,L_{-1}] \mathcal{O}_1) \mathcal{O}_2-k_1 \mathcal{O}_1([L_1,L_{-1}]\mathcal{O}_2)\nonumber\\
        &= 2(k_2 k_1 \calO_1 \calO_2 - k_1 k_2 \calO_1 \calO_2) = 0\\
    \bar{L}_1[\calO_1\calO_2]_{0,1;z}
        &=k_2([\bar{L}_1,L_{-1}] \mathcal{O}_1) \mathcal{O}_2-k_1 \mathcal{O}_1([\bar{L}_1,L_{-1}]\mathcal{O}_2)
        = 0\,,
\end{align}
so that $[\calO_1\calO_2]_{0,1;z}$ satisfies the Carrollian limit of the 3D primary condition (and similarly for $[\calO_1\calO_2]_{0,1;\bar{z}}$). Note again this would be true with or without the restriction $k_i=\bar k_i$.

However, unlike before, if we consider
\begin{align}
    P_{-\frac{1}{2},\frac{1}{2}}[\calO_1\calO_2]_{0,1;z}
        &=k_2([P_{-\frac{1}{2},\frac{1}{2}},L_{-1}] \mathcal{O}_1) \mathcal{O}_2-k_1 \mathcal{O}_1([P_{-\frac{1}{2},\frac{1}{2}},L_{-1}]\mathcal{O}_2)
        = 0\,,\\
    P_{\frac{1}{2},-\frac{1}{2}}[\calO_1\calO_2]_{0,1;z}
        &=k_2([P_{\frac{1}{2},-\frac{1}{2}},L_{-1}] \mathcal{O}_1) \mathcal{O}_2-k_1 \mathcal{O}_1([P_{\frac{1}{2},-\frac{1}{2}},L_{-1}]\mathcal{O}_2)\nonumber\\
        &= k_2(P_{-\frac{1}{2},-\frac{1}{2}}\mathcal{O}_1) \mathcal{O}_2-k_1 \mathcal{O}_1 (P_{-\frac{1}{2},-\frac{1}{2}}\mathcal{O}_2)\,.
\end{align}
The operator does \textit{not} satisfy the full Poincar\'e primary conditions, with an obstruction proportional to the chiral Carrollian scaling dimensions. However, the last line is proportional to $[\calO_1\calO_2]_{0,1;u}$ for scalars, so the different components are in a non-trivial multiplet of the Carrollian boosts.

We can also compute the action of $L_0$ and $\bar L_0$ for the conglomerates, to find:
\begin{alignat}{3}
    L_0[\calO_1\calO_2]_{0,1;u}
        &= (k_1+k_2+\tfrac{1}{2}) [\calO_1\calO_2]_{0,1;u}\,,\quad
    &\bar{L}_0[\calO_1\calO_2]_{0,1;z}
        &= (\bar{k}_1+\bar{k}_2+\tfrac{1}{2}) [\calO_1\calO_2]_{0,1;z}\,,\\
            L_0[\calO_1\calO_2]_{0,1;z}
        &= (k_1+k_2+1) [\calO_1\calO_2]_{0,1;z}\,,\quad
    &\bar{L}_0[\calO_1\calO_2]_{0,1;z}
        &= (\bar{k}_1+\bar{k}_2) [\calO_1\calO_2]_{0,1;z}\,,\\
    L_0[\calO_1\calO_2]_{0,1;\bz}
        &= (k_1+k_2) [\calO_1\calO_2]_{0,1;\bz}\,,\quad
    &\bar{L}_0[\calO_1\calO_2]_{0,1;\bz}
        &=(\bar{k}_1+\bar{k}_2+1) [\calO_1\calO_2]_{0,1;\bz}\,.
\end{alignat}

By the discussion above, if we start with a complex massless scalar in the bulk, the extrapolate field is $\Phi := \Phi_{(\frac{1}{2},\frac{1}{2})}$, and the Carrollian conglomerate operator
\be
    :\!\bar \Phi \p_u \Phi-\p_u\bar \Phi\Phi\!: 
\ee
is a Carrollian (actually, Poincar\'e) primary of the form~\eqref{carprim} with weights
\be
k=\frac{3}{2}\,, \quad ~~\bar k=\frac{3}{2}\,.
\ee
More generally, we see that the operators
\be
\{
    :\!\bar \Phi\overset{\leftrightarrow}{\p_u} \Phi\!:,
    ~:\!\bar \Phi\overset{\leftrightarrow}{\p}\Phi\!:, 
    ~:\!\bar \Phi\overset{\leftrightarrow}{\bar \p} \Phi\!: \}
\ee
are in a non-trivial multiplet under the Carrollian boosts. As a result, we see that we need more general multiplets under the Carroll boosts~\cite{Bagchi:2019xfx} for the multiparticle operators than the representations \eqref{carprim} considered in~\cite{Donnay:2022aba,Donnay:2022wvx} which capture the single particle fields.

\paragraph{$\mathbf{n=1}$, $\mathbf{\ell=0}$} Now at $n=1$, $\ell=0$ we have the CFT primary
\be\label{O10}
\left[\mathcal{O}_1 \mathcal{O}_2\right]_{1,0}=
\frac{\Delta_1}{2 \Delta_1+2-d}\left(\partial^2 \mathcal{O}_1\right) \mathcal{O}_2-\partial_\mu \mathcal{O}_1 \partial^\mu \mathcal{O}_2+\frac{\Delta_2}{2 \Delta_2+2-d} \mathcal{O}_1\left(\partial^2 \mathcal{O}_2\right)
\ee
where for us $d=3$. We now have terms that scale differently under the Carrollian limit
\be\badat{3}\label{car10prim}
    \left[\mathcal{O}_1 \mathcal{O}_2\right]_{1,0}&=-\frac{1}{c^2}\left[
    \frac{\Delta_1}{2 \Delta_1-1}\left(\partial_u^2 \mathcal{O}_1\right) \mathcal{O}_2-\partial_u\mathcal{O}_1 \partial_u \mathcal{O}_2+\frac{\Delta_2}{2 \Delta_2-1} \mathcal{O}_1\left(\partial_u^2 \mathcal{O}_2\right)\right]\\
    &+2\left[
    \frac{\Delta_1}{2 \Delta_1-1}\left(\p\bar\p \mathcal{O}_1\right) \mathcal{O}_2-\partial\mathcal{O}_1 \bar\p \mathcal{O}_2-\bar\partial\mathcal{O}_1 \p \mathcal{O}_2+\frac{\Delta_2}{2 \Delta_2-1} \mathcal{O}_1\left(\partial\bar \p  \mathcal{O}_2\right)\right].
\eadat\ee
As in the last example, it is instructive to let $k_i$ and $\bar k_i$ be independent during our intermediate manipulations. Since the Carrollian limit will not mix the leading term with another, we expect from the AdS uplift that the first line will be annihilated by $\{P_{\frac12,\frac12},L_1,\bar{L}_1\}$. Indeed, from our discussion of $u$-descendants again, we know that the first line is a Poincar\'e primary with weights
\be
k=k_1+k_2+1, ~~\bar k=\bar{k}_1 + \bar{k}_2 + 1,
\ee
so let us focus on the second line.

Define, for notational ease, the operator
\begin{equation}
    \begin{aligned}
        O(\alpha_1,\alpha_2,\alpha_3,\alpha_4) 
            &:= \alpha_1\, (L_{-1} \bar{L}_{-1} \calO_1) \calO_2 + \alpha_2\, (L_{-1} \calO_1) (\bar{L}_{-1} \calO_2) \\
            &+ \alpha_3\, (\bar{L}_{-1} \calO_1) ({L}_{-1} \calO_2) + \alpha_4\, \calO_1 (L_{-1} \bar{L}_{-1} \calO_2)\,,
    \end{aligned}
\end{equation}
then a straightforward exercise shows that
\begin{align}
    L_1 O(\{\alpha_i\}) 
        &= 2(\alpha_1 k_1+\alpha_3 k_2)\bar{L}_{-1}\calO_1\calO_2 + 2(\alpha_2 k_1 + \alpha_4 k_2)\calO_1 \bar{L}_{-1}\calO_2)\,,\\
    \bar{L}_1 O(\{\alpha_i\}) 
        &= 2(\alpha_1 \bar{k}_1+\alpha_2 \bar{k}_2)L_{-1}\calO_1\calO_2 + 2(\alpha_3 \bar{k}_1 + \alpha_4 \bar{k}_2)\calO_1 L_{-1}\calO_2)\,.
\end{align}
As a result, we see that only (a scalar multiple of) the operator
\begin{equation}
    O(k_2 \bar{k}_2, -\bar{k}_1 k_2, -k_1 \bar{k}_2, k_1 \bar{k}_1)
\end{equation}
is annihilated by $L_1$ and $\bar{L}_1$. A similar exercise shows that
\begin{equation}\label{n0l1}
    P_{\frac{1}{2},\frac{1}{2}}O(\{\alpha_i\}) = \alpha_1\, (P_{-\frac{1}{2},-\frac{1}{2}} \calO_1) \calO_2 + \alpha_4\, \calO_1 (P_{-\frac{1}{2},-\frac{1}{2}}\calO_2)\,.
\end{equation}

By the algebra above, we see that a composite primary built from radiative scalars is \textit{not} annihilated by $P_{\frac{1}{2},\frac{1}{2}}$, but is proportional to the $u$-primary-descendant $P_{-\frac{1}{2},-\frac{1}{2}}[\calO_1 \calO_2]_{0,0}$. i.e. the composite primary above does \textit{not} satisfy the 3D CFT primary conditions inherited by the Carrollian theory. This can be understood in the AdS uplift. There, the translations $P_{\frac{1}{2},\frac{1}{2}}$ and $P_{-\frac{1}{2},-\frac{1}{2}}$ do not commute, but rather give a term proportional to $L_0$ and $\bar L_0$. There will thus be a cancellation between the first and second lines of~\eqref{car10prim} at this subleading order in the Carrollian limit.

\vspace{1em}

Now we see something that should give us pause. In the case of the $n=1$, $\ell=0$ CFT operators, the naive flat $c\to0$ limit of the conglomerated primaries is isolating terms with only $u$ derivatives, and not giving us operators with larger celestial spin. However, the OPE of operators with finite angular separation suggests that our theory should have composite operators ``on the right hand side of the OPE'' with larger celestial spin.
Indeed, there seems to be a tension: as pointed out in~\eqref{eq:PP} above, two finite energy massless particles with finite angular separation have non-trivial $P^2$. Meanwhile, any collinear particles on $\mathcal{I}^+$ will have $P^2=0$. We can see this by writing
\be\badat{3}
:\!\Phi(u,0,0) \Phi(0,0,0)\!: \,\,=\, \sum_n \frac{1}{n!}:\!\p_u^n \Phi(0,0,0) \Phi(0,0,0)\!:
\eadat
\ee
where $\p_u^n \Phi(0,0,0)\propto P_{-\frac{1}{2},-\frac{1}{2}}^n \Phi(0,0,0)$. Then the fact that the momenta commute implies this operator will be annihilated by
\be
    \{P_{\frac{1}{2},\frac{1}{2}},P_{\frac{1}{2},-\frac{1}{2}},P_{-\frac{1}{2},\frac{1}{2}}\}
\ee
and hence
\be\label{eq:P2}
    P^2=-P_{\frac{1}{2},\frac{1}{2}}P_{-\frac{1}{2},-\frac{1}{2}}+P_{\frac{1}{2},-\frac{1}{2}}P_{-\frac{1}{2},\frac{1}{2}}
\ee
will vanish. From the perspective of the previous subsection, we see that the non-trivial mass squared requires a nontrivial representation of the translations of the $ISO(2)$ in~\eqref{siso2}. In the upcoming work~\cite{cmp25} we will see how this can be probed with light-ray operators that end at the bulk point configuration. These operators pass through spatial infinity in the flat limit.

In the following section we will turn to the celestial basis. While these operators are non-local in $u$, the operators on the celestial sphere do not obviously evade the problems we've encountered above. Indeed one manifestation of these subtleties appears when considering the ``usual'' massless OPE expressions in~\cite{Pate:2019lpp} and checks Poincar\'e invariance on both sides. The single particle operators are massless while the two particle combination with finite angular separation is not.\footnote{We thank Yangrui Hu for discussions on this point.} This appears in the manipulations of~\cite{Himwich:2021dau}, where they bootstrap the celestial OPEs using only a chiral half of Poincar\'e. Despite these concerns for purely massless OPEs, the fact that the treatment of massive operators at timelike infinity and massless operators at null infinity can both readily be mapped to the celestial sphere makes it straightforward to transform momentum space partial waves for multiparticle states to a celestial basis. We address this further in Sections \ref{sec:unitarity} and \ref{sec:Weinberg}.

\subsection{From Carrollian to Celestial CFT}\label{sec:CarToCCFT}
As we transition over to a purely Celestial treatment of multiparticle operators let us close this section with a small aside that shows how the dimensional reduction/change of basis story extends beyond the single particle states. Namely, let us directly compare the Carrollian and Celestial bases for any primary that transforms as in~\eqref{carprim}, composite or otherwise (see also \cite{Donnay:2022aba,Donnay:2022wvx, Banerjee:2024yir, Banerjee:2024hvb}). Recall our conventions from Section \ref{sec:carrollianPrimary} where we introduced Carrollian primaries that transformed as in~\eqref{carprim}
\be
    \delta_\xi\Phi_{(k,\bar k)}(u,z,\bz)
        = [(f+u\alpha)\p_u+X_{k,\bar{k}}]\Phi_{(k,\bar k)}(u,z,\bz)\,,
\ee
where $\alpha := \frac{1}{2}(\p Y+\bar \p \bar Y)$ and $X_{k,\bar{k}} := Y\p+k\p Y +\bar Y{\bar\p}+\bar k\bar \p \bar Y$.  Now define the following smeared Carrollian primary along null-infinity
\be\label{dimred}
    Z_{k,\bar{k}}^\nu(z,\bz):=\Gamma[\nu]\int_{-\infty}^\infty du\,  u^{-\nu} \Phi_{(k,\bar k)}(u,z,\bz)\,.
\ee
Varying $Z_{k,\bar{k}}^\nu$, with appropriate $u$-falloffs for $\Phi$, we find
\begin{align}
    \delta_\xi Z_{k,\bar{k}}^\nu 
        &=\Gamma[\nu]\int_{-\infty}^\infty du\,  u^{-\nu} \delta_{\xi}\Phi_{(k,\bar k)}(u,z,\bz)\,\\
        &=f Z_{k,\bar{k}}^{\nu+1}+\left(\alpha \nu-\alpha+X_{k,\bar{k}}\right) Z_{k,\bar{k}}^\nu\\
        &=f Z_{k,\bar{k}}^{\nu+1}+X_{h,\bar{h}} Z_{k,\bar{k}}^\nu\,,
\end{align}
where
\be\label{nuweight}
    h(\nu,k)=\frac{1}{2}(\nu-1+2k)\,,~~\bar h(\nu,\bar{k})=\frac{1}{2}(\nu-1+2\bar k)\,.
\ee
We note that the first term is proportional to the global translations and shifts $\nu \mapsto \nu+1$, which is how we expect translations to act on celestial primaries (see also Section \ref{sec:SingleParticle}). Meanwhile, the second term is the transformation of a conformal primary under the Lorentz group of weight $h(\nu,k)$ and $\bar{h}(\nu,\bar{k})$. Thus, the integral transform $Z_{k,\bar{k}}^\nu$ defines a celestial primary with
\begin{equation}
    \Delta = k+\bar{k}+\nu-1\,,\quad J = k-\bar{k}\,.
\end{equation}
Most importantly, this true for any Carrollian primary, not just single particle operators. As a result, all the conglomerate Carrollian primaries constructed in
Section \ref{sec:MultiCarroll} will define additional new operators on the celestial sphere.

\section{Celestial CFT, Unitarity, and Multiparticle States 
}\label{sec:Celestial}
In Section \ref{sec:ExtrapolateDictionary}, we saw how to construct celestial operators from bulk fields. Specifically, we saw in \eqref{eq:CCFTOp1}--\eqref{eq:CCFTOp2} that celestial operators creating massless states are obtained by a Mellin transform of the usual creation/annihilation operators. Such a transformation also exists for massive states, see e.g. \cite{Pasterski:2017ylz}. In this section, we will revisit the definition of celestial operators from a representation-theoretic point of view (often focusing on massless states for simplicity, although arguments are easily generalized), to better understand multiparticle states in celestial holography.

The goal of Section \ref{sec:HilbertSpaces} is to set up some representation theoretic machinery that we will apply to better understand both the single and multiparticle celestial states in the following sections. We will start by reviewing Wigner's usual construction of the UIRs of the Poincar\'e group and the meaning of ``wavefunctions'' in the expansion of a bulk field. We then repeat the analysis for the Celestial CFT case, discussing the induction from a distinguished subgroup. We also discuss the reduction from Poincar\'e to the Lorentz subgroup and introduce the critical notion of ``channel,'' which plays a role in our construction of multiparticle states. 

In Section~\ref{sec:SingleParticle} we revisit the one-particle states in the celestial basis using our results of Section \ref{sec:HilbertSpaces}. We will start by explicitly working out the reduction/decomposition of Poincar\'e representations into $SL(2,\bbC)$ representations for massless representations $V_{0,\sigma}$. Similar results have appeared very recently in \cite{Iacobacci:2024laa}, and our section offers a complimentary (and compatible) perspective on the issues discussed there. In particular, we revisit the meaning of translation operators acting on celestial primaries $\calO_{\Delta,J}(z,\bz)$. Along the way, we will comment on some subtleties of distributions and highest weight states that have appeared in the literature. 

We then examine two approaches to constructing/identifying multiparticle celestial states. The first employs the OPE block in Section~\ref{sec:multOPE}. In Section \ref{sec:unitarity} we call attention to the (often underappreciated) point that celestial CFTs can be thought of as unitary CFTs in the sense of Euclidean $SO(1,d)$ unitarity. While they are almost never reflection positive, we demonstrate that standard tools of unitary Euclidean CFT representation theory are extremely useful for the objects appearing in CCFT, especially the OPE block perspective on CFT. In Section \ref{sec:MultiCelestial} we use the previously developed tools to highlight the celestial OPE block expansion as the obvious choice of expansion for CCFT from the perspective of 2D (E)CFT, in comparison to the usual Celestial OPE. We use this formalism to define continuous collections of multiparticle states, as well as a discrete series of multiparticle states. We comment on previous work on conformal partial wave and block decompositions, and how issues there can be resolved with the celestial OPE block. Finally, Section~\ref{sec:Weinberg} follows a Weinbergian approach by mapping momentum space multiparticle states to the Celestial basis.

Almost all of the sections require an understanding of $SL(2,\bbC)$ representation theory, of which we give a lightning fast summary in Appendix \ref{sec:SL2CReps}.

\subsection{Some Useful Representation Theory}\label{sec:HilbertSpaces}
The Hilbert space of a unitary theory in Minkowski spacetime should carry some unitary representation of the Poincar\'e group $P := \bbR^4 \rtimes SO^+(1,3)$.\footnote{Whether we should call $\bbR^4 \rtimes SO^+(1,3)$, $\bbR^4 \rtimes SO(1,3)$, $\bbR^4 \rtimes O(1,3)$, etc. ``the Poincar\'e group'' is not critical for our discussion here. It will only add or remove some $\bbZ_2$ signs to the discussion above, so we will mostly ignore it. From here out, we will also refer to $\tP$ as ``the Poincar\'e group'' and no confusion should arise as we will usually always be referring to $\tP$ anyway.\label{footnote:Z2s}} Since we care about projective representations, we will actually study the universal cover
\begin{equation}
    \tP := \bbR^4 \rtimes SL(2,\bbC)\,.
\end{equation}
One particle states correspond to (vectors in) Unitary Irreducible Representations (UIRs) of $\tP$, which are labeled by eigenvalues of the (quadratic and quartic) Casimirs as well as some additional discrete data labeling the ``sign of energy'' (see \cite{Weinberg:1995mt} for the standard treatment for physicists or \cite{rideau1966reduction, hai1969harmonic, hai1971harmonic} for an excellent systematic treatment with an eye towards harmonic analysis).  For example, we expect the Hilbert space of a free bosonic mass $m$ spin-$s$ field to be the Fock space: 
\begin{equation}\label{eq:FreeHilbertSpace}
    \calH = \bbC \oplus V_{m,s} \oplus \mathrm{Sym}^2 (V_{m,s}) \oplus \dots\,.
\end{equation}

The symmetric tensor products of representations above define the ``multiparticle states'', and can be decomposed into direct integrals over ``effective'' one-particle states in the coupled basis. The tensor product of two Poincar\'e UIRs in (3+1)D decomposes as follows \cite{lomont1960decomposition, lomont1967reduction, raczka1986theory}:
    \begin{align}
        V_{0,\sigma_1} \otimes V_{0,\sigma_2}
            &\cong \int^{\oplus}_{M \geq 0} dM\, \bigoplus_{S = \abs{\sigma_1-\sigma_2}}^\infty \tilde{\mu}_{\sigma_1,\sigma_2}(M,S) \,V_{M,S}\,, \label{00lm}\\
        V_{m,s} \otimes V_{0,\sigma} 
            &\cong \int^{\oplus}_{M \geq m} dM\, \bigoplus_{\ell = \sigma}^\infty \bigoplus_{S=\abs{\ell-s}}^{\ell+s} \tilde{\mu}_{m,s,\ell,\sigma}(M,S) \,V_{M,S}\,,\\
        V_{m_1,s_1} \otimes V_{m_2,s_2} 
            &\cong \int^{\oplus}_{M \geq m_1+m_2} \,\lquad\lquad\lquad dM\, \bigoplus_{\ell=0}^\infty \bigoplus_{s = \abs{s_1-s_2}}^{s_1 + s_2} \bigoplus_{S=\abs{\ell-s}}^{\ell+s} \tilde{\mu}_{m_i,s_i,\ell,s}(M,S) \,V_{M,S}\,,\label{slm}
    \end{align}
where we have introduced the direct integral measures $\tilde{\mu}(M,S)$. This is the basis of the partial wave expansion, and will play an important role for our discussion in Section \ref{sec:Weinberg}. A good state with well-defined quantum numbers, whether it be single- or multiparticle, is thus a vector in a UIR $V_{m,s}$. We assume $m^2 \geq 0$, although much of the following discussion can be repeated without this restriction.

To construct the Poincar\'e UIR $V_{m,s}$, a standard strategy is to induce the rep from a subgroup of $\tP$. It is a non-trivial fact that all UIRs of semi-direct product groups can be obtained this way \cite{mackey1951induced, mackey1952induced, mackey1953induced, Mackey1968InducedRO}.\footnote{Generally, given Lie groups $H < G$ and a rep $D:H \to GL(W)$ of $H$, one induces a rep of $G$ on the space of sections $\Gamma(E)$ of the homogeneous vector bundle $G \times_D W \to G/H$. We are suppressing many details, but this roughly presents the states in the representation as $L^2$-normalizable (possibly twisted) $W$-valued wavefunctions on $G/H$ (see e.g. \cite{Farrill_Lectures, Oblak:2016eij} for more details, definitions of inner products, unitarity properties, etc.). Generally, it is far from guaranteed that an induced representation will actually be irreducible.} In the Poincar\'e case, all non-trivial physical UIRs of the Poincar\'e group $V_{m,s}$ are infinite dimensional and are \textit{not} highest weight-representations.

\subsubsection{Wigner's Momentum Basis} 
As discussed above, the UIRs $V_{m,s}$, are the building blocks for a perturbative theory. In Wigner's construction of the UIRs, we induce from a subgroup of the form $H=\bbR^4 \rtimes S$. First, we pick a UIR of $\bbR^4$ (equivalent to picking a reference momentum 4-vector $k$), then we pick a UIR of the stabilizer subgroup $S = \mathrm{Stab}_{SL(2,\bbC)}(k)$, and finally we induce from the combined rep of $H$. For example, for a massive momentum vector $k$, the stabilizer is an $SU(2) \subset SL(2,\bbC)$, and states are labeled by an $SU(2)$ spin and an on-shell momenta in the mass hyperboloid $H_3^+ \cong SL(2,\bbC)/SU(2) \cong G/H$. In inducing representations, we naively expect more representations of the big group $\tP$ as we lift from the subgroup $H$, however, many reps become identified as orbits under the full group action. Thus all momentum vectors on the same mass-shell hyperboloid will lead to the same representation of $\tP$. 

\paragraph{Momentum Space Wavefunctions} From the point of view of induced representations, wavefunctions like $\epsilon^\alpha_I(q)e^{-i p\cdot x}$ or $v^\alpha(\mathbf{p},\sigma) e^{-i p\cdot x}$ appearing in fields are just matrix coefficients, or a $G\times H$-bitensor, converting $H$-rep indices to $G$-rep indices. Similarly, a mass-shell equation like the Klein-Gordon equation $(\partial^2-m^2)\phi(x) = 0$ is a projector onto a particular representation. We will expand on this construction in the next section when we discuss the analogous wavefunctions in CCFT. 

\paragraph{$\delta$-Normalizable States} When we explicitly construct the representation $V_{m,s}$, the basis we choose depends on which operators we want to diagonalize. If we construct $V_{m,s}$ as an induced representation, then different bases are more natural than others. In Wigner's basis, the momenta $P_\mu$ are naturally diagonalized,
\begin{equation}
    P_\mu \ket{p,\sigma} = p_\mu \ket{p,\sigma}\,.
\end{equation}
This basis makes momentum conservation clear. Moreover, since $P^2$ is a Casimir of $H = \bbR^4 \rtimes S$, the connection between the quantum numbers of states in $V_{m,s}$ and the original rep of $H$ is also clear.

Since the operator $P_\mu$ is unbounded, it will only ever be defined on a dense subspace $\calS$ of the Hilbert space $V_{m,s}$. Eigenstates of $P_\mu$ are not actually genuine states in the Hilbert space, but distribution-valued ``states'' in a suitable dual $\calS^\prime$. For example, $P_\mu$ is only truly-defined on the subspace of states
\begin{equation}
    \ket{f} = \int \frac{d^3 \mathbf{p}}{(2\pi)^3} \frac{1}{2E_{\mathbf{p}}} f(\mathbf{p}) \ket{\mathbf{p}}
\end{equation}
such that $E_{\mathbf{p}} f(\mathbf{p})$ is square-integrable (given our choice of measure).\footnote{Recall this occurs even for the quantum mechanics of a free particle on a line. In this case, the Hilbert space is the Hilbert space of $L^2$-normalizable functions on $\bbR$, i.e. $\calH = L^2(\bbR,dx)$. So only a proper $L^2$-normalizable function $f$ on $\bbR$ gives a vector in the Hilbert space. However, there are many interesting unbounded (self-adjoint) operators of interest, like $\hat{x}$, defined by $\hat{x} f(y) = y f(y)$ or $\hat{p} = -i\partial_x$. At most, unbounded operators $\hat\calO$ can only be defined on a dense subspace of $\calH$, so that $\hat{\calO}f$ is not actually a vector in $\calH$ for all $f\in \calH$; this is clearly the case for an operator like $\hat{x}$. A dual problem is that there are many nice functions which are not in the Hilbert space. For example, the plane waves $\psi_k(x) = e^{ikx}$ are of physical interest and are eigenfunctions of $\hat{p}$, but are not $L^2$-normalizable. 

How do we make sense of eigenvectors of $\hat{x}$ and $\hat{p}$? The resolution is that inside of $\calH$ is the Schwartz space $\calS(\bbR)$ of smooth functions which decay faster than any polynomial at infinity, and all of whose derivatives decay faster than polynomials at infinity. $\calS(\bbR)$ famously forms a dense subspace of $L^2(\bbR)$, is mapped to itself under Fourier transforms (because Fourier transforms swap $\partial$ and $x$), and $\hat{x}$ and $\hat{p}$ are clearly both defined on $\calS(\bbR)$. The \textit{topological} dual of $\calS(\bbR)$ is the space of all tempered distributions $\calS'(\bbR)$. Together $\calS \subset \calH \subset \calS'$ form a ``rigged Hilbert space'' or ``Gel'fand Triple,'' the upshot of the construction is that $\calS'$ gives us a large collection of objects which act like ``eigenfunctions'' $\ket{x}$ for the operator $\hat{x}$. The plane waves are of course the usual $\braket{x}{k} = e^{ikx}$.} The fact that $P_\mu$ generates a non-compact subgroup is reflected by the continuous spectrum. 

\paragraph{Multiparticle States} In the Hilbert space of our QFT, multiparticle states are vectors in a (symmetrized or anti-symmetrized) tensor product of the representations $V_{m,s}$. Once we decompose multiparticle states into partial waves using the tensor product decompositions in \eqref{00lm}--\eqref{slm}, the representations in the direct integral will be describable in the same momentum basis as above. This decomposition allows us to apply the celestial changes of bases to the multiparticle states in the theory, as we will discuss further in Section \ref{sec:Weinberg}.

\subsubsection{The Celestial or Lorentz Basis} In the CCFT scenario, we want to study a Lorentz basis for scattering states. These states should diagonalize the $K_3$/dilatation and $J_3$/rotation operators (equivalently, $L_0$ and $\bar{L}_0$). There is a long history of developing such Lorentz bases for the Poincar\'e group, using a basis of $SL(2,\bbC)$ covariant states with definite $K_3$ and $J_3$ eigenvalues (see e.g. \cite{shapiro1956expansion, zastavenko1960integral, Joos:1962qq}). We have provided a review of the representation theory of $SL(2,\bbC)$ and the Gel'fand-Naimark $z$-basis in Appendix \ref{sec:SL2CReps}. In summary, any UIR of $SL(2,\bbC)$ can be realized on $L^2(\bbC)$, and each rep is spanned by the dense set of $\delta$-normalizable ``states'' $\ket{\Delta,J;z,\bar{z}}$. 

\paragraph{Celestial Primaries}
In a Lorentz-invariant QFT, the Hilbert space can be decomposed into representations of the Lorentz group and a state like $\ket{\Delta,J;z,\bar{z}}$ defines the  ``local operator'' $\calO_{\Delta,J}(z,\bar{z})$ for us by
\begin{equation}\label{eq:SL2Cdefn}
    \ket{\Delta,J;z,\bz} = \calO_{\Delta,J}(z,\bz)\ket{0}\,.
\end{equation}
In our case, we will eventually think of $\ket{0}$ as the Poincar\'e invariant vacuum state of the larger Minkowski scattering theory. In Section~\ref{sec:SingleParticle} we will revisit aspects of the single particle states in more detail.

\paragraph{Inducing Boost Eigenstates} In the Wigner basis, we pick a momentum vector (which corresponds to a representation of $\bbR^4$), and induce representations from $\bbR^4$ and a group stabilizing that momentum. The representations are $L^2$-normalizable functions on (the coset spaces) the mass hyperboloid $\mathbb{H}_3^+$ in the massive case, and the forward null cone $\Lambda^+$ in the massless case (with positive energy).

For the celestial/boost basis, we instead pick a null ray and a group stabilizing the null-direction \cite{Banerjee:2018gce}. A null ray can be identified with a representation label for the group of translations preserving a co-dimension 1 lightlike hyperplane, so we construct an induced representation from a group of the form:
\begin{equation}
    S = \bbR^{d-1} \rtimes IGO(d-2)
\end{equation}
where $IGO(d-2) := ISO(d-2)\cdot\bbR^*$ and $\bbR^*$ acts by dilatations. 

Concretely, if we fix the particular reference null-momentum $k^\mu = \omega(1,0,0,1)$ on the north pole of the celestial sphere, the inducing subgroup is generated by (in our conventions) the elements:
\begin{equation}
    P_1,~P_2,~P_0-P_3,~L_0,~\bar{L}_0~,~L_1,~\bar{L}_1 \,.
\end{equation}
We recognize these generators as precisely the ones in \eqref{prim} and \eqref{l0}. Representations in this case are defined on $\mathbb{R} \times \mathbb{P}\Lambda^+$ where $\bbR$ is interpreted as the transverse null direction to the lightlike hyperplane and $\mathbb{P}\Lambda^+$ is the \textit{projective} (forward) null cone, which is identified with the celestial sphere.\footnote{Warning: we are trying to be slightly more careful about the obvious $\bbZ_2$ factors for incoming/outgoing radiation here, a la Footnote \ref{footnote:Z2s}.}

\paragraph{Celestial Wavefunctions} We can also study how these states (or, rather, the operators creating them) are embedded into bulk field operators and the associated celestial primary wavefunctions, as captured classically in \eqref{eq:3Dmu}. 

The celestial primary wavefunctions connect the operators $\calO_{\Delta,J}$ labeled by a UIR of $SL(2,\bbC)$ to the bulk fields $\phi^s(x)$. Recall that we usually take bulk field components $\phi^s$ to transform in a finite-dimensional non-unitary $s:=(j_1,j_2)$-representation of the Lorentz group, but particles transform in a UIR of the Poincar\'e group. If we consider a rep $s$ of the Lorentz group, we can ask how it induces a representation of Poincar\'e: the induced representation is not unitary and not irreducible. Even in the scalar case $s = (0,0)$ (where it coincidentally is unitary), we generically lift to a highly reducible representation by functions $\phi:G/H \to \bbR$ where $G/H \cong \bbR^{1,3}$. To produce a UIR, one uses the bulk equations of motion
\begin{equation}
    (\partial^2 - m^2)\phi(x) = 0\,,
\end{equation}
which we interpret as a projector onto the $(m,0)$ UIR of Poincar\'e in the scalar case here. 

For any fixed particle/UIR of the Poincar\'e group, it is a standard exercise to understand when they can be embedded into different fields \cite{Weinberg:1995mt}. Relativistic wavefunctions are the matrix coefficients which connect the field operator $\phi^s$ to the UIRs of the Poincar\'e group. i.e. they are solutions to the bulk equations of motion which translate the representation theory of the Lorentz group into the representation theory of the Poincar\'e group. 

Typically we pick some basis set and expand the bulk field into wavefunctions whose coefficients can then be canonically quantized. Classically, we ``extract'' the Fourier coefficients $a^\dagger_p$ in front of wavefunctions using the Klein-Gordon inner product, as in \eqref{eq:3Dmu}. Quantum mechanically, this is the overlap
\begin{equation}
    \mel{0}{\phi^s(x)}{p,\sigma}\,.
\end{equation}
As mentioned above, for massless particles, in a momentum basis, the wavefunctions for bulk fields are schematically of the form $\epsilon^\alpha(p) e^{i p \cdot x}$, and ensure that the states have the right mass-energy relation and helicity degrees of freedom to transform appropriately.

By this logic, the defining rule for a spin-$s$ celestial conformal primary wavefunction $\Phi^{s}_{\Delta, J}(x;z,\bz)$ should be that it act as a matrix coefficient intertwining the $SL(2,\bbC)$ action of Lorentz transformations on bulk spacetime coordinates $x \in\bbR^{1,3}$, with the $SL(2,\bbC)$ action on the celestial sphere by fractional linear transformations. Essentially by definition, such an object transforms in the tensor product\footnote{Whether or not it transform in the spin-$s$ representation $D^{s}$, or the opposite representation $D^{s\,\dagger}$ depends on the conventions for the Klein-Gordon inner product.} representation $\pi_{\Delta,J}\otimes D^{s\,\dagger}$ of $SL(2,\bbC)$, i.e. a spin-$s$ primary wavefunction with conformal weights $(\Delta,J)$ transforms like:
\begin{equation}
    \Phi^{s}_{\Delta,J}\left(\Lambda^\mu_\nu x^\nu; \tfrac{a z + b}{c z + d}, \tfrac{\bar{a} \bar{z} + \bar{b}}{\bar{c} \bar{z} + \bar{d}}\right) = (cz+d)^{\Delta+J}(\bar{c}\bar{z}+\bar{d})^{\Delta-J} D^{s}(\Lambda) \Phi^{s}_{\Delta,J}(x^\mu; z, \bar{z})\,.
\end{equation}
The diagonal action is required for coherence of the Lorentz transformation properties of all objects in the inner product \eqref{eq:3Dmu}. It should also satisfy the appropriate bulk equation of motion to ensure that it extracts the correct $(m,s)$-rep of Poincar\'e. Viewing $\Phi^s_{\Delta,J}(x;\,\cdot\,)$ as an $L^2$-function on $\bbC$, a spin-$s$ conformal primary wavefunction transforms (see Appendix \ref{sec:SL2CReps} for conventions):
\begin{equation}
    \pi_{\Delta,J}(g^{-1}) \Phi^{s}_{\Delta,J}(x;z,\bz) = D^{s}(\Lambda)\Phi^{s}_{\Delta,J}(\Lambda^{-1}\cdot x; z,\bz)\,.
\end{equation}
It is know that the principal continuous series of representations appears in CCFT as a complete basis of wavefunctions for input to the Klein-Gordon inner product: i.e. every classical (on-shell) wavepacket can be decomposed as a linear combination of wavepackets in the principal continuous series \cite{Pasterski:2017kqt}.

\paragraph{Decomposing Poincar\'e with $SL(2,\bbC)$} A unitary representation of the Poincar\'e group decomposes only into unitary principal series representations of the Lorentz group $SL(2,\bbC)$ \cite{Joos:1962qq, mukunda1968zero, chakrabarti1968lorentz, ruhl1969convolution, chakrabarti1971lorentz, macdowell1972reduction, cantoni1975structure}. At the level of representations, we can write these decompositions as:
\begin{align}
    V_{0,\sigma} \label{eq:PoincareSL2C}
        &\cong \int^{\oplus}_{\lambda\in\bbR} d\lambda\, \mu_{0,\sigma}(\lambda) W_{1+i\lambda,\sigma}\,,\\
    V_{m,s} \label{eq:PoincareSL2CMass}
        &\cong \int^{\oplus}_{\lambda\in\bbR} d\lambda\, \bigoplus_{J = -s}^{s} \mu_{m,s}(\lambda,J) \, W_{1+i\lambda,J}\,,
\end{align}
for some measures $\mu_{0,\sigma}$ and $\mu_{m,s}$ (which we will just call $\mu_{m,s}$ from here on). 

The measures highlight an important point about multiplicities of representations. For example, we know that the trivial representation $\bbC$ of Poincar\'e decomposes into just the trivial representation $W_{0,0} \cong \bbC$ of $SL(2,\bbC)$. But, for essentially every other representation $V_{m,s}$ of Poincar\'e, the rep $W_{1+i\lambda,J}$ appears with weight $\mu$ in the decomposition. This point is underscored in the Hilbert space of a QFT, which involves a direct sum over many representations of Poincar\'e (even in a free theory, see \eqref{eq:FreeHilbertSpace}), each of which decompose individually over the entire $SL(2,\bbC)$ principal series. As a result, we need a way to deal with these multiplicities.

The finite dimensional analogy is instructive, in which case all $\mu$ become counting measures. Consider the decomposition of an $SO(3)$ vector and singlet under some distinguished $U(1) \subset SO(3)$, then $\mathbf{3} \oplus \mathbf{1} \cong_{\varphi} \mathbf{-1} \oplus \mathbf{0}^2 \oplus \mathbf{1}$. In this case, one needs to add ``channels'' for the two $\mathbf{0}$ reps that appear, and the isomorphism $\varphi$ encodes information about the recombination of $U(1)$ reps into $SO(3)$ reps.\footnote{A simpler (but less interesting) example is to decompose any non-trivial rep of any group into reps of the trivial subgroup.} In the Poincar\'e case, this is even more pronounced as there are (very roughly) a ``continuum'' of channels for each representation, based on the measures $\mu$. For example, for the Hilbert space of a free massive scalar, the isomorphism $\varphi$ encodes that some very particular element of $W_{1,0}$ is part of a ``one-particle state'' in the Lorentz decomposition of $\mathcal{H}$ in \eqref{eq:FreeHilbertSpace}.

From a CCFT point of view, forgetting the data in $\varphi$ altogether is a huge loss of information, equivalent to forgetting Poincar\'e symmetry for only Lorentz symmetry. It is not immediately clear what the best representation of $\varphi$ is, but one piece of data is that we dress each of the conformal primaries with an additional index $\calO^m_{\Delta,J}$ which ``remembers'' that the state $\calO^m_{\Delta,J} \ket{0}$ was part of a Poincar\'e multiplet of mass $m$. Practically speaking, this mass channel index $m$ declares that the corresponding state (recall \eqref{eq:SL2Cdefn})
\begin{equation}
    \calO^m_{\Delta,J}(z,\bz) \ket{0} = \ket{\Delta,J,m;z,\bz}
\end{equation}
 only has non-zero matrix elements
\begin{equation}
    D^{m,\Delta,J}_{m,s} := \braket{p,s}{\Delta,J,m;z,\bz}\,,
\end{equation}
with $\ket{p,s}$ where $p^2 = -m^2$. This will play a very important in role in both our OPE block and partial wave approaches to multiparticle states in Sections \ref{sec:multOPE} and \ref{sec:Weinberg}. The massive vectors also need a discrete index describing the recombination of $SL(2,\bbC)$ $J$'s into Poincar\'e spins. 

As a result, we generally expect that the CFT spectrum of a 2D CCFT contains the entire principal series, and so ``has many operators at each $\Delta$.''Just as in a non-compact CFT.

\subsection{Revisiting Celestial Single Particle States}
\label{sec:SingleParticle}
Let us now apply some of our representation theoretic tools to revisit the single particle states in the celestial basis. The main takeaways are:
\begin{itemize}
    \item The ``state'' $\ket{\Delta,J;z,\bar{z}}$ is never a normalizable state in the Hilbert space, analogous to the state $\ket{x}$ in quantum mechanics, and embeds itself into distributions on the Hilbert space. 
    \item We discuss how, even when the rep $\pi_{\Delta,J}$ is \textit{not} a highest-weight module for $\mathfrak{sl}(2,\bbC)$, the states $\ket{\Delta,J;z,\bar{z}}$ can look like highest-weight states. 
     \item We discuss how translations act on the $z$-basis, showing how the fact that the $P_\mu$ connect different UIRs of Lorentz into a larger Poincar\'e multiplet can be reconciled with the claim that the scattering data is captured by the principal series.
\end{itemize}
The first two points are new to this work and resolve a confusion present in the literature. The third has some overlap with \cite{Iacobacci:2024laa}. Let us now turn to the reduction of Poincar\'e representations into $SL(2,\bbC)$ representations for massless particles.

\paragraph{Massless Particles to $SL(2,\bbC)$} 
We are  now interested in reducing a rep of Poincar\'e to Lorentz, working in the $z$-basis.\footnote{Most earlier work in the literature is done in the (discrete) ``canonical'' $SU(2)$ basis \cite{mukunda1968zero, chakrabarti1968lorentz, chakrabarti1971lorentz} (see (B.4) of \cite{Mizera:2022sln} and \cite{Conrady:2010sx} for additional references in different bases). In \cite{macdowell1972reduction}, MacDowell and Roskies reduce representations of the Poincar\'e group with respect to the Lorentz group in the $z$-basis for massive scalar representations (see also the earlier work of \cite{novozhilov1969representations}). Actually, MacDowell and Roskies go further and compute the general form of the $\calT$-matrix in the $z$-basis and try to find solutions, and also transform scattering amplitudes from momentum space to the $z$-basis via Mellin transforms.  All missing cases were recently understood in \cite{Iacobacci:2024laa}.} 
We will use the massless case as an example.
Start with a standard momentum eigenstate $\ket{p,\sigma}$ ``in'' the rep $V_{0,\sigma}$. The inner product for such Poincar\'e distributional states is:
\begin{equation}
    \braket{p_1,\sigma_1}{p_2,\sigma_2} = (2\pi)^3 (2 E_{\vec{p}_1}) \delta^{(3)}(\vec{p}_1-\vec{p}_2)\delta_{\sigma_1\sigma_2}\,.
\end{equation}
Since we are interested in massless representations' reduction to $SL(2,\bbC)$, we parameterize the null momentum $p$ by
\be
    p^\mu = \omega q^\mu(w,\bar{w})= \omega (1+w\bw,w+\bw,i(\bw-w),1-w\bw)\,,
\ee
so that the Poincar\'e inner product becomes
\begin{equation}
    \braket{\omega_1,z_1,\bar{z}_1,\sigma_1}{\omega_2,z_2,\bar{z}_2,\sigma_2} = \frac{\delta(\omega_1-\omega_2)}{\omega_1}\delta^{(2)}(z_1-z_2)\delta_{\sigma_1\sigma_2}\,.
\end{equation}
Under a Lorentz transform $g\in SL(2,\bbC)$, the states transform as
\begin{equation}\label{eq:LorentzMomentum}
    U(g) \ket{\omega,w,\bar{w},\sigma}
        = \left(\frac{\partial w'}{\partial w}\right)^{+\frac{\sigma}{2}} \left(\frac{\partial \bar w'}{\partial \bar w}\right)^{-\frac{\sigma}{2}} \ket{\omega',w',\bar{w}',\sigma}\,,
\end{equation}
where $\omega \mapsto \omega' = \abs{c w + d}^2 \omega$, and $z \mapsto z' = g\cdot z$ and $\bar{z} \mapsto \bar{z}' = \bar{g}\cdot \bar{z}$ are the usual fractional linear transformations.

Next we use the distributional $z$-basis for the principal series representation $W_{\Delta,J}$ (see Appendix \ref{sec:SL2CReps}). The inner product is
\begin{equation}
    \braket{\Delta_1,J_1;z_1,\bar{z}_1}{\Delta_2,J_2;z_2,\bar{z}_2} = \delta(\Delta_1-\Delta_2)\delta^{(2)}(z_1-z_2)\delta_{J_1J_2}\,.
\end{equation}
Given a matrix $g\in SL(2,\bbC)$, we know that a $z$-basis state transforms as:
\begin{equation}\label{eq:LorentzzBasis}
    U(g) \ket{\Delta,J;z,\bar{z}}
        = \left(\frac{\partial z'}{\partial z}\right)^{\frac{\Delta+J}{2}} \left(\frac{\partial \bar z'}{\partial \bar z}\right)^{\frac{\Delta-J}{2}} \ket{\Delta,J;z',\bar{z}'}\,,
\end{equation}

The matrix element intertwining the representations is
\begin{equation}
    D_{0,\sigma}^{\Delta,J}(\omega,w,\bar{w};z,\bar{z}) = \braket{\omega,w,\bar{w},\sigma}{\Delta,J;z,\bar{z}}
\end{equation}
and so transforms under Lorentz transformations according to \eqref{eq:LorentzMomentum} and \eqref{eq:LorentzzBasis}:
\begin{equation}
    D_{0,\sigma}^{\Delta,J}(\omega,w,\bar{w};z,\bar{z}) 
        = \left(\frac{\partial w'}{\partial w}\right)^{-\frac{\sigma}{2}} \left(\frac{\partial \bar w'}{\partial \bar w}\right)^{+\frac{\sigma}{2}}\left(\frac{\partial z'}{\partial z}\right)^{\frac{\Delta+J}{2}} \left(\frac{\partial \bar z'}{\partial \bar z}\right)^{\frac{\Delta-J}{2}} D_{0,\sigma}^{\Delta,J}(\omega',w',\bar{w}';z',\bar{z}') \,.
\end{equation}
We can use this transformation rule to derive the matrix element $D^{\Delta,J}_{0,\sigma}$. However, there is an ambiguity following from the shadow transform: the overlap with the state $\ket{\Delta,J;z,\bar{z}}$ and the shadow basis state $\ket{\Delta,J;z,\bar{z}}_{sh}$, will both transform appropriately under Lorentz transformations. The two solutions are thus:
\begin{align}
    D_{0,\sigma}^{\Delta,J}(\omega,w,\bar{w};z,\bar{z}) 
        &= \omega^{\Delta-2} \delta^{(2)}(z-w) \delta_{\sigma}^J\,,\label{eq:overlap}\\
    \tilde{D}_{0,\sigma}^{\tilde{\Delta},\tilde{J}}(\omega,w,\bar{w};z,\bar{z}) 
        &= K_{h,\bar{h}}\, \omega^{-\tilde{\Delta}} (z-w)^{-\tilde{\Delta}-\tilde{J}} (\bar{z}-\bar{w})^{-\tilde{\Delta}+\tilde{J}} \delta_{\sigma}^J \,.\label{eq:shadowlap}
\end{align}
One can check that shadow transforming \eqref{eq:overlap} gives \eqref{eq:shadowlap}. In the usual celestial CFT conventions, we define states so that \eqref{eq:overlap} is the matrix element we desire. It is an easy sanity check to ensure that both of these expressions transform appropriately under the $SL(2,\bbC)$ action. Finally, note the $\delta_{\sigma}^J$ terms, which we expect since $J$ and $\sigma$ are the eigenvalue of the same rotation operator $M_{12}$ in bulk Minkowski space and on the celestial CFT.

\paragraph{Highest Weight vs Principal Series}
The unitary principal series representations $\pi_{\Delta,J}$ of $SL(2,\bbC)$ are neither highest-weight nor lowest-weight representations: they are weight-modules for $\mathfrak{sl}(2,\bbC)$ which have no ``primary vector.'' However, when we turn to field theory, we still talk about operators $\mathcal{O}_{\Delta,J}$ creating a state $\ket{\Delta,J}$ when acting on the vacuum $\ket{0}$, and build a module of descendants by acting with raising/lowering operators. This appears to be a contradiction: it looks as if the operator $\mathcal{O}_{\Delta,J}$ creates an $\mathfrak{sl}(2,\bbC)$ Verma module (or Weyl module), with highest-weight state $\ket{\Delta,J}$, embedded inside the irreducible unitary principal series representation $W_{\Delta,J}$. 

Said more colloquially, the tension is as follows: the celestial states look like they are annihilated by $L_1$ and $\bar L_1$ making them highest weight. But principal series representations are not highest weight. So how can these be principal series representations?

We have already seen the resolution to this apparent tension above: the Vermas are not actually inside the representation space $L^2(\bbC)$, but are being embedded in some extended space of distributions $\calS^\prime \supset L^2(\bbC)$ as more general linear functionals on the algebra of operators. Practically speaking, the state created by a single operator
\begin{equation}
    \calO_{\Delta,J}(z,\bar{z})\ket{0}:=\ket{\Delta,J; z,\bar{z}} = \int_{\bbC} d^2w f(w) \ket{\Delta,J;w,\bar{w}}\,,
\end{equation}
has a $\delta$-function matrix coefficient $f(w)=\delta^{(2)}(w-z)$. The exact space of distributional matrix coefficients $\calS'$ that are allowed is a choice that depends on the operators we diagonalize. For this reason, some caution should be used when studying, e.g. the action of Witt-generators
\begin{equation}
    L_n^{(z)} = -z^{n+1}\partial_z
\end{equation}
on unsmeared states. 

Note that this module built from descendants of the delta function is a rep of the Lie algebra, but it won't integrate to a rep of the group without an integrality condition on the weights, which is compatible with our understanding of HWRs of $\mathfrak{sl}(2,\bbC)$ and UIRs of $SL(2,\bbC)$.\footnote{We thank Davide Gaiotto for pointing out this phenomena and for useful discussions on this point.} Generically, an operator that creates a state which diagonalizes a non-compact generator will have to embed itself into distributions \cite{itzykson1969group}.

\paragraph{Translation Operators} The interesting question now is: how do (3+1)D translations act on the $z$-basis? We can anticipate the answer from the physical intuition of the previous sections. A point of 2D CCFT and the celestial holography dictionary is that scattering amplitudes become like Euclidean ``correlation functions'' on the 2D celestial sphere. But from the correspondence with 3D Carrollian CFT, we expect the 2D scaling dimension $\Delta$ to be a simple integral transform ($\calB^{\pm}$-transform) of the null coordinate $u$ on $\scri^\pm$ (see also Section \ref{sec:CarToCCFT}). From this, we anticipate that the derivative along the null direction induces a shift in the scaling dimension, and so, by Lorentz invariance, we expect that all momentum generators will induce a shift $\Delta \to \Delta + 1$ of the $SL(2,\bbC)$ conformal scaling dimension. So the $P_\mu$ connect different UIRs of Lorentz into one large Poincar\'e multiplet. 

More formally, the dense space of vectors in the representations $V_{0,\sigma}$ of Poincar\'e upon which $P_\mu$ are well-defined are those vectors of the form
\begin{equation}
    \ket{f,\sigma} = \int \frac{d^3 \mathbf{p}}{(2\pi)^3} \frac{1}{2E_{\mathbf{p}}} f_\sigma(\mathbf{p}) \ket{\mathbf{p},\sigma}\,,
\end{equation}
where $E_\mathbf{p} f(\mathbf{p})$ is square integrable for $d^3\mathbf{p}/2E_{\mathbf{p}}$. Switching to null-coordinates, we can re-write this equivalently as:
\begin{equation}
    \ket{f,\sigma} = \int d\omega \,d^2w \,\omega\, f_\sigma(\omega,w,\bar{w}) \ket{\omega, w,\bar{w},\sigma}\,.
\end{equation}
In the Lorentz $z$-basis, using the non-shadowed matrix element, these are the functions
\begin{align}
    \tilde{f}_{\sigma}(\Delta,z,\bar{z}) 
        &:= \braket{\Delta,z,\bar{z},\sigma}{f,\sigma}\\
        &= \int d\omega\, d^2w \,\omega\, f_\sigma(\omega,w,\bar{w}) \braket{\Delta, \sigma;z,\bar{z}}{\omega, w,\bar{w},\sigma}\\
        &= \int d\omega \, \omega^{\Delta^*-1} f_\sigma(\omega,z,\bar{z})\,.
\end{align}
In other words, we recover from this bottom-up perspective that momentum operators are well-defined on states whose coefficients $\tilde{f}_{\sigma}(\Delta,z,\bz)$ are Mellin transforms of $L^2$-normalizable wavepackets on the forward null cone $\Lambda^+ \sim \bbR_{+} \times \bbC$.

The analytic properties of Mellin transforms are well-understood \cite{flajolet1995mellin, bertrand1995mellin}. For an integrable function $f_\sigma(\omega,z,\bz)$, the Mellin transform $\tilde{f}_\sigma(\Delta,z,\bz)$ is analytic on a complex-$\Delta$ strip $a \leq \mathrm{Re}(\Delta) \leq b$, controlled by the asymptotic expansion of $f_\sigma$ as $\omega \to 0$ and $\omega \to\infty$ (see also Appendix A of \cite{Guevara:2019ypd}). In particular, for our general class of functions, $\tilde{f}_\sigma$ will be analytic to the right of $a = 1$. The matrix element of $P_\mu$ can be computed
\begin{align}
    \mel{\Delta',J';z',\bz'}{P_\mu}{\Delta,J;z,\bz} 
        &= \int d\omega\, d\omega' \, \omega^{\prime \Delta^{\prime*}-1} \omega^{\Delta-1} \mel{\omega',z',\bz',J'}{P_\mu}{\omega,z,\bz,J}\\
        &= \int d\omega \, \omega^{\Delta-\Delta'} q_\mu(z,\bz) \delta^{(2)}(z'-z)\delta_{J'J}\\
        &= 2\pi \,q_\mu(z,\bz) \delta(\Delta-\Delta'+1) \delta^{(2)}(z'-z)\delta_{J'J}\,,
\end{align}
and indeed we see it looks as if $\Delta' = \Delta + 1$ for a non-zero matrix element.

To repeat the points from the previous sections: the fact that the translation operators map us off the principal series $\Delta$ values is not a problem, not even for unitarity, since the states $\ket{\Delta,J;z,\bz}$ were never in the Hilbert space to begin with. Likewise, $\ket{\Delta+1,J;z,\bz}$ is not in the Hilbert space either, and the $P_\mu$ operator was never defined on such states. $P_\mu$ was only defined on linear combinations whose coefficient wavefunctions admit suitable analytic continuations in a strip and integrability conditions. The state with weight $\Delta+1$ can be interpreted as some new distribution on the space of analytic functions on the complex-$\Delta$ strip. See \cite{Iacobacci:2024laa} for a compatible discussion in terms of finite translations.

\subsection{An OPE Block Approach to Multiparticle Primaries}\label{sec:multOPE}
In this section we will employ the Poincar\'e representation decomposition introduced in Section~\ref{sec:HilbertSpaces} to the OPE block formulation of CCFT. We start by reviewing this formalism and recalling the notion of ``unitary Euclidean CFTs.''

\subsubsection{Unitary Euclidean CFTs}\label{sec:unitarity}
In any holographic correspondence we expect the bulk and boundary Hilbert space to be literally the same space. If CCFT is holographic correspondence, then we expect this to remain the case, and indeed it is: the celestial description is just a change of basis from plane-waves to conformal primary wavefunctions. 

This also guarantees that the CCFT Hilbert space is unitary. In other words, the CCFT has a natural vacuum vector $\ket{0}$, vector space of states equipped with a unitary inner product and conjugation $\dagger$ operation, and even normal ordering with respect to the Poincar\'e vacuum $\ket{0}$. We will call this the ``Poincar\'e quantization'' of the CCFT. Recalling \eqref{eq:CCFTOp1} and \eqref{eq:CCFTOp2}:
\begin{equation}
	\calO_{\Delta}^{+}(z,\bar{z}) 
	= \int_0^\infty d\omega\,\omega^{\Delta-1} a_p^{\dagger}\,,\qquad
	\calO_{\Delta}^{-}(z,\bar{z}) 
	= \int_0^\infty d\omega\,\omega^{\Delta-1} b_p\,,
\end{equation}
we see that Poincar\'e normal ordering in CCFT moves all $\calO^+_{\Delta}$ and $\calO^{-\dagger}_{\Delta}$ to the left, while time-ordering moves all $\calO_{\Delta}^{+}$ and $\calO_{\Delta}^{+\dagger}$ to the left. This points to potentially interesting connections to the out-of-time ordered ``correlators,''\footnote{The quotations are to emphasize that we do not explicitly study the example of bulk dynamical gravity, and that the conformally covariant boundary data thus do not necessarily form correlation functions of a local QFT with a conserved stress tensor.} studied in \cite{Caron-Huot:2023vxl}.

\paragraph{CCFTs are Euclidean Unitary CFTs} So what is meant when it is said that Celestial CFTs are not unitary? When we call a $(d-1)+1$-dimensional Lorentzian CFT, with conformal symmetry group $SO(d,2)$, ``unitary,'' we mean that only unitary representations of $SO(d,2)$ appear in the spectrum. On the other hand, when we call a $d$-dimensional Euclidean CFT, with conformal symmetry group $SO(d+1,1)$, ``unitary,'' we often \textit{do not} mean that the representations appearing in the spectrum correspond to unitary representations of $SO(d+1,1)$. We often mean that if we analytically continue to $SO(d,2)$, that the representations appearing in the spectrum will be unitary reps of $SO(d,2)$. For this reason, it is clearer to call such a Euclidean CFT ``reflection positive'' or ``physical'' following \cite{Gadde:2017sjg}. Hence there is no contradiction with the holographic expectation nor the mathematics of the reduction from Poincar\'e to the Lorentz group: \textit{the Poincar\'e quantization of Celestial CFTs gives a natural unitary inner product, and the Hilbert space decomposes entirely into unitary representations of the Euclidean conformal group. But they are not ``reflection positive'' 2D CFTs, i.e. unitary when Wick-rotated.}

\paragraph{Tensoring Principal Series Representations} Let us turn away from CCFT specifically for a moment and to general Euclidean conformal representation theory. As explained in \cite{Gadde:2017sjg}, the tensor product of two principal series representations decomposes into principal series representations (as we would guess for compatibility with our understanding of Poincar\'e), and the Clebsch-Gordan Coefficients are kinematic conformal 3-point functions \cite{dobrev1986lecture, Czech:2016xec, Gadde:2017sjg}. That is, given vectors $\ket{R_i;z_i, \bar{z}_i}$, $i=1,2$, in principal series representations $W_{R_{i}} := W_{\Delta_{i},J_{i}}$, the tensor product is
\begin{equation}
    \ket{R_1; z_1,\bar{z}_1} \otimes \ket{R_2; z_2,\bar{z}_2}
        = \int dR \, \int d^2z_P\,  \mathcal{K}^{R_1,R_2\tilde{R}}(\{z_i,\bar{z}_i\};z_P,\bar{z}_P) \ket{R; z_P,\bar{z}_P}\,,
\end{equation}
where the integral over representations is performed with the Plancherel measure (see Appendix \ref{sec:SL2CReps})
\begin{equation}\label{eq:SL2CMeasure}
    \int dR\, := \sum_J \int_{1 - i\infty}^{1+i\infty} d\Delta\, \rho_J(\Delta)\,,
\end{equation}
and the Clebsch-Gordan coefficient is
\begin{equation}\label{eq:Kinematic3Pt}
    \mathcal{K}^{R_1,R_2,R_3}(z_1,z_2,z_3)
        = \frac{\alpha}{
        \abs{z_{12}}^{\Delta_1+\Delta_2-\Delta_3}
        \abs{z_{23}}^{\Delta_2+\Delta_3-\Delta_1}
        \abs{z_{31}}^{\Delta_3+\Delta_1-\Delta_2}}\,.
\end{equation}
In other words, up to normalization, this is just the three-point function $\expval{\calO_1(z_1)\calO_2(z_2)\calO_3(z_3)}$ with the dynamical data $C_{R_1, R_2, R_3}$ stripped off. We will take advantage of this fact below.

\paragraph{The OPE Block} The \textit{OPE block} perspective, described in \cite{Czech:2016xec, Gadde:2017sjg}, suggests that the OPE of a unitary Euclidean CFT is most naturally written 
\begin{equation}\label{eq:OPEBlock}
    \calO_{R_1}(z_1,\bar{z}_1) \calO_{R_2}(z_2,\bar{z}_2)
        = \int dR\, C_{R_1, R_2, R} \int d^2z_P\, \mathcal{K}^{R_1,R_2,\tilde{R}}(z_1,z_2,z_P) \calO_{R}(z_P)\,,
\end{equation}
where $\tilde{R}$ crucially denotes the shadow of the representation $R$. This is in contrast to the typical form of the OPE
\begin{equation}\label{eq:OPETypical}
    \calO_{R_1}(z_1,\bar{z}_1) \calO_{R_2}(z_2,\bar{z}_2)
        = \sum_R C_{R_1,R_2,R} \,C(z_{1P},z_{2P},\partial_P)\calO_R(z_P)\,,
\end{equation}
where $\sum_R$ is a sum over conformal primaries and $C(x,y,\partial)$ is an infinitely long differential operator capturing the sum over descendants of $\calO_R(z_P)$. The expansion in \eqref{eq:OPEBlock} differs from the usual OPE in \eqref{eq:OPETypical} in two independent, but essential, ways:
\begin{enumerate}
    \item Rather than sum over a discrete set of representations, $\sum_R$, we have an integral over the unitary principal series representations. In typical CFTs, the scaling dimensions $\Delta_R$ of the primaries are real and bounded below, corresponding to Lorentzian unitary representations appearing in the spectrum of the theory. In our case, the $\Delta_R$ in the integral are along the principal series of Euclidean unitary representations. This makes the OPE more naturally reflect the tensor product structure of principal series representations (which is less obvious with the usual OPE).
    \item Rather than present the Hilbert space of the CFT as being spanned by the dense collection of operators $\calO_R(z)$ and their descendants \textit{at fixed $z$}, as in standard radial quantization around a point, we have used a basis $\calO_R(z_P)$ which uses \textit{only} primaries $R$, but \textit{smeared over all of $\bbC$}.\footnote{These are sometimes called the \textit{local basis} and \textit{global basis} respectively \cite{Czech:2016xec}, and it is straightforward to confirm that for a fixed $R$ one can recover the usual $C(z_{1P},z_{2P},\partial_P)$ term by term by a careful analysis of the kernel $\mathcal{K}^{R_1, R_2, \tilde{R}}$ and using $\calO_R(z_P) = e^{-i z_P \cdot P} \calO_R(0) e^{i z_P \cdot P}$.} This will be an advantageous in the celestial sphere so that we never have to think of derivatives of celestial primaries $\partial^m \bar{\partial}^n \calO_R(z,\bz)$.
\end{enumerate}

\subsubsection{Multiparticle States and the OPE}\label{sec:MultiCelestial}
The discussions in the previous sections strongly suggest that it is natural, from a 2D CFT point of view, to use the OPE block description in celestial CFTs. Especially since the candidate local primaries naturally live on the principal continuous series and create states in the $z$-basis. As we will see, this also gives us a (potential) connection between the usual celestial OPE and the Euclidean OPE block described above, and equips us with a definition of multiparticle operators mimicking the discussion in Section \ref{sec:Conglom}.

\paragraph{The Celestial OPE Block} Thus far, we have not used any extra special properties of celestial CFTs, like enhanced Poincar\'e invariance. The inclusion of Poincar\'e invariance in the OPE block expansion was performed for non-identical scalars to one massive scalar in Appendix B of \cite{Guevara:2021tvr}, but the arguments are quite general as we will see. Start with the OPE block \eqref{eq:OPEBlock} and absorb the channel indices (described in Section \ref{sec:HilbertSpaces}) into the $R_i$, the OPE block is then 
\begin{equation}\label{eq:celestialOPE}
    \calO_{R_1}(z_1,\bz_1)\calO_{R_2}(z_2,\bz_2) = \int dR\, d^2z_P \, \expval{\calO_{R_1}(z_1,\bz_1)\calO_{R_2}(z_2,\bz_2)\calO_{\tilde{R}}(z_P,\bz_P)} \calO_R(z_P,\bz_P)\,.
\end{equation}
Again, we emphasize that here the measure $dR$ is not just the one in \eqref{eq:SL2CMeasure}, it also has integrals over channel indices. Now we use the translation generators
\begin{align}
    (P_1^\mu + P_2^\mu)&\calO_{R_1}(z_1,\bz_1)\calO_{R_2}(z_2,\bz_2)\nonumber\\ 
        &= \int dR\, d^2z_P \, (\mathcal{D}_{P_1}^\mu + \mathcal{D}_{P_2}^\mu)\expval{\calO_{R_1}(z_1,\bz_1)\calO_{R_2}(z_2,\bz_2)\calO_{\tilde{R}}(z_P,\bz_P)} \calO_R(z_P,\bz_P)\label{eq:PPOPE1}\\
        &= \int dR\, d^2z_P \, \mathcal{D}_{P_P}^\mu\expval{\calO_{R_1}(z_1,\bz_1)\calO_{R_2}(z_2,\bz_2)\calO_{\tilde{R}}(z_P,\bz_P)} \calO_R(z_P,\bz_P)\label{eq:PPOPE2}\\
        &= \int dR\, d^2z_P \, \expval{\calO_{R_1}(z_1,\bz_1)\calO_{R_2}(z_2,\bz_2)\calO_{\tilde{R}}(z_P,\bz_P)} P_P^\mu\calO_R(z_P,\bz_P)\label{eq:PPOPE3}
\end{align}
where we introduced $\mathcal{D}_{P_i}^\mu$ for the differential action on correlators. We pass from \eqref{eq:PPOPE1} to \eqref{eq:PPOPE2} by using the translation Ward identities, and we pass from \eqref{eq:PPOPE2} to \eqref{eq:PPOPE3} by (formal) Hermiticity of $P^\mu$ on the $\calO$. In doing so, we have showed that the OPE block is appropriately Poincar\'e covariant, and dub this the \textit{celestial OPE block} expansion. In summary, the Euclidean OPE block is the celestial OPE block so long as one uses the appropriately Poincar\'e-constrained celestial three point functions for the integral kernel and $P$ is Hermitian.

The Poincar\'e constraints on celestial three-point functions are known from direct integral transforms of amplitudes and by generally considering Ward identities (see e.g. \cite{Lam:2017ofc, Law:2019glh, Chang:2022seh}). Since a generic three-point function which appears in the integral kernel is non-vanishing, the integral $\int dR$ is truly necessary. For example, since $P^2\ket{p_1,p_2} = -4\omega_1\omega_2\abs{z_{12}}^2\ket{p_1,p_2}$, we expect two celestial operators $\calO_{\Delta_1}^0$ and $\calO_{\Delta_2}^0$, creating bulk massless scalars, to couple via the celestial OPE block to all $\calO_{\Delta}^m$ with $m > 0$ and $\Delta \in 1+i\bbR$, as seen in \cite{Guevara:2021tvr}. The two operators can also couple to a third massless scalar with $\delta$-function support in the soft and collinear regions \cite{Chang:2022seh}.

\paragraph{Continuous Two-Particle Primaries} The celestial OPE block decomposition is also compatible with Poincar\'e representation theory and our discussion of multiparticle states in Section \ref{sec:Conglom}. Consider a massless scalar for simplicity, we define a two-particle state as an element of the (symmetrized) tensor product of the Poincar\'e UIRs $V_{0,0}$. Recall the tensor product of \eqref{00lm}:
\begin{equation}\label{eq:PoincareTensor}
    V_{0,0} \otimes V_{0,0} \cong \int_{m\geq 0}^\oplus dm\, \bigoplus_{s=0}^{\infty} 
 \tilde{\mu}(m,s) \, V_{m,s}\,,
\end{equation}
 Decomposing the right hand side of \eqref{eq:PoincareTensor} into Lorentz representations, we see that
\begin{equation}\label{eq:PoincareTensorSL2C}
    V_{0,0} \otimes V_{0,0} 
        \cong \int_{m,\lambda}^\oplus dm \, d\lambda\, \bigoplus_{s=0}^{\infty} \bigoplus_{J = -s}^{s} \, \tilde{\mu}(m,s) \mu_{m,s}(\lambda,J) \, W_{1+i\lambda,J}\,.
\end{equation}
The celestial OPE block correctly reflects this tensor product structure, where the combination $(m,s)$ are our additional channel indices. Moreover, we expect most Clebsch-Gordan coefficients taking states from $(V_{0,0})^{\otimes 2} \to V_{0,s}$ to be $0$, since two finite energy massless states only create another massless state in the collinear limit. This is reflected in the celestial OPE block by the fact that the celestial three point functions $\expval{\calO_{R_1}\calO_{R_2}\calO_{\tilde{R}}}$ only have support in the collinear and soft limits.  The results of \cite{Guevara:2021tvr} were isolating very particular channels $(V_{0,0})^{\otimes 2} \to V_{m,0}$ channels of this full OPE block.

Our first definition/result is therefore that: \textit{every celestial operator $\calO_{R}$ with non-trivial celestial three-point function $\expval{\calO_{R_1}\calO_{R_2}\calO_{\tilde{R}}}$ is an $\calO_{R_1}\calO_{R_2}$-two particle operator}. This may seem extreme given that this includes a continuous number of primaries, but it is important to note that our celestial CFTs do not have a discrete spectrum where it is even possible to talk about individual operators at particular $\Delta$ and $J$ in an easy way. We will refine this definition/result below.

\paragraph{Conformal Partial Wave Decompositions} In the usual AdS/CFT discussion of multiparticle states (see Section \ref{sec:Conglom} and~\cite{Fitzpatrick:2011dm}), starting with a free theory and turning on bulk interactions gives anomalous dimensions to the operators appearing in the OPE of two CFT primaries. This shifting of the anomalous dimensions is neatly encoded in the poles and residues of the conformal partial wave expansion of a correlator. While in Section~\ref{sec:Conglom} we were applying this framework to CFT$_3$, since CCFTs are just 2D (E)CFTs with additional Poincar\'e symmetry, we can use the conformal partial wave expansion, to track the deformation of the CCFT from a (celestial) MFT.

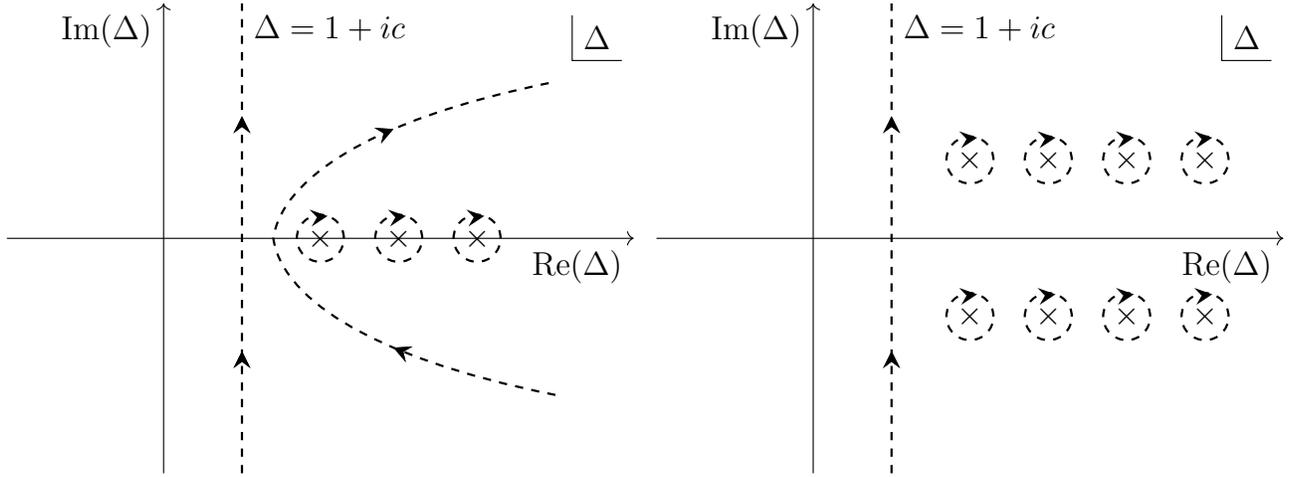
\begin{figure}
\begin{minipage}{.5\textwidth}
\centering
\resizebox{\columnwidth}{!}{
    \begin{tikzpicture}
      \draw[->] (-2, 0) -- (6, 0) node[below left] {$\mathrm{Re}(\Delta)$};
      \draw[->] (0, -3) -- (0, 3) node[below left] {$\mathrm{Im}(\Delta)$};
    
    \node[below left] at (6, 3) {\tikz[baseline]{\node[draw, rectangle, draw=none] (box) {$\Delta$}; \draw (box.north west) -- (box.south west) -- (box.south east);}};
    
      \draw[thick, dashed, ->- = 0.26 rotate 0, ->- = 0.76 rotate 0] (1, -3) -- (1, 3) node[below right] {$\Delta = 1 + i c$};
    
      \foreach \x in {2, 3, 4} {
        \node at (\x, 0) {$\times$};
        \draw[thick, dashed, ->- = 0.2 rotate 205] (\x, 0) circle [radius=0.3];
      }

  \draw[thick, dashed, ->- = 0.25 rotate 0, ->- = 0.75 rotate 0] 
    (5, -2) 
  .. controls (2.5, -1.5) and (1.5, -0.7) 
  .. (1.4,0)
  .. controls (1.5, 0.7) and (2.5, 1.5) 
  .. (5, 2);
    \end{tikzpicture}
}
\end{minipage}%
\begin{minipage}{.5\textwidth}
\centering
\resizebox{\columnwidth}{!}{
    \begin{tikzpicture}
      \draw[->] (-2, 0) -- (6, 0) node[below left] {$\mathrm{Re}(\Delta)$};
      \draw[->] (0, -3) -- (0, 3) node[below left] {$\mathrm{Im}(\Delta)$};
    
    \node[below left] at (6, 3) {\tikz[baseline]{\node[draw, rectangle, draw=none] (box) {$\Delta$}; \draw (box.north west) -- (box.south west) -- (box.south east);}};
    
      \draw[thick, dashed, ->- = 0.26 rotate 0, ->- = 0.76 rotate 0] (1, -3) -- (1, 3) node[below right] {$\Delta = 1 + i c$};
    
      \foreach \x in {2, 3, 4, 5} {
        \node at (\x, +1) {$\times$};
        \node at (\x, -1) {$\times$};
        \draw[thick, dashed, ->- = 0.2 rotate 205] (\x, -1) circle [radius=0.3];
        \draw[thick, dashed, ->- = 0.2 rotate 205] (\x, +1) circle [radius=0.3];
      }
      
    \end{tikzpicture}
}
\end{minipage}%
\caption{Left, the conformal partial wave expansion of correlators along the principal continuous series can be deformed into a sum over residues along the poles of $I_{\Delta,J}/n_{\Delta,J}$. The locations of the poles correspond to scaling dimensions of exchange operators in the OPE, and the residues correspond to (products of) three-point functions between external and exchanged operators. Right, the celestial OPE block expansion is performed along the principal continuous series, which can be deformed to pick up the poles of $\expval{\calO_{R_1}\calO_{R_2}\calO_{\tilde{R}}}$, giving a discrete version of the celestial OPE block expansion (in any fixed channel). Depicted are the poles of the two massless, one massive three point function in \eqref{eq:3ptPoles}.}
\label{fig:CPW_vs_OPE}
\end{figure}

For example, given a (stripped) four-point function of primary operators $f(z,\bz)$, i.e. depending only on conformal cross ratios $(z,\bz)$, the partial wave expansion is a decomposition of the correlator into an orthogonal basis of harmonic functions for $SL(2,\bbC)$
\begin{equation}\label{eq:PartialWave}
    f(z,\bz) = \sum_{J} \int_{1-i\infty}^{1+i\infty} \frac{d\Delta}{2\pi i} \,  \frac{I_{\Delta,J}}{n_{\Delta,J}}\Psi_{\Delta,J}(z,\bz) + (\text{non-norm.})\,,
\end{equation}
where $n_{\Delta,J}$ is a known normalization factor for the partial waves while all dynamical data is encoded in $I_{\Delta,J}$ (see e.g. \cite{Simmons-Duffin:2017nub, Mazac:2018qmi}). In this case, given sufficient fall-off conditions in the $\Delta$-plane, the contour integral over the principal series can be deformed into a sum over poles to the right. The locations of poles in the partial wave expansion encode the scaling dimensions of exchanged operators in the usual OPE, while the residues encode three-point function coefficients, see Figure \ref{fig:CPW_vs_OPE} (left).

In flat space, if we start with a free theory and turn on bulk interactions, the $\calS$-matrix takes the form $\calS = \mathds{1} + i \calT$. For massive $2\to 2$ scattering, it is known that the appropriate bulk-to-celestial transform of $\mathds{1}$ turns the scattering amplitude $\mathds{1}_{2\to 2}(p_i)$ into the four-point function of a (complex) generalized free theory, up to some subtleties with contact terms in $z$ and $\bz$ \cite{ShuHengTalk} e.g. if one transforms the two incoming particles with weight $\Delta_i$, and the two outgoing particles to have weight $\Delta_o$, then
\begin{equation}
    \mathds{1}_{2\to 2}(p_i) \,\,\mapsto\,\, \delta(\Delta_{io})^2 \frac{\mathcal{N}(\Delta_i,\Delta_o)}{\abs{z_{12}}^{2\Delta_i} \abs{z_{34}}^{2\Delta_o}}\left[(z\bz)^{\Delta_i} + \left(\frac{z\bz}{(1-z)(1-\bz)}\right)^{\Delta_i}\right] + (\text{contact terms})\,.
\end{equation}
We will call this the Celestial Mean Field Theory. A similar result holds for the transform of massless scalars, but one must use a shadow basis to avoid pure contact terms in $z$ and $\bz$.  Therefore, just as in AdS/CFT, interactions from the $\calT$ matrix deform the CCFT from the CMFT. 

It is also known in CCFT that if one performs the partial wave decomposition of four-point celestial correlators using \eqref{eq:PartialWave}, that the principal series contour can be deformed into a sum over poles off the principal series in the $\Delta$-plane, and that the residues can sometimes be identified with known celestial $3$-point function coefficients \cite{Lam:2017ofc, Garcia-Sepulveda:2022lga, Fan:2023lky}. In particular, in Section 5 of \cite{Garcia-Sepulveda:2022lga}, the authors consider the CCFT associated with the 3D $O(N)$ model and consider the partial wave decomposition of the $4$-scalar correlator with scaling dimension $\Delta_{\phi}$. The poles come in two semi-infinite towers, one which depends on $\Delta_\phi$ and one which does not:
\begin{equation}\label{eq:towers}
    \Delta_{\calO} = 2n+1\,,\quad 
    \Delta_{\calO} = 2\Delta_{\phi} + n -\tfrac{1}{2}\,,\quad
    n = 0, 1, 2, \dots\,.
\end{equation}
The residues of the poles that do \textit{not} depend on $\Delta_\phi$ were matched to celestial three-point functions coming from three-point interactions in the bulk theory, up to an unexplained $\Delta_\phi$-dependent pre-factor, which also appeared in the CCFT conformal block expansion studied in \cite{Atanasov:2021cje}. On the other hand, the residues of $\Delta_\phi$-dependent towers could not be matched to any three-point functions. Our celestial OPE block suggests resolutions to these issues. From the celestial OPE block, these $\Delta_{\phi}$-dependent phases can be explained as coming from the measures in expressions like \eqref{eq:PoincareTensorSL2C}, which are not generically $\delta$-function supported, reflecting the ``many'' operators with the same scaling dimension at any particular exchange $\Delta$. Likewise, the $\Delta_\phi$ dependent towers should be matched to two-particle operators in the spectrum.

\paragraph{Deforming the Celestial OPE Contour} The above discussion suggests a connection to the celestial OPE block. In the celestial OPE block, we present the OPE with a kernel $\expval{\calO_{R_1}(z_1,\bz_1)\calO_{R_2}(z_2,\bz_2)\calO_{\tilde{R}}(z_P,\bz_P)}$. We generically expect the kernel to have infinite towers of poles in the $\Delta$ plane and, mimicking the conformal partial wave decomposition, we can deform the integral $\int dR$ with an infinite sum over poles of the kernel.  

Consequently, as a second definition/result which is a variant of the one above, we might choose to focus only on this deformed contour so that: \textit{every pole that appears in the kernel $\expval{\calO_{R_1}\calO_{R_2}\calO_{\tilde{R}}}$ in the celestial OPE block of $\calO_{R_1}$ and $\calO_{R_2}$ is an $\calO_{R_1}\calO_{R_2}$-two particle operator.}\footnote{Recall that the integration kernel $dR$ has some slight non-trivial $\Delta$ and $J$ dependence in it from \eqref{eq:SL2CPlan}, which will soften apparent poles.} Since we generically expect the kernel to have a discrete collection of poles in $\Delta$, we expect there to be a discrete collection of two-particle operators for fixed $R_1$, $R_2$, and decomposition channel $m$, more closely analogous to CFTs with a discrete spectrum.

For example, the three point function of two celestial operators $\calO_{\Delta_i}(z_i,\bz_i)$ which create bulk massless scalars with a third celestial operator $\calO_{\Delta}^m(z_P,\bz_P)$ creating a bulk massive scalar is fixed by Poincar\'e symmetry to have the form \cite{Lam:2017ofc,Law:2019glh, Guevara:2021tvr}: 
\begin{align}
    \expval{\calO_{\Delta_1}(z_1,\bz_1)\calO_{\Delta_2}(z_2,\bz_3)\tilde{\calO}^m_{\Delta}(z_P,\bz_P)} 
        &= \frac{1}{\Delta-1}\expval{\calO_{\Delta_1}(z_1,\bz_1)\calO_{\Delta_2}(z_2,\bz_3)\tilde{\calO}^m_{1-\Delta}(z_P,\bz_P)} \\
        &=
        \frac{\mathcal{N}}{\Delta-1} \mathcal{K}^{\Delta_1,\Delta_2,\tilde{\Delta}}(z_1,z_2,z_P)B\left(\tfrac{\Delta_{12} + 1 - \Delta}{2},\tfrac{\Delta_{21}+1-\Delta}{2}\right)
\end{align}
where $\mathcal{N}$ contains normalization constants, dependence on $g$, $m$, $J$, etc. but no dependence on $\Delta$. Closing the $\Delta$-contour, the pole at $\Delta=1$ is removed by the double zero in the $SL(2,\bbC)$ Plancherel measure in \eqref{eq:SL2CPlan} and the kinematic three-point function $\mathcal{K}^{\Delta_1,\Delta_2,\tilde{\Delta}}$ is analytic in $\Delta$, as can be seen from \cite{Guevara:2021tvr}. As a result, all poles inside the contour come from the Euler beta function and are single poles. If we write $\Delta_1 = 1+i\nu_1$ and $\Delta_2=1+i\nu_2$ then $\Delta_{12} = i \nu_{12}$ and the poles and residues of the Euler beta function are
\begin{equation}\label{eq:3ptPoles}
    \Delta_{\pm} = 1 + 2k \pm i \nu_{12} \,,\quad
    \underset{\hphantom{{}_{\pm}}\Delta_{\pm}}{\mathrm{Res}}\, B\left(\tfrac{\Delta_{12} + 1 - \Delta}{2},\tfrac{\Delta_{21}+1-\Delta}{2}\right) = \frac{(-1)^{k+1}\,\Gamma(-k\mp i\nu)}{k!\, \Gamma(-2k \mp i\nu)}\,,
\end{equation}
where $k \in \bbZ_{\geq 0}$. Based on our second definition, this defines (two) semi-infinite towers of $\Delta$-dependent operators at locations $\Delta_{\pm}$ whose three-point functions are given by the residue above and the various normalization constants, we depict this in Figure \ref{fig:CPW_vs_OPE} (right).

Note that it is also not unreasonable for an (direct) integral of unitary representations to be replaced by a (direct) sum over these non-unitary representations, since the operators only created distributional states. This simply means that these different presentations of distributional states evaluate to the same thing on suitable spaces of test functions. A similar phenomena was noticed already in \cite{chakrabarti1968lorentz}. In principle, one could also compute the analogous scaling dimensions and residues for other channels, and even replace the integral over $\int d^2z_P$ by the more familiar tower of descendants $C(z_{1P},z_{2P},\partial_P)$ as in \eqref{eq:OPETypical}.

\subsection{A Weinbergian Approach to Multiparticle Primaries}\label{sec:Weinberg} 
In the end, our space of states will be in representations of the full Poincar\'e group, with objects of definite $m^2$ and $s$. In Section \ref{sec:MultiCarroll}, we saw how taking the naive flat limit of conglomerate CFT primaries only landed us on $m^2=0$ states. Since the tensor product of massless states will generally have non-trivial $P^2$, this is clearly not sufficient. However, in the previous subsections, we saw how the celestial basis naturally sidesteps these subtleties by treating massive and massless states more uniformly in the map to the celestial sphere. Indeed, in Section~\ref{sec:multOPE} we constructed multiparticle states from the point of view of the Lorentz subgroup, treating objects of all masses and spins, at the price of introducing many channels for the $SL(2,\bbC)$ states -- encoding the overarching Poincar\'e representation theory.

In this section we describe an alternative approach to multiparticle operators, where we employ the momentum space partial wave expansion.\footnote{See~\cite{Law:2020xcf} for an approach that is somewhat intermediate to this and the previous sections. While not directly interested in multiparticle states, they construct Poincare invariant partial waves to decompose CCFT 4-point functions labeled by both the Poincar\'e Casimirs and Celestial data.}  By decomposing two-particle states $\ket{p_1,p_2}$ into the coupled basis, we can then use the usual ``one-particle'' bulk-to-boundary propagators of \cite{Pasterski:2017kqt,Law:2020tsg} to map the individual partial wave components to the celestial sphere.

\paragraph{Momentum Space Partial Waves}
In \eqref{00lm}--\eqref{slm} we gave the direct integral decomposition of the tensor product of Poincar\'e representations. If we work with the usual momentum basis for the representations, the overlap of states in the tensor product with states in the direct integral are the usual momentum space partial waves, i.e. Clebsch-Gordan coefficients for the decomposition.

In particular, if we take two non-identical (to avoid symmetry factors) massive spinning particles, with masses $m_1$ and $m_2$, and spins $s_1$ and $s_2$, then a two particle state is $\ket{\mathbf{p}_1,\sigma_1; \mathbf{p}_2,\sigma_2}$, where $\sigma_1$ and $\sigma_2$ are $z$-components of the spin. The usual partial wave expansion (see (3.7.5) of \cite{Weinberg:1995mt}) says the overlap with a state in the coupled basis (on the RHS of \eqref{slm}) is given by:
\begin{equation}\label{eq:PWD}
    \begin{aligned}
        \braket*{\mathbf{p}_1,\sigma_1;\mathbf{p}_2,\sigma_2}{E,\mathbf{p},S,\sigma;\ell,s} 
            &= \left(\left|\mathbf{p}_1\right| E_1 E_2 / E\right)^{-1 / 2} \delta^{(3)}\left(\mathbf{p}-\mathbf{p}_1-\mathbf{p}_2\right) \\
            & \quad \times \delta\left(E-\sqrt{\mathbf{p}_1^2+M_1^2}-\sqrt{\mathbf{p}_2^2+M_2^2}\right) \\
            & \quad \times \sum_{m_s, m_\ell} C_{s_1 s_2 }\left(s, m_s; \sigma_1, \sigma_2\right) C_{\ell s}(S, \sigma ; m_\ell, m_s) Y_{\ell}^{m_\ell}\left(\hat{\mathbf{p}}_1\right)\,,
    \end{aligned}
\end{equation}
where $(E,\mathbf{p})$ is the total four-momentum. 

One advantage of the partial wave decomposition is that the various continuous and discrete labels become one continuous label (center of mass energy), describing the effective mass of the exchange particle, and all other labels are discrete. In particular, we have a collection of ``orbital'' angular momenta $\ell$ with three-component $m_{\ell}$ that describes the spherical harmonics around the direction $\hat{\mathbf{p}}_1$. The two internal spins $s_1$ and $s_2$ with three components $\sigma_1$ and $\sigma_2$ have been added to the internal spin $s$ with three-component $m_s$, using the usual Clebsch-Gordan coefficients for $SU(2)$. Finally, the orbital angular momentum $\ell$ and total internal spin $s$ are added to $S$ with three-component $\sigma$. As in the previous sections, note that all of these additional discrete labels still appear in our states as (discrete) channel indices.

In the case of two scalar particles, the spin Clebsch-Gordan coefficients drop out, and the total angular momentum $S$ is the orbital angular momentum. For two massless scalars, it reduces further and the partial wave becomes
\begin{equation}\label{12p}
    \begin{aligned}
    \langle\omega_1,z_1,\bz_1;\omega_2,z_2,\bz_2|M,y,z,\bz,S,\sigma\rangle 
            &= \left(\omega_1^2 \omega_2 / E\right)^{-1/2} 
            Y_{S}^\sigma\left(z_1,\bz_1\right)\\
            &\times \delta^{(4)}(M\hat{p}^\mu(y,z,\bz)-\omega_1 q^\mu(z_1,\bz_1)-\omega_2q^\mu(z_2,\bz_2))
    \end{aligned}
\end{equation}
where $E=\sqrt{M^2 + \mathbf{p}^2}$ is the energy of the massive state, whose four momentum is parametrized
\begin{equation}
    m\hat{p}^\mu(y,z,\bz) = \frac{m}{2y}(1+y^2+\abs{z}^2,z + \bz, i(\bz-z),1-y^2 -\abs{z}^2)\,.
\end{equation}
For the case of two massless scalars, let us denote the overlap by
\begin{equation}
    C_{0,0}^{M,S}(y,z,\bz,\sigma;\{\omega_i,z_i,\bz_i\}) = \braket{\omega_1,z_1,\bz_1;\omega_2,z_2,\bz_2}{M,y,z,\bz,S,\sigma}\,.
\end{equation}

\paragraph{Bulk-to-boundary Propagating Multiparticle States} Celestial primary wavefunctions are wavepackets designed so that scattering states transform covariantly under the Lorentz group, and are the standard method for extracting celestial amplitudes from their momentum space counterparts \cite{Pasterski:2017ylz, Schreiber:2017jsr}. As explained previously, they are constructed as intertwiners between the momentum and celestial bases \cite{Pasterski:2016qvg, Pasterski:2017kqt}. For massive states, the intertwiners that map an on-shell momentum to a point on the celestial sphere are constructed from the Euclidean AdS bulk-to-boundary propagator, taking us from a point on the mass hyperboloid $\mathbb{H}_3^+$ to the conformal boundary. The 4D spinning massless and zero-spin massive wavefunctions were constructed in \cite{Pasterski:2017kqt}, fermionic wavefunctions in \cite{Muck:2020wtx, Narayanan:2020amh}, and massive spinning wavefunctions in \cite{Law:2020tsg}. 

Now we can apply the results of \cite{Law:2020tsg} to bulk-boundary-propagate our partial wave states. The massive spinning conformal primary wavefunctions take the form
\begin{equation}\label{spwf}
    \phi_{\pm, M, \Delta, J}^{\mu_1 \ldots \mu_S}(X ; w, \bar{w})=\int_0^{\infty} \frac{d y}{y^3} \int d^2z\, \sum_{\sigma=-S}^S G_{\Delta,J,\sigma}^{S}(w, \bar{w} ; y, z, \bar{z}) \epsilon_\sigma^{\mu_1 \ldots \mu_S} e^{ \pm i m \hat{p}^\nu X_\nu}.
\end{equation}
The $G^{S}_{\Delta,J,\sigma}$ are $(2S+1)\times (2S+1)$ matrices with components $J,\sigma = -s,\dots,+s$ relating the $SO(3)$ spin $\sigma$ to the $SO(2)$ spin $J$ (recall \eqref{eq:PoincareSL2CMass}). The explicit forms for $S=0,1,2$, as well as orthonormality conditions and other properties, can be found in \cite{Law:2020tsg} (see also \cite{Iacobacci:2024laa}). In particular, they are related to the usual spinning bulk-to-boundary propagators~\cite{Costa:2011mg} made out of homogeneous polynomials on embedding space by performing the sum over $\sigma$ in~\eqref{spwf}.

Now we can use this bulk-to-boundary construction to map our intermediate partial waves to the celestial sphere. This brings us to our third and final definition of two-particle operator in CCFT: \textit{a two-particle operator $\calO_{\Delta, J}^{M,S}$ is defined as an operator which creates any intermediate state in the partial-wave expansion of two one-particle states in the celestial basis.}

Let us now apply this definition to obtain a formal analog of the conglomeration kernel. Starting with a massive celestial state
\be\label{Oms}
 \calO^{M,S}_{\Delta,J}(z,\bz) \ket{0} = \int_{\mathbb{H}_3^{+}} [d^3\hat{p}]\, G_{\Delta,J,\sigma}^{S}(\hat{\mathbf{p}};z,\bz) \ket{M,S;\mathbf{p},\sigma}\,,
\ee
we can insert the identity operator in the momentum basis 
\be
\mathbb{1}=\sum_{n=0}^\infty \prod_{i=1}^n 
\int \!\frac{d^3p_i}{(2\pi)^3 2p^0_i}\ket*{p_1,\ldots, p_n}\!\!\!\bra*{ p_1,\ldots, p_n}\,,
\ee
to obtain
\begin{align}
    \calO^{M,S}_{\Delta,J}(z,\bz) \ket{0} 
        &= \int_{\mathbb{H}_3^{+}} [d^3\hat{p}]\, G_{\Delta,J,\sigma}^{S}(\hat{\mathbf{p}};z,\bz) \ket{M,S;\mathbf{p},\sigma}\\
        &= \prod_i^2
    \int \!\frac{d^3p_i}{(2\pi)^3 2p^0_i} \int_{\mathbb{H}_3^+} [d^3\hat{p}]\, G_{\Delta,J,\sigma}^{S}(\hat{\mathbf{p}};z,\bz)  \,C_{0,0}^{M,S}(\hat{\mathbf{p}},\sigma;\{\omega_i,z_i,\bz_i\}) \ket{p_1,p_2}\\
            &= \prod_i^2
    \int \!\frac{d^3p_i}{(2\pi)^3 2p^0_i} \int_{\mathbb{H}_3^+} [d^3\hat{p}]\, G_{\Delta,J,\sigma}^{S}(\hat{\mathbf{p}};z,\bz)  \,C_{0,0}^{M,S}(\hat{\mathbf{p}},\sigma;\{\omega_i,z_i,\bz_i\}) \nonumber \\
    &~~\times  \int_{\mathrm{PCS}} \frac{d\Delta_2}{2\pi i} \frac{d\Delta_1}{2\pi i} \, \omega_1^{-\Delta_1} \omega_2^{-\Delta_2}\,\calO_{\Delta_2}(z_2,\bz_2) \calO_{\Delta_1}(z_1,\bz_1) \ket{0}\,.
\end{align}
In the last line we have used the inverse of the celestial transform for massless states \eqref{p12delta}. We see that our CCFT will contain a set of two-particle operators labeled by $M$ and $S$. For more complicated situations than massless scalars, the two-particle operators will have additional discrete channel labels corresponding to the additional indices in \eqref{eq:PWD}. A similar procedure extends to higher particle number states, with additional discrete labels capturing the relative orientations of the constituent momenta which form a compact space. 

One final point is worth mentioning. Now that we are equipped with our understanding of the bulk-to-boundary construction of spinning massive celestial primary wavefunctions from momentum space ones, we can prescribe a multiparticle analog of the `Celestial basis transform': starting with a momentum space partial wave amplitude we can use~\eqref{Oms} to write the corresponding celestial correlator.iI.e. the map~\cite{Pasterski:2016qvg,Law:2020tsg}
\begin{equation}
\tilde{\cal A}^{S_i}_{\Delta_i, J_i}=\left(\prod_{i=1}^j \int \frac{d y_i}{y_i^3} d z_i d \bar{z}_i \sum_{\sigma_i=-S_i}^{S_i} G_{\Delta_i J_i \sigma_i}^{S_i}\right) \mathcal{A}_{\sigma_i}
\end{equation}
extends beyond the usual case of scattering amplitudes labeled by external single particle states, to momentum space partial wave amplitudes as well.

\section{Further Connections} 
Having explored multiparticle states for the flat hologram from different directions, let us close with some natural future directions. The preceding discussions raise a number of comments and conjectures to connect our results to the broader literature. 

\paragraph{Expanded Set of Carrollian Primaries} In our Carrollian investigations in Section~\ref{sec:Carrollian}, the Poincar\'e primary conditions followed from the Carrollian limit for massless single particle states, but we also found that more general representations appeared in the flat limit of multiparticle primaries. We used the first few double trace primary operators to illustrate this point, but it would be worth a more systematic cataloging. 

We would also like to understand how to represent massive states on null infinity. Although massive operators should not reach $\scri$, surely it makes sense to examine correlators of Carrollian primaries in states with nonzero mass. Some recent work in this direction includes \cite{Chen:2023naw,Have:2024dff} and the upcoming \cite{cmp25}. Note that we side stepped this issue entirely in the celestial picture, by working with massive representations whose transformation under Poincar\'e is different from their massless counterparts~\cite{Law:2019glh}.

\paragraph{Beyond the $\cal S$-Matrix}  As mentioned in Section~\ref{sec:multOPE}, the Poincar\'e normal ordering moves all the $\calO^+_{\Delta}$ and $\calO^{-\dagger}_{\Delta}$ to the left in expectation values, while time-ordering moves all $\calO_{\Delta}^{+}$ and $\calO_{\Delta}^{+\dagger}$ to the left. This is a manifestation of the manner in which the flat hologram really involves both an ``in'' and an ``out'' boundary component. Celestial holography should be able to capture more than $\cal S$-matrix elements~\cite{Kim:2023qbl, Kraus:2024gso, Jain:2023fxc, Jorstad:2024yzm}, and it would be interesting to further explore the connections to the more general asymptotic observables introduced in \cite{Caron-Huot:2023vxl}. 

\paragraph{The Celestial OPE} Now that we better understand the role of multiparticle states in celestial CFT from a kinematical perspective, it would behoove us to compute OPE coefficients between them. Some progress along these lines has been made in~\cite{Ball:2023sdz,Guevara:2024ixn} using multi-collinear limits of amplitudes but only for composite primaries with no transverse derivatives. It should be illuminating to understand what the bulk renormalization group and unitarity constraints look like from this perspective.  

Another potentially interesting route that was omitted from our discussions here is to start from the 4D OPE of the free theory, restrict operators to a null hyperplane and dimensionally reduce to the celestial basis.  This is closer in spirit to the conformal collider literature~\cite{Hofman:2008ar} as one is using a 4D CFT as the starting point rather than a flat limit of the 3D CFT at the boundary of AdS$_4$. While a bit further afield, it would be very interesting to try to identify any useful applications to describing jet physics, since this is a natural context where we need to deal with many collinear particles.

It would also be useful to finally pin down and cleanly show exactly how the ``all massless'' single-particle OPEs extracted from splitting functions, taking the form
    \cite{Pate:2019lpp, Banerjee:2020kaa,  Himwich:2021dau, Jiang:2021csc, Adamo:2021zpw, Costello:2022upu, Bittleston:2022jeq, Ball:2022bgg, Ball:2023sdz}
    \begin{equation}
        \calO_{1}(z,\bz)\calO_{2}(0,0) \sim \sum_{k} C_{12k}(z,\bz,\partial) \calO_{k}(0,0)\,
    \end{equation}
    without any delta functions in $z$ or $\bz$ are consistent with full Poincar\'e covariance. Recall that the operators on the left-hand side and right-hand side both create bulk massless particles; then applying the mass Casimir $P^2$ to both sides will lead to an inconsistency as the RHS is $0$ while the LHS is not (except when $z=\bz=0$ in Euclidean signature or one of the two vanishes in the analytic continuation to split signature). If the usual celestial OPE is just one particular channel of our celestial OPE block expansion, then Poincar\'e covariance could in principle be appropriately restored in the other channels, reflecting the non-trivial coupling to massless two-particle states. 

    \paragraph{Bootstrapping from Celestial CFT} The approach of decomposing Poincar\'e into $SL(2,\mathbb{C})$ reps and viewing the celestial correlators as an even more constrained 2D CFT has some implications for how one might try to bootstrap CCFTs. Since the celestial OPE block is identical to the usual OPE block, but with appropriate Poincar\'e constrained three-point functions, the usual Euclidean CFT crossing equations (see e.g. (2.14) of \cite{Gadde:2017sjg}) will apply to CCFT. The additional functional restrictions on the three-point functions should be extremely constraining. For example, we expect that a generic correlator of 3 current operators $JJJ$ in a 2D Euclidean CFT will be only moderately constrained. However, if we also declare that the current operators describing an interacting massless spin-1 theory in the bulk, then we might expect the bootstrap to essentially recover only Yang-Mills theory.

    \paragraph{Handling Theories with Memory}  As discussed in Section \ref{sec:SMatrix}, in theories with interacting massless particles, non-free CFTs, or theories with nontrivial topological features \cite{Jain:2014nza, Mehta:2022lgq, Minwalla:2022sef,Gabai:2022snc}, the meaning of the $\calS$-matrix is not as clear. In an interacting scale invariant theory, it is not clear how one would define a set of scattering wavepackets, and non-scale invariant theories with massless particles coupled to charged fields famously experience infrared divergences. These IR divergences are related to the existence of memory effects and soft hair \cite{Strominger:2014pwa,Pasterski:2015tva,Hawking:2015qqa,Hawking:2016msc,Hawking:2016sgy,Satishchandran:2019pyc,Prabhu:2022zcr} and are linked to the lack of a standard Fock space of asymptotic in/out states containing all the states of the theory \cite{Kulish:1970ut} (see also e.g. \cite{Hannesdottir:2019opa, Hannesdottir:2019umk, Prabhu:2022zcr} and references within).\footnote{In some cases, the massless $\calS$-matrix can be computed despite all these subtleties using, e.g. integrability techniques in 2D \cite{Zamolodchikov:1992zr, Fendley:1993xa}.}

    A standard strategy for dealing with these ``memory-effect type'' IR divergences is to ask questions which are insensitive to the IR subtleties, e.g. only calculating inclusive scattering cross-sections involving hard particles. This is a reasonable response if one wants to restrict their attention from questions involving soft particles, like the memory effect. But this is ultimately unsatisfying if the memory effect is experimentally observable \cite{Hubner:2019sly,Mitman:2020pbt}. A common proposal for producing an IR finite $\calS$-matrix is the Fadeev-Kulish dressing procedure \cite{Kulish:1970ut} for massive QED. However, as explained in \cite{Prabhu:2022zcr}, the Fadeev-Kulish dressing and alternatives have conceptual issues and fail in more complicated scenarios like massless QED, Yang-Mills, gravity, etc. As emphasized in \cite{Prabhu:2022zcr}: the issue with defining in/out states in massless theories with charged particles is \textit{not} with defining in/out states that accommodate scattering with memory; that can be done. The problem is trying to realize all the states in a single separable Hilbert space \cite{kibble1968coherent1,kibble1968coherent2,kibble1968coherent3,kibble1968coherent4}. In theories with memory there is is a continuum of superselection sectors, so separability is lost in a naive direct integral. In other words, memory-effect type IR divergences reflect the inability to define a good separable Hilbert space of states.
    
    In light of this, it would be interesting to develop a celestial or holographic approach to the algebraic-$\calS$ matrix defined in \cite{Prabhu:2018gzs} to accommodate theories with memory. It may be illuminating to better understand how the Lorentz reduction of Section \ref{sec:HilbertSpaces} appears in theories with memory, and how e.g. BMS particles \cite{Oblak:2016eij, Bekaert:2024jxs} decompose into $SL(2,\bbC)$ and/or Virasoro representations. 

\section*{Acknowledgements}
We would like to thank Jacob Abajian, Jan Albert, Adam Ball, Jacqueline Caminiti, Luca Ciambelli, Scott Collier, Diego Delmastro, Rajeev Erramilli, Laurent Freidel, Davide Gaiotto, Yangrui Hu, Carmen Jorge-Diaz, Eivind J{\o}rstad, Hare Krishna, Rob Myers, Joshua Sandor, Atul Sharma, and Daniel Weiss for useful conversations. JK was funded via various NSERC programs for major durations of this work. SP is supported by the Celestial Holography Initiative at the Perimeter Institute for Theoretical Physics and by the Simons Collaboration on Celestial Holography. Research at the Perimeter Institute is supported by the Government of Canada through the Department of Innovation, Science and Industry Canada and by the Province of Ontario through the Ministry of Colleges and Universities.

\appendix
\section{Symmetries of Flat Space Limits}\label{sec:FlatAlg}
In the main text, we take advantage of the fact that flat Minkowski spacetime is the large radius (vanishing cosmological constant) limit of Anti-de Sitter space to understand our holographic setup. The conformal boundary of AdS space is then naturally identified with asymptotic infinity in flat spacetimes. In particular, $\scri^{\pm}$ is a null manifold, supporting a Carrollian CFT, as applied in Section \ref{sec:Carrollian}. Likewise, the celestial sphere (and some features of celestial holography) can be understood by a particular ``dimensional reduction'' of this null direction, as explained in Section \ref{sec:CarToCCFT}.

In this Appendix, we consistently tabulate the various defining commutation relations and kinematics for the algebras described in the main text; discuss the $\Lambda \to 0$ \.In\"on\"u-Wigner contraction of $\AdS_{d+1} \to \mathrm{Mink}_{d+1}$ and its holographically dual contraction $\CFT_{d} \to \mathrm{CarCFT}_{d}$; and record the explicit change of bases between algebras used in the text. To put it differently, our goal in this appendix will be to explicitly describe all the algebras in Figure \ref{fig:IWContraction} and the particular isomorphisms and contractions between them.

\begin{figure}
\begin{adjustbox}{scale=1.00,center}{
    \begin{tikzcd}[sep=large]
    	{\begin{matrix}\mathbb{R}^{2,d}\\ \mathfrak{so}(2,d) \\ \{J_{AB}\} \end{matrix}} & {\begin{matrix}\mathrm{AdS}_{d+1}\\ \mathfrak{so}(2,d) \\ \{M_{\mu\nu}, P_\mu\}\end{matrix}} & {\begin{matrix}\mathrm{Mink}_{d+1}\\ \mathfrak{iso}(1,d) \cong \mathfrak{so}(1,d) \ltimes \mathbb{R}^{d+1}\\ \{M_{\mu\nu}, P_\mu\}\end{matrix}} \\
    	& {\begin{matrix}\mathrm{CFT}_{d}\\ \mathfrak{so}(2,d) \\ \{D, K_a, P_a, M_{ab}\}\end{matrix}} & {\begin{matrix}\mathrm{CCA}_{d} \cong \mathrm{ECFT}_{d-1} \ltimes \mathbb{R}^{d+1}\\ \mathfrak{iso}(1,d) \cong \mathfrak{so}(1,d) \ltimes \mathbb{R}^{d+1} \\ \{D, K_i, P_i, J_{ij}, K_0, P_0, B_i\} \end{matrix}}
    	\arrow[squiggly, leftrightarrow, from=1-2, to=2-2]
    	\arrow[squiggly, leftrightarrow, from=1-3, to=2-3]
    	\arrow["c\to0", from=2-2, to=2-3]
    	\arrow["{\Lambda\to 0}", from=1-2, to=1-3]
    	\arrow[hook', from=1-2, to=1-1]
    \end{tikzcd}
    }
\end{adjustbox}
\caption{$\mathfrak{so}(2,d)$ is the algebra of isometries of $\AdS_{d+1}$ and the algebra of conformal isometries of $\Mink_{d}$, which can be naturally understood by the embedding space formalism as the rotations/boosts of $\mathbb{R}^{2,d}$. The flat space limit ($R\to \infty$ or $\Lambda \to 0$) of $\AdS_{d+1}$ is $\Mink_{d+1}$, whose algebra of isometries is the standard Poincar\'e algebra $\mathfrak{iso}(1,d)$. This algebra is obtained by a standard \.In\"on\"u-Wigner contraction of the $\mathfrak{so}(2,d)$ algebra. For the CFT dual side, this amounts to taking the ultra-relativistic limit ($c\to0$) in the $\CFT_d$; in this case $\mathrm{iso}(1,d)$ is interpreted as the Carrollian conformal symmetries of the Carrollian manifold $\scri^{\pm} \simeq S^{d-1} \times \mathbb{R}$.}
\label{fig:IWContraction}
\end{figure}

\subsection{\texorpdfstring{$\mathfrak{so}(2,d)$}{so(2,d)} and \texorpdfstring{$\AdS_{d+1}/\CFT_{d}$}{AdS(d+1)/CFT(d)}}\label{sec:so2dalgebra}
\paragraph{Rotations of Embedding Space} One particularly useful presentation of the Lie algebra generators of $\mathfrak{so}(2,d)$ is as rotations $\{J_{AB}\}$ of a $(d+2)$-dimensional embedding space $\mathbb{R}^{2,d}$ with standard coordinates $\{X^A\}$ and metric $G = \mathrm{diag}(-1,+1,\dots,+1,-1)$, labeled by capital Roman indices $A,B \in \{0,1,\dots,d,d+1\}$. Clearly there are $(d+1)(d+2)/2$ independent generators. A differential representation of these rotations is:
\begin{equation}
    J_{AB} := \alpha( X_A \partial_B - X_B \partial_A )\,,
\end{equation}
with commutation relations
\begin{equation}
    [J_{AB},J_{CD}]
        =2\alpha(G_{B[D}J_{C]A}-G_{A[D}J_{C]B})\,,
\end{equation}
and
\begin{equation}
    \tr(J_{AB}J_{CD}) = -2\alpha^2(G_{AC}G_{BD}-G_{AD}G_{BC})\,.
\end{equation}
Here, we leave the constant $\alpha$ in for ease of comparison to other results.

\paragraph{Isometries of $\mathrm{AdS}_{d+1}$} We can relabel our generators of $\mathfrak{so}(2,d)$ as 
\begin{equation}
    M_{\mu\nu} := J_{\mu\nu}\,,\qquad
    P_{\mu} := \beta J_{\mu,d+1}\,, 
\end{equation}
with Greek indices $\mu,\nu\in \{0,1,\dots, d\}$ and $\eta_{\mu\nu} := G_{\mu\nu} = \mathrm{diag}(-1,+1,\dots,+1)$. We introduce the constant $\beta$ in order to track useful rescalings, but it can be set to $1$. In this new basis, the commutation relations become 
\begin{align}\
    [M_{\mu\nu},M_{\lambda\rho}]
        &=2\alpha (\eta_{\nu[\rho}M_{\lambda]\mu}-\eta_{\mu[\rho}M_{\lambda]\nu})\,,\label{eq:AdSMM}\\
    [M_{\mu\nu},P_{\rho}]
        &= \alpha \, 2\eta_{\rho[\nu}P_{\mu]}\,,\label{eq:AdSMP}\\
    [P_{\mu},P_{\nu}]
        &= \alpha \beta^2 M_{\mu\nu}\,.\label{eq:AdSPP}
\end{align}
The generators $\{M_{\mu\nu}, P_\mu\}$ are identified with isometries of $\AdS_{d+1}$, viewed as the hyperboloid
\begin{equation}
    X_A X^A = - (X^{0})^2 + \vec{X}^2 -(X^{d+1})^2= -R^2
\end{equation}
embedded in $\mathbb{R}^{2,d}$. Indeed, one can check in standard global AdS coordinates that $P_0 = \alpha \beta \partial_t$. Note that the ``translations'' $P_\mu$ do not commute thanks to the intrinsic curvature of $\AdS_{d+1}$.

\paragraph{Conformal Isometries of $\Mink_d$} Our final presentation of $\mathfrak{so}(2,d)$ is as conformal isometries of $d$-dimensional Minkowski space $\Mink_{d}$. In this case, we will label the generators $\{\mathbf D,\mathbf P_a, \mathbf K_a,\mathbf M_{ab}\}$, with lowercase Roman indices $a,b\in\{2,3,\dots,d+1\}$, which obey the following algebra:
\begin{align}
    [\mathbf D,\mathbf P_a] &= +i\mathbf P_a\,,\\
    [\mathbf D,\mathbf K_a] &= -i\mathbf K_a\,,\\
    [\mathbf K_a,\mathbf P_b] &= -2i(\eta_{ab} \mathbf D + \mathbf M_{ab})\,,\\
    [\mathbf M_{ab},\mathbf P_c] &= i(\eta_{ac}\mathbf P_b - \eta_{bc} \mathbf P_a)\,,\\
    [\mathbf M_{ab},\mathbf K_c] &= i(\eta_{ac}\mathbf K_b - \eta_{bc} \mathbf K_a)\,,\\
    [\mathbf M_{ab},\mathbf M_{cd}] &= -2i(\eta_{b[d}\mathbf M_{c]a}-\eta_{a[d}\mathbf M_{c]b})\,.
\end{align}
Here, we have the Minkowski metric $\eta_{ab} := G_{ab} = \mathrm{diag}(+1,+1,...,-1)$. In the main text we use a more natural indexing, but we stick with this choice in this appendix for comparison of results.

To relate this to the other presentations, we simply identify:
\begin{align}
    \mathbf{D} 
        &= \frac{i}{\alpha} J_{01}\,,\\
    \mathbf{P}_a 
        &= p (J_{0a}+J_{1a})\,,\\
    \mathbf{K}_a 
        &= k (J_{0a}-J_{1a})\,,\\
    \mathbf{M}_{ab} 
        &= -i \frac{1}{\alpha} J_{ab}\,.
\end{align}
This satisfies the above algebra when $pk=-1/\alpha^2$. A convenient choice to therefore relate all the algebras as real Lie algebras, is to have originally chosen $\alpha = \pm i$.

On (spinning) conformal primary operators $\calO_I(x)$, the conformal generators act (in our conventions) by:
\begin{align}
    [\mathbf D, \calO_I(x)] &= i (x\cdot\partial + \Delta) \calO_I(x)\label{eq:Conf1}\\
    [\mathbf P_a, \calO_I(x)] &= i \partial_a \calO_I(x)\,,\\
    [\mathbf K_a, \calO_I(x)] &= i(x^2 \partial_a - 2 x_a x\cdot \partial - 2 \Delta x_a)\calO_I(x)+2(S_{ab})^J_I x^b \calO_J(x)\,,\\
    [\mathbf M_{ab}, \calO_I(x)] &= i(x_a \partial_b - x_b \partial_a) \calO_I(x)-(S_{ab})^J_I\calO_J(x)\,.\label{eq:Conf4}
\end{align}
Here $S_{ab}$ is the appropriate spin-representation matrix for $\calO_I(x)$.\footnote{In our conventions, the abstract generator acts by minus the differential representation
\begin{equation}
    [Q,\calO_I(x)] := -\mathscr{D}_Q\calO_I(x)\,,
\end{equation}
so that the commutation relations of the abstract generators and the differential operators are the same.}

At risk of stating the obvious, we note that ``all three'' of these algebras are the same: they are all just different bases for $\mathfrak{so}(2,d)$. It is simply convenient to call them the different names since each basis makes different actions more obvious.

\subsection{\texorpdfstring{$\mathfrak{iso}(1,d)$}{iso(1,d)} and \texorpdfstring{$\mathrm{\Mink_{d+1}}/\CarCFT_{d}$}{Mink(d+1)/CarCFT(d)}}\label{sec:iso1dalgebra}
\paragraph{Isometries of $\Mink_{d+1}$}
The algebra of isometries of $(d+1)$-dimensional Minkowski spacetime is $\mathfrak{iso}(1,d)\cong\mathfrak{so}(1,d)\ltimes\mathbb{R}^{d+1}$. The $\mathfrak{so}(1,d)$ subalgebra corresponds to the Lorentz subalgebra, while the $\mathbb{R}^{d+1}$ is the translation subalgebra. A convenient basis is given by $\{M_{\mu\nu},P_{\mu}\}$, subject to the commutation relations
\begin{align}\
    [M_{\mu\nu},M_{\lambda\rho}]
        &=2\alpha (\eta_{\nu[\rho}M_{\lambda]\mu}-\eta_{\mu[\rho}M_{\lambda]\nu})\,,\\
    [M_{\mu\nu},P_{\rho}]
        &= \alpha \, 2\eta_{\rho[\nu}P_{\mu]}\,,\\
    [P_{\mu},P_{\nu}]
        &= 0\,.
\end{align}
Clearly these are the same commutation relations as those of the isometries of $\AdS_{d+1}$ in (\ref{eq:AdSMM}--\ref{eq:AdSPP}), except that the translations $P_{\mu}$ commute in flat space. To understand this modification, simply re-instate factors of the AdS radius by rescaling (setting $\beta = R$ in the previous section):
\begin{equation}
    M_{\mu\nu} \mapsto M_{\mu\nu}\,\qquad P_\mu \mapsto RP_\mu\,.
\end{equation}
The flat space limit is obtained by taking $R\to\infty$ (or $\Lambda \to 0$). This is a generic example of an \'In\"on\"u-Wigner contraction \cite{inonu1953contraction}: given a Lie algebra $\mathfrak{g}$, fix a subalgebra $\mathfrak{h} < \mathfrak{g}$ and Abelianize the generators in $\mathfrak{g}\backslash\mathfrak{h}$ (this is set-complement, not a quotient). The resulting algebra is $\mathfrak{g}' \cong \mathfrak{h} \ltimes \mathfrak{a}$, where $\mathfrak{a}$ is an Abelian Lie algebra constructed from the (now-mutually commuting) generators in $\mathfrak{g}\backslash\mathfrak{h}$. We use the notation to emphasize that the algebra does not necessarily split as a direct sum.

\paragraph{Conformal Isometries of a $d$-Dimensional Carrollian Manifold} For our purposes, it will be most useful (and easiest) to define the conformal Carroll algebra $\mathfrak{cca}_d \cong \mathfrak{iso}(d,1)$ in the ``traditional'' way: as an ultra-relativistic limit ($c\to0$) of the $d$-dimensional conformal algebra $\mathfrak{so}(d,2)$, contracting one timelike coordinate.\footnote{Carrollian manifolds (and their conformal isometries) can also be defined in an intrinsic way, independent of ultra-relativistic limits (see e.g. \cite{Ciambelli:2019lap, Bagchi:2019clu} and references within for reviews). This point of view also removes confusions about dynamics that can arise in the degenerate limit, see e.g. \cite{Ciambelli:2023xqk}.} In this case, the generators are $\{\mathbf D,\mathbf P_i, \mathbf K_i,\mathbf M_{ij}, \mathbf{H}, \mathbf{K}, \mathbf{B}_i \}$, with lowercase Roman indices $i,j\in\{2,\dots,d\}$, which obey the following algebra:
\begin{alignat*}{5}
    [\mathbf{D},\mathbf{H}]
        &= i \mathbf{H}\,,\quad
    &[\mathbf{K},\mathbf{P}_i]
        &= 2i\mathbf{B}_i\,,\quad
    &[\mathbf{M}_{ij},\mathbf{P}_k]
        &= i(\delta_{ik}\mathbf{P}_j-\delta_{jk}\mathbf{P}_i)\,,
    \\
    [\mathbf{D},\mathbf{P}_i]
        &= i \mathbf{P}_i\,,\quad
    &[\mathbf{K}_i,\mathbf{H}]
        &= -2i\mathbf{B}_i\,,\quad
    &[\mathbf{M}_{ij},\mathbf{K}_k]
        &= i(\delta_{ik}\mathbf{K}_j-\delta_{jk}\mathbf{K}_i)\,,
    \\
    [\mathbf{D},\mathbf{K}]
        &= -i \mathbf{K}\,,\quad
    &[\mathbf{B}_i,\mathbf{K}_j]
        &= i\delta_{ij}\mathbf{K}\,,
    &[\mathbf{M}_{ij},\mathbf{B}_k]
        &= i(\delta_{ik}\mathbf{B}_j-\delta_{jk}\mathbf{B}_i)\,,
    \\
    [\mathbf{D},\mathbf{K}_i]
        &= -i \mathbf{K}_i\,,\quad
    &[\mathbf{B}_i,\mathbf{P}_j]
        &= i\delta_{ij}\mathbf{H}\,,
    &[\mathbf{M}_{ij},\mathbf{M}_{kl}]
        &= -2i(\delta_{b[d}\mathbf M_{c]a}-\delta_{a[d}\mathbf M_{c]b})\,,
    &&
    \\
    &
    &[\mathbf{K}_i,\mathbf{P}_j]
        &= -2i(\delta_{ij}\mathbf{D} + \mathbf{M}_{ij})\,.
\end{alignat*}

To obtain this, we have split off the spatial components of the standard Lorentzian $\CFT_d$ commutation relations, $\mathbf P_i$ and $\mathbf K_i$, from the temporal components $\mathbf H:=\mathbf{P}_{d+1}$ and $\mathbf K:= \mathbf{K}_{d+1}$, as well as the boosts $\mathbf B_i:=\mathbf{M}_{i,d+1}$. To relate this to the Poincar\'e generators, we identify:
\begin{alignat}{3}
    \mathbf{D} 
        &= \frac{i}{\alpha} M_{01}\,,\\
    \mathbf{P}_i 
        &= p (M_{0i}+M_{1i})\,,
    \quad
    &\mathbf{H}
        &= \frac{p}{\beta}(P_0 + P_1)\,,\\
    \mathbf{K}_i
        &= k (M_{0i}-M_{1i})\,,
    \quad
    &\mathbf{K}
        &= \frac{k}{\beta}(P_0 - P_1)\,,\\
    \mathbf{M}_{ij} 
        &= -i \frac{1}{\alpha} M_{ij}\,,
    \quad
    &\mathbf{B}_i
        &= -\frac{i}{\alpha\beta}P_i\,,
\end{alignat}
where $pk=-1/\alpha^2$ as before.

We can also identify the individual pieces of $\mathfrak{cca}_d \cong \mathfrak{so}(1,d) \ltimes \mathbb{R}^{d+1}$. In particular, the $\mathfrak{so}(1,d)$ subalgebra is generated by $\{\mathbf D,\mathbf P_i, \mathbf K_i,\mathbf M_{ij}\}$, while the $\mathbb{R}^{d+1}$ is generated by $\{\mathbf{H}, \mathbf{K}, \mathbf{B}_i \}$. Null infinity of asymptotically flat spacetime has the topology $\scri^{\pm} \simeq S^{d-1} \times \mathbb{R}$, where $S^{d-1}$ is the celestial sphere (a Euclidean manifold) and $\mathbb{R}$ is null. We see that the $\mathfrak{so}(1,d)$ subalgebra can be interpreted as the algebra of conformal isometries acting on the $S^{d-1}$, with additional symmetries coming from the null direction fibered over the circle. In Section \ref{sec:CarToCCFT}, we compactify this null direction and exchange it for the conformal dimension $\Delta$ in the celestial CFT.

\subsubsection{Disambiguation of BMS}\label{app:BMS}
The isometry algebra of $\mathrm{Mink}_{d}$ spacetime is the Poincar\'e algebra $\mathfrak{so}(1,d) \ltimes \bbR^{d-1,1}$. Specializing to $d=4$ for simplicity of notation (the general case is not significantly different), we consider vector fields acting on null infinity of the form:
\begin{equation}
    \xi = f(z,\bz)\partial_u + Y^A(z,\bz)\partial_A\,.
\end{equation}
With this, there are three slightly different scenarios where ``BMS algebras'' typically appear in the literature, controlled by how arbitrary the sphere killing vectors $Y(z,\bz)$ are allowed to be:
\begin{enumerate}
    \item \textbf{BMS Algebra} $\mathfrak{bms}_{d}$ \cite{Bondi:1962px}. Here, $Y^A(z,\bz)$ are restricted to only globally defined conformal Killing vectors on $S^{d-2}$. E.g. for $S^2$, these are the usual:
    \begin{equation}
        Y(z,\bz) \quad \sim \quad 
        \begin{matrix}
        \partial_z\,,& z \partial_z\,,& z^2 \partial_z\,,\\
        \partial_{\bz}\,,& \bz \partial_{\bz}\,,& {\bz}^2 \partial_{\bz}\,.
        \end{matrix}
    \end{equation}
    \item \textbf{Extended BMS Algebra} \cite{Barnich:2011mi}. Here we allow the $Y(z,\bz)$ to extend to conformal killing vectors which are only locally defined. E.g. in the case of $d=2$ this is the inclusion of all $z^n \partial_z$ and $\bz^n \partial_{\bz}$, so the $\mathfrak{sl}(2,\bbC)$ action is enhanced to two copies of the Witt algebra.
    \item \textbf{General BMS Algebra} \cite{Campiglia:2020qvc}. Alternatively, we can allow $Y(z,\bz)$ to be any arbitrary diffeomorphism of $S^{d-2}$.
\end{enumerate}
In all three cases, the BMS algebra is infinite-dimensional because of the $S^{d-2}$-dependent (but $u$-constant) null-translations $f(z,\bz)$, generating the subalgebra of ``supertranslations'' on null infinity. In particular, it is actually the BMS algebra above that is equivalent to the conformal isometries of a ($d-1$)-dimensional null CFT \cite{Duval:2014uoa, Duval:2014uva, Duval:2014lpa}, i.e. $\mathfrak{bms}_{d} \cong \mathfrak{ccar}_{d-2}$.

We could view the Poincar\'e algebra as being case $0$ in the list above: where $Y^A(z,\bz)$ is allowed to be an arbitrary conformal killing vector for $S^{d-2}$, but $f(z,\bz)$ is restricted to be (covariantly) constant on $S^{d-2}$. This ``Global BMS algebra'' (Poincar\'e) is actually what we use in the text.

\section{Representations of \texorpdfstring{$SL(2,\bbC)$}{SL(2,C)}}\label{sec:SL2CReps}
Here we recount the representation theory of $SL(2,\bbC)$ that will be important for understanding the construction of induced representations and for understanding, especially, the reduction of Poincar\'e reps to the Lorentz group in Section \ref{sec:HilbertSpaces}.

$SL(2,\bbC)$ is simply connected, so we can look directly at its UIRs. A UIR is characterized by the eigenvalues of the two Casimirs of $SL(2,\bbC)$:
\begin{align}
    \frac{1}{2}L_{\mu\nu} L^{\mu\nu} &= [J^2 + (\Delta-1)^2 - 1]\mathds{1}\\
    \frac{1}{4!} \epsilon^{\mu\nu\rho\sigma} {L_{\mu\nu} L_{\rho\sigma}} &= [J(\Delta-1)]\mathds{1}\,.
\end{align}
In addition to the trivial representation, with $(\Delta,J)=(0,0)$, there are two continuous families of UIRs:
\begin{itemize}
    \item \textbf{Principal Continuous Series}. $\Delta = 1+i\lambda$ with $\lambda\in\bbR$ and $J \in \frac{1}{2}\bbZ$. These are an infinite family of infinite-dimensional tempered representations of $SL(2,\bbC)$.\footnote{The $J=0$ principal series representations are often called ``spherical principal series'' representations.} The representations $(\lambda,J)$ and $(-\lambda,-J)$ are unitarily equivalent.
    \item \textbf{Complementary Series}. $\Delta \in (0,2)\backslash\{1\}$ and $J=0$. These are not tempered.
\end{itemize}
These form all the UIRs of $SL(2,\bbC)$. Note: there is no discrete series of UIRs for the group $SL(2,\bbC)$, as guaranteed by a theorem of Harish-Chandra (see \cite{harish1966discrete}, Theorem 13). The only tempered UIRs of $SL(2,\bbC)$, and therefore ($\delta$-function) normalizable, are the principle continuous series representations i.e. only they appear in the decomposition of the regular representation $L^2(SL(2,\bbC))$.\footnote{More generally, representation theory of $SO(1,d)$ also generically contains a point set of (untempered) UIRs, called ``exceptional representations,'' that appear as subrepresentations of reducible representations at some particular distinguished integral $\Delta$ and $J$ (see e.g. \cite{dobrev1986lecture, Sun:2021thf} for more information on these representations). This suggests possible missed UIRs in our $SL(2,\bbC)$ discussion. However, in low dimensions, a series of exceptional isomorphisms either find them to be trivial or identify them with other non-trivial representations we have already identified. This fact is compatible with physical intuition of the corresponding massless spinning particles in $dS_3$ e.g. the duality between a massless scalar and spin-1 field, or the non-existence of propagating degrees of freedom for gravity \cite{Penedones:2023uqc}. We thank Zimo Sun for an enlightening discussion on this point.} Further note that the trivial representation is not tempered.\footnote{This fact is well-known in the conformal bootstrap, and is the reason one must remove the identity piece of (stripped) four-point functions before performing the conformal partial wave decomposition. See Section 5 of \cite{Garcia-Sepulveda:2022lga} for some comments in CCFT.}

\subsection{The \texorpdfstring{$z$}{z}-Basis}
Given some fixed $(\Delta,J)$ irrep of $SL(2,\bbC)$, let $W_{\Delta,J}$ be the associated representation space and $\pi_{\Delta,J}$ the corresponding group action. There are many different bases used in the literature to describe the irrep $W_{\Delta,J}$ (see Appendix B of \cite{Mizera:2022sln} for a list of partial references).\footnote{Perhaps the most common is the ``canonical basis,'' aka Joos basis \cite{Joos:1962qq}, which is induced from the maximal compact $SU(2)$. In this orthonormal basis, states are simultaneous eigenstates of the two Casimirs as well as $\vec{J}^2$ and $J_3$ in $SL(2,\bbC)$. i.e. it is an infinite discrete basis of states $\ket{j_0,m}_{\Delta,J}$ where $j_0=J,J+1,\dots,\infty$ and $m=-j,-j+1,\dots,j$. When decomposing an $SL(2,\bbC)$ irrep into $SU(2)$ irreps, $j_0$ is the smallest spin eigenvalue that appears, see \cite{Gaiotto:2023hda}.} In the Gel'fand-Naimark ``$z$-basis,'' $W_{\Delta,J} = L^2(\bbC)$ and an element $g\in SL(2,\bbC)$ acts on complex functions by
\begin{align}
    \pi_{\Delta,J}(g^{-1}) f(z) 
        &= \abs{c z + d}^{-2\Delta} \left(\frac{cz+d}{\abs{c z + d}}\right)^{-2J} f(g\cdot z)\\
        &= (cz+d)^{-\Delta-J}(\bar{c}\bar{z}+\bar{d})^{-\Delta+J}f(g\cdot z)\,,
\end{align}
where $g\cdot z$ denotes the usual fractional linear action of $g\in SL(2,\bbC)$ on $\bbC$.

For the principal continuous series, the Hilbert space norm on $W_{\Delta,J}$ is simply the standard inner product:
\begin{equation}
    \braket{f}{g} = \int_{\bbC} dz\, \bar{f}(z)g(z)\,.
\end{equation}
For the complementary series, the Hilbert space is still $L^2(\bbC)$, but with a non-local inner product (see \cite{knapp2001representation}, see also \cite{bars1972operator} for an interesting unified treatment):
\begin{equation}
    \braket{f}{g} = \int_{\bbC} \int_{\bbC} dz \,dw\, \frac{\bar{f}(z)g(w)}{\abs{z-w}^{2(2-\Delta)}} \,.
\end{equation}

In the $z$-basis for principal series representations we therefore use the ``states'' $\ket{\Delta,J;z,\bar{z}}$ as a basis for $L^2(\bbC)$, with normalization:
\begin{equation}
    \braket{\Delta,J;z,\bar{z}}{\Delta',J';z',\bar{z}'} = \delta(\Delta-\Delta')\delta^{(2)}(z-z')\delta_{JJ'}\,.
\end{equation}
See \cite{bars1972operator} for a construction of the $Z$ and $\bar{Z}$ operators diagonalized by these states. Note that there are two largely independent distributional features of this inner product:
\begin{enumerate}
    \item Given a fixed representation $\pi_{\Delta,J}$, the representation space $W_{\Delta,J}$ is the infinite dimensional space $L^2(\bbC)$. And, in the $z$-basis, a function is decomposed as a linear combination
    \begin{equation}
        \ket{f} = \int_{\bbC} d^2z\, f(z) \ket{\Delta,J;z,\bz}\,.
    \end{equation}
    The fact that the states spanning the rep $W_{\Delta,J}$ are distributions in $(z,\bar{z})$ reflects the continuous basis we chose.
    \item The $\delta(\Delta-\Delta')\delta_{JJ'}$ term reflects the fact that the matrix element functions $\pi_{\Delta,J}$ and $\pi_{\Delta',J'}$ are orthonormal in the sense of harmonic analysis on $SL(2,\bbC)$. i.e. it corresponds to the fact that the principal continuous representations of $SL(2,\bbC)$ are $\delta$-normalizable, as described by the Plancherel measure on the space of (tempered) UIRs. Since the principal series representations are the only tempered representations, they are the only representations with this property. The Plancherel measure on $SL(2,\bbC)$ is
    \begin{equation}\label{eq:SL2CPlan}
        \rho_J(\Delta) = -\frac{1}{8\pi^3}(\Delta - J - 1) (\Delta + J - 1) = -\frac{1}{8\pi^3} (2h-1)(2\bar{h}-1)
    \end{equation}
    in the normalization conventions of \cite{Karateev:2018oml}.
\end{enumerate}

\subsection{The Shadow Transform}
Given any representation $W_{\Delta,J}$, the ``shadow representation'' is the representation $W_{\tilde{\Delta},-J}$, with $\tilde{\Delta} = 2-\Delta$, i.e. $(\tilde{h},\tilde{\bar{h}}) = (1-h, 1-\bar h)$. The shadow transform is the name for the explicit intertwining map describing the isomorphism between these two representations. It is the reason why the principal series reps are identified under $(\lambda,J)\mapsto (-\lambda,-J)$. An explicit formula for the shadow transform is \cite{Osborn:2012vt, Gadde:2017sjg, Karateev:2018oml, Sun:2021thf}
\begin{equation}
    \tilde{f}(z,\bar{z}) = K_{h,\bar{h}} \int_{\bbC} d^2y \frac{1}{(z-y)^{2-2h}(\bar{z}-\bar{y})^{2-2\bar{h}}} f(y,\bar{y})\,,
\end{equation}
where the normalization is chosen so that
\begin{equation}
    \expval{\calO_{h,\bar{h}}(z_1,\bar{z}_1)\calO_{h',\bar{h}'}(z_2,\bar{z}_2)} 
        = \frac{\delta_{h_1 h_2}\delta_{\bar{h}_1 \bar{h}_2}}{z_{12}^{h_1+h_2} \bar{z}_{12}^{\bar{h}_1+\bar{h}_2}}\,,\quad
    \expval*{\tilde{\calO}_{\tilde{h},\tilde{\bar{h}}}(z_1,\bar{z}_1)\tilde{\calO}_{\tilde{h}',\tilde{\bar{h}}'}(z_2,\bar{z}_2)} 
        = \frac{\delta_{\tilde{h}_1 \tilde{h}_2}\delta_{\tilde{\bar{h}}_1 \tilde{\bar{h}}_2}}{z_{12}^{\tilde{h}_1+\tilde{h}_2} \bar{z}_{12}^{\tilde{\bar{h}}_1+\tilde{\bar{h}}_2}}\,.
\end{equation}
In CFT terms, the shadow operator is non-local with respect to the original operator, and is ``formal'' in that it is only defined up to integrating it against the original operator in correlation functions, i.e. it has $\delta$-function two-point function with the un-shadowed operator.

\bibliographystyle{utphys}
\bibliography{references}

\providecommand{\href}[2]{#2}\begingroup\raggedright\begin{thebibliography}{100}

\bibitem{tHooft:1993Dmi}
G.~'t~Hooft, ``{Dimensional reduction in quantum gravity},'' {\em Conf. Proc.
  C} {\bfseries 930308} (1993) 284--296,
  \href{http://arxiv.org/abs/gr-qc/9310026}{{\ttfamily arXiv:gr-qc/9310026}}.

\bibitem{Susskind:1994vu}
L.~Susskind, ``{The World as a hologram},''
  \href{http://dx.doi.org/10.1063/1.531249}{{\em J. Math. Phys.} {\bfseries 36}
  (1995) 6377--6396}, \href{http://arxiv.org/abs/hep-th/9409089}{{\ttfamily
  arXiv:hep-th/9409089}}.

\bibitem{Strominger:2014pwa}
A.~Strominger and A.~Zhiboedov, ``{Gravitational Memory, BMS Supertranslations
  and Soft Theorems},'' \href{http://dx.doi.org/10.1007/JHEP01(2016)086}{{\em
  JHEP} {\bfseries 01} (2016) 086},
  \href{http://arxiv.org/abs/1411.5745}{{\ttfamily arXiv:1411.5745 [hep-th]}}.

\bibitem{Pasterski:2015tva}
S.~Pasterski, A.~Strominger, and A.~Zhiboedov, ``{New Gravitational
  Memories},'' \href{http://dx.doi.org/10.1007/JHEP12(2016)053}{{\em JHEP}
  {\bfseries 12} (2016) 053}, \href{http://arxiv.org/abs/1502.06120}{{\ttfamily
  arXiv:1502.06120 [hep-th]}}.

\bibitem{Pasterski:2015zua}
S.~Pasterski, ``{Asymptotic Symmetries and Electromagnetic Memory},''
  \href{http://dx.doi.org/10.1007/JHEP09(2017)154}{{\em JHEP} {\bfseries 09}
  (2017) 154},
\href{http://arxiv.org/abs/1505.00716}{{\ttfamily arXiv:1505.00716 [hep-th]}}.

\bibitem{Strominger:2017zoo}
A.~Strominger, {\em {Lectures on the Infrared Structure of Gravity and Gauge
  Theory}}.
\newblock {Princeton University Press}, 2018.
\newblock
\href{http://arxiv.org/abs/1703.05448}{{\ttfamily arXiv:1703.05448 [hep-th]}}.
\newblock

\bibitem{Strominger:2021mtt}
A.~Strominger, ``{$w_{1+\infty}$ Algebra and the Celestial Sphere: Infinite
  Towers of Soft Graviton, Photon, and Gluon Symmetries},''
  \href{http://dx.doi.org/10.1103/PhysRevLett.127.221601}{{\em Phys. Rev.
  Lett.} {\bfseries 127} no.~22, (2021) 221601}.

\bibitem{Guevara:2021abz}
A.~Guevara, E.~Himwich, M.~Pate, and A.~Strominger, ``{Holographic Symmetry
  Algebras for Gauge Theory and Gravity},''
  \href{http://arxiv.org/abs/2103.03961}{{\ttfamily arXiv:2103.03961
  [hep-th]}}.

\bibitem{Himwich:2021dau}
E.~Himwich, M.~Pate, and K.~Singh, ``{Celestial operator product expansions and
  w$_{1+\infty}$ symmetry for all spins},''
  \href{http://dx.doi.org/10.1007/JHEP01(2022)080}{{\em JHEP} {\bfseries 01}
  (2022) 080}, \href{http://arxiv.org/abs/2108.07763}{{\ttfamily
  arXiv:2108.07763 [hep-th]}}.

\bibitem{Mago:2021wje}
J.~Mago, L.~Ren, A.~Y. Srikant, and A.~Volovich, ``{Deformed $w_{1+\infty}$
  Algebras in the Celestial CFT},''
  \href{http://dx.doi.org/10.3842/SIGMA.2023.044}{{\em SIGMA} {\bfseries 19}
  (2023) 044}, \href{http://arxiv.org/abs/2111.11356}{{\ttfamily
  arXiv:2111.11356 [hep-th]}}.

\bibitem{Bu:2022iak}
W.~Bu, S.~Heuveline, and D.~Skinner, ``{Moyal deformations, W$_{1+\infty}$ and
  celestial holography},''
  \href{http://dx.doi.org/10.1007/JHEP12(2022)011}{{\em JHEP} {\bfseries 12}
  (2022) 011}, \href{http://arxiv.org/abs/2208.13750}{{\ttfamily
  arXiv:2208.13750 [hep-th]}}.

\bibitem{Drozdov:2023qoy}
P.~Drozdov and T.~Kimura, ``{Structure of deformed w1+\ensuremath{\infty}
  symmetry and topological generalization in Celestial CFT},''
  \href{http://dx.doi.org/10.1016/j.physletb.2023.138272}{{\em Phys. Lett. B}
  {\bfseries 847} (2023) 138272},
  \href{http://arxiv.org/abs/2306.11693}{{\ttfamily arXiv:2306.11693
  [math-ph]}}.

\bibitem{Mason:2023mti}
L.~Mason, R.~Ruzziconi, and A.~Yelleshpur~Srikant, ``{Carrollian amplitudes and
  celestial symmetries},''
  \href{http://dx.doi.org/10.1007/JHEP05(2024)012}{{\em JHEP} {\bfseries 05}
  (2024) 012}, \href{http://arxiv.org/abs/2312.10138}{{\ttfamily
  arXiv:2312.10138 [hep-th]}}.

\bibitem{Costello:2022upu}
K.~Costello and N.~M. Paquette, ``{Associativity of One-Loop Corrections to the
  Celestial Operator Product Expansion},''
  \href{http://dx.doi.org/10.1103/PhysRevLett.129.231604}{{\em Phys. Rev.
  Lett.} {\bfseries 129} no.~23, (2022) 231604},
  \href{http://arxiv.org/abs/2204.05301}{{\ttfamily arXiv:2204.05301
  [hep-th]}}.

\bibitem{Costello:2022wso}
K.~Costello and N.~M. Paquette, ``{Celestial holography meets twisted
  holography: 4d amplitudes from chiral correlators},''
  \href{http://arxiv.org/abs/2201.02595}{{\ttfamily arXiv:2201.02595
  [hep-th]}}.

\bibitem{Costello:2023vyy}
K.~J. Costello, ``{Bootstrapping two-loop QCD amplitudes},''
  \href{http://arxiv.org/abs/2302.00770}{{\ttfamily arXiv:2302.00770
  [hep-th]}}.

\bibitem{Zeng:2023qqp}
K.~Zeng, ``{Twisted Holography and Celestial Holography from Boundary Chiral
  Algebra},'' \href{http://dx.doi.org/10.1007/s00220-023-04917-0}{{\em Commun.
  Math. Phys.} {\bfseries 405} no.~1, (2024) 19},
  \href{http://arxiv.org/abs/2302.06693}{{\ttfamily arXiv:2302.06693
  [hep-th]}}.

\bibitem{Garner:2023izn}
N.~Garner and N.~M. Paquette, ``{Twistorial monopoles \& chiral algebras},''
  \href{http://dx.doi.org/10.1007/JHEP08(2023)088}{{\em JHEP} {\bfseries 08}
  (2023) 088}, \href{http://arxiv.org/abs/2305.00049}{{\ttfamily
  arXiv:2305.00049 [hep-th]}}.

\bibitem{Fernandez:2024tue}
V.~E. Fern\'andez, N.~M. Paquette, and B.~R. Williams, ``{Twisted holography on
  AdS$_3 \times S^3 \times$ K3 \& the planar chiral algebra},''
  \href{http://dx.doi.org/10.21468/SciPostPhys.17.4.109}{{\em SciPost Phys.}
  {\bfseries 17} no.~4, (2024) 109},
  \href{http://arxiv.org/abs/2404.14318}{{\ttfamily arXiv:2404.14318
  [hep-th]}}.

\bibitem{Garner:2024tis}
N.~Garner and N.~M. Paquette, ``{Scattering off of Twistorial Line Defects},''
  \href{http://arxiv.org/abs/2408.11092}{{\ttfamily arXiv:2408.11092
  [hep-th]}}.

\bibitem{Fernandez:2024qnu}
V.~E. Fern\'andez and N.~M. Paquette, ``{Associativity is enough: an all-orders
  2d chiral algebra for 4d form factors},''
  \href{http://arxiv.org/abs/2412.17168}{{\ttfamily arXiv:2412.17168
  [hep-th]}}.

\bibitem{Ciambelli:2018wre}
L.~Ciambelli, C.~Marteau, A.~C. Petkou, P.~M. Petropoulos, and K.~Siampos,
  ``{Flat holography and Carrollian fluids},''
  \href{http://dx.doi.org/10.1007/JHEP07(2018)165}{{\em JHEP} {\bfseries 07}
  (2018) 165}, \href{http://arxiv.org/abs/1802.06809}{{\ttfamily
  arXiv:1802.06809 [hep-th]}}.

\bibitem{Ciambelli:2018ojf}
L.~Ciambelli and C.~Marteau, ``{Carrollian conservation laws and Ricci-flat
  gravity},'' \href{http://dx.doi.org/10.1088/1361-6382/ab0d37}{{\em Class.
  Quant. Grav.} {\bfseries 36} no.~8, (2019) 085004},
  \href{http://arxiv.org/abs/1810.11037}{{\ttfamily arXiv:1810.11037
  [hep-th]}}.

\bibitem{Bagchi:2019xfx}
A.~Bagchi, A.~Mehra, and P.~Nandi, ``{Field Theories with Conformal Carrollian
  Symmetry},'' \href{http://dx.doi.org/10.1007/JHEP05(2019)108}{{\em JHEP}
  {\bfseries 05} (2019) 108}, \href{http://arxiv.org/abs/1901.10147}{{\ttfamily
  arXiv:1901.10147 [hep-th]}}.

\bibitem{Bagchi:2019clu}
A.~Bagchi, R.~Basu, A.~Mehra, and P.~Nandi, ``{Field Theories on Null
  Manifolds},'' \href{http://dx.doi.org/10.1007/JHEP02(2020)141}{{\em JHEP}
  {\bfseries 02} (2020) 141}, \href{http://arxiv.org/abs/1912.09388}{{\ttfamily
  arXiv:1912.09388 [hep-th]}}.

\bibitem{Gupta:2020dtl}
N.~Gupta and N.~V. Suryanarayana, ``{Constructing Carrollian CFTs},''
  \href{http://dx.doi.org/10.1007/JHEP03(2021)194}{{\em JHEP} {\bfseries 03}
  (2021) 194}, \href{http://arxiv.org/abs/2001.03056}{{\ttfamily
  arXiv:2001.03056 [hep-th]}}.

\bibitem{Freidel:2022bai}
L.~Freidel and P.~Jai-akson, ``{Carrollian hydrodynamics from symmetries},''
  \href{http://arxiv.org/abs/2209.03328}{{\ttfamily arXiv:2209.03328
  [hep-th]}}.

\bibitem{Ciambelli:2019lap}
L.~Ciambelli, R.~G. Leigh, C.~Marteau, and P.~M. Petropoulos, ``{Carroll
  Structures, Null Geometry and Conformal Isometries},''
  \href{http://dx.doi.org/10.1103/PhysRevD.100.046010}{{\em Phys. Rev. D}
  {\bfseries 100} no.~4, (2019) 046010},
  \href{http://arxiv.org/abs/1905.02221}{{\ttfamily arXiv:1905.02221
  [hep-th]}}.

\bibitem{Ciambelli:2023xqk}
L.~Ciambelli, ``{Dynamics of Carrollian Scalar Fields},''
  \href{http://arxiv.org/abs/2311.04113}{{\ttfamily arXiv:2311.04113
  [hep-th]}}.

\bibitem{Maldacena:1997re}
J.~M. Maldacena, ``{The Large N limit of superconformal field theories and
  supergravity},'' \href{http://dx.doi.org/10.1023/A:1026654312961}{{\em Adv.
  Theor. Math. Phys.} {\bfseries 2} (1998) 231--252},
  \href{http://arxiv.org/abs/hep-th/9711200}{{\ttfamily arXiv:hep-th/9711200}}.

\bibitem{Witten:1998qj}
E.~Witten, ``{Anti-de Sitter space and holography},''
  \href{http://dx.doi.org/10.4310/ATMP.1998.v2.n2.a2}{{\em Adv. Theor. Math.
  Phys.} {\bfseries 2} (1998) 253--291},
\href{http://arxiv.org/abs/hep-th/9802150}{{\ttfamily arXiv:hep-th/9802150
  [hep-th]}}.

\bibitem{Susskind:1998vk}
L.~Susskind, ``{Holography in the flat space limit},''
  \href{http://dx.doi.org/10.1063/1.1301570}{{\em AIP Conf. Proc.} {\bfseries
  493} no.~1, (1999) 98--112},
  \href{http://arxiv.org/abs/hep-th/9901079}{{\ttfamily arXiv:hep-th/9901079}}.

\bibitem{Giddings:1999jq}
S.~B. Giddings, ``{Flat space scattering and bulk locality in the AdS / CFT
  correspondence},'' \href{http://dx.doi.org/10.1103/PhysRevD.61.106008}{{\em
  Phys. Rev. D} {\bfseries 61} (2000) 106008},
  \href{http://arxiv.org/abs/hep-th/9907129}{{\ttfamily arXiv:hep-th/9907129}}.

\bibitem{Polchinski:1999ry}
J.~Polchinski, ``{S matrices from AdS space-time},''
  \href{http://arxiv.org/abs/hep-th/9901076}{{\ttfamily arXiv:hep-th/9901076}}.

\bibitem{Penedones:2010ue}
J.~Penedones, ``{Writing CFT correlation functions as AdS scattering
  amplitudes},'' \href{http://dx.doi.org/10.1007/JHEP03(2011)025}{{\em JHEP}
  {\bfseries 03} (2011) 025}, \href{http://arxiv.org/abs/1011.1485}{{\ttfamily
  arXiv:1011.1485 [hep-th]}}.

\bibitem{Paulos:2016fap}
M.~F. Paulos, J.~Penedones, J.~Toledo, B.~C. van Rees, and P.~Vieira, ``{The
  S-matrix bootstrap. Part I: QFT in AdS},''
  \href{http://dx.doi.org/10.1007/JHEP11(2017)133}{{\em JHEP} {\bfseries 11}
  (2017) 133}, \href{http://arxiv.org/abs/1607.06109}{{\ttfamily
  arXiv:1607.06109 [hep-th]}}.

\bibitem{Hijano:2019qmi}
E.~Hijano, ``{Flat space physics from AdS/CFT},''
  \href{http://dx.doi.org/10.1007/JHEP07(2019)132}{{\em JHEP} {\bfseries 07}
  (2019) 132}, \href{http://arxiv.org/abs/1905.02729}{{\ttfamily
  arXiv:1905.02729 [hep-th]}}.

\bibitem{Komatsu:2020sag}
S.~Komatsu, M.~F. Paulos, B.~C. Van~Rees, and X.~Zhao, ``{Landau diagrams in
  AdS and S-matrices from conformal correlators},''
  \href{http://dx.doi.org/10.1007/JHEP11(2020)046}{{\em JHEP} {\bfseries 11}
  (2020) 046}, \href{http://arxiv.org/abs/2007.13745}{{\ttfamily
  arXiv:2007.13745 [hep-th]}}.

\bibitem{Compere:2020lrt}
G.~Comp\`ere, A.~Fiorucci, and R.~Ruzziconi, ``{The $\Lambda$-BMS$_4$ Charge
  Algebra},'' \href{http://arxiv.org/abs/2004.10769}{{\ttfamily
  arXiv:2004.10769 [hep-th]}}.

\bibitem{Li:2021snj}
Y.-Z. Li, ``{Notes on flat-space limit of AdS/CFT},''
  \href{http://dx.doi.org/10.1007/JHEP09(2021)027}{{\em JHEP} {\bfseries 09}
  (2021) 027}, \href{http://arxiv.org/abs/2106.04606}{{\ttfamily
  arXiv:2106.04606 [hep-th]}}.

\bibitem{PipolodeGioia:2022exe}
L.~Pipolode~Gioia and A.-M. Raclariu, ``{Eikonal Approximation in Celestial
  CFT},'' \href{http://arxiv.org/abs/2206.10547}{{\ttfamily arXiv:2206.10547
  [hep-th]}}.

\bibitem{Bagchi:2023fbj}
A.~Bagchi, P.~Dhivakar, and S.~Dutta, ``{AdS Witten Diagrams to Carrollian
  Correlators},'' \href{http://arxiv.org/abs/2303.07388}{{\ttfamily
  arXiv:2303.07388 [hep-th]}}.

\bibitem{deGioia:2023cbd}
L.~P. de~Gioia and A.-M. Raclariu, ``{Celestial sector in CFT: Conformally soft
  symmetries},'' \href{http://dx.doi.org/10.21468/SciPostPhys.17.1.002}{{\em
  SciPost Phys.} {\bfseries 17} no.~1, (2024) 002},
  \href{http://arxiv.org/abs/2303.10037}{{\ttfamily arXiv:2303.10037
  [hep-th]}}.

\bibitem{Pasterski:2016qvg}
S.~Pasterski, S.-H. Shao, and A.~Strominger, ``{Flat Space Amplitudes and
  Conformal Symmetry of the Celestial Sphere},''
  \href{http://dx.doi.org/10.1103/PhysRevD.96.065026}{{\em Phys. Rev.}
  {\bfseries D96} no.~6, (2017) 065026},
\href{http://arxiv.org/abs/1701.00049}{{\ttfamily arXiv:1701.00049 [hep-th]}}.

\bibitem{Raclariu:2021zjz}
A.-M. Raclariu, ``{Lectures on Celestial Holography},''
  \href{http://arxiv.org/abs/2107.02075}{{\ttfamily arXiv:2107.02075
  [hep-th]}}.

\bibitem{Pasterski:2021rjz}
S.~Pasterski, ``{Lectures on celestial amplitudes},''
  \href{http://dx.doi.org/10.1140/epjc/s10052-021-09846-7}{{\em Eur. Phys. J.
  C} {\bfseries 81} no.~12, (2021) 1062},
  \href{http://arxiv.org/abs/2108.04801}{{\ttfamily arXiv:2108.04801
  [hep-th]}}.

\bibitem{Donnay:2023mrd}
L.~Donnay, ``{Celestial holography: An asymptotic symmetry perspective},''
  \href{http://dx.doi.org/10.1016/j.physrep.2024.04.003}{{\em Phys. Rept.}
  {\bfseries 1073} (2024) 1--41},
  \href{http://arxiv.org/abs/2310.12922}{{\ttfamily arXiv:2310.12922
  [hep-th]}}.

\bibitem{Bittleston:2024rqe}
R.~Bittleston, G.~Bogna, S.~Heuveline, A.~Kmec, L.~Mason, and D.~Skinner, ``{On
  AdS$_{4}$ deformations of celestial symmetries},''
  \href{http://dx.doi.org/10.1007/JHEP07(2024)010}{{\em JHEP} {\bfseries 07}
  (2024) 010}, \href{http://arxiv.org/abs/2403.18011}{{\ttfamily
  arXiv:2403.18011 [hep-th]}}.

\bibitem{Hofman:2008ar}
D.~M. Hofman and J.~Maldacena, ``{Conformal collider physics: Energy and charge
  correlations},'' \href{http://dx.doi.org/10.1088/1126-6708/2008/05/012}{{\em
  JHEP} {\bfseries 05} (2008) 012},
  \href{http://arxiv.org/abs/0803.1467}{{\ttfamily arXiv:0803.1467 [hep-th]}}.

\bibitem{Hu:2022txx}
Y.~Hu and S.~Pasterski, ``{Celestial Conformal Colliders},''
  \href{http://arxiv.org/abs/2211.14287}{{\ttfamily arXiv:2211.14287
  [hep-th]}}.

\bibitem{Hu:2023geb}
Y.~Hu and S.~Pasterski, ``{Detector operators for celestial symmetries},''
  \href{http://dx.doi.org/10.1007/JHEP12(2023)035}{{\em JHEP} {\bfseries 12}
  (2023) 035}, \href{http://arxiv.org/abs/2307.16801}{{\ttfamily
  arXiv:2307.16801 [hep-th]}}.

\bibitem{Costello:2022jpg}
K.~Costello, N.~M. Paquette, and A.~Sharma, ``{Top-down holography in an
  asymptotically flat spacetime},''
  \href{http://arxiv.org/abs/2208.14233}{{\ttfamily arXiv:2208.14233
  [hep-th]}}.

\bibitem{Costello:2023hmi}
K.~Costello, N.~M. Paquette, and A.~Sharma, ``{Burns space and holography},''
  \href{http://dx.doi.org/10.1007/JHEP10(2023)174}{{\em JHEP} {\bfseries 10}
  (2023) 174}, \href{http://arxiv.org/abs/2306.00940}{{\ttfamily
  arXiv:2306.00940 [hep-th]}}.

\bibitem{Bittleston:2024efo}
R.~Bittleston, K.~Costello, and K.~Zeng, ``{Self-Dual Gauge Theory from the Top
  Down},'' \href{http://arxiv.org/abs/2412.02680}{{\ttfamily arXiv:2412.02680
  [hep-th]}}.

\bibitem{Donnay:2022aba}
L.~Donnay, A.~Fiorucci, Y.~Herfray, and R.~Ruzziconi, ``{A Carrollian
  Perspective on Celestial Holography},''
  \href{http://arxiv.org/abs/2202.04702}{{\ttfamily arXiv:2202.04702
  [hep-th]}}.

\bibitem{Bagchi:2022emh}
A.~Bagchi, S.~Banerjee, R.~Basu, and S.~Dutta, ``{Scattering Amplitudes:
  Celestial and Carrollian},''
  \href{http://dx.doi.org/10.1103/PhysRevLett.128.241601}{{\em Phys. Rev.
  Lett.} {\bfseries 128} no.~24, (2022) 241601},
  \href{http://arxiv.org/abs/2202.08438}{{\ttfamily arXiv:2202.08438
  [hep-th]}}.

\bibitem{Donnay:2022wvx}
L.~Donnay, A.~Fiorucci, Y.~Herfray, and R.~Ruzziconi, ``{Bridging Carrollian
  and Celestial Holography},''
  \href{http://arxiv.org/abs/2212.12553}{{\ttfamily arXiv:2212.12553
  [hep-th]}}.

\bibitem{Ciambelli:2022vot}
L.~Ciambelli, ``{From Asymptotic Symmetries to the Corner Proposal},''
  \href{http://dx.doi.org/10.22323/1.435.0002}{{\em PoS} {\bfseries Modave2022}
  (2023) 002}, \href{http://arxiv.org/abs/2212.13644}{{\ttfamily
  arXiv:2212.13644 [hep-th]}}.

\bibitem{Pate:2019lpp}
M.~Pate, A.-M. Raclariu, A.~Strominger, and E.~Y. Yuan, ``{Celestial Operator
  Products of Gluons and Gravitons},''
  \href{http://arxiv.org/abs/1910.07424}{{\ttfamily arXiv:1910.07424
  [hep-th]}}.

\bibitem{Banerjee:2020kaa}
S.~Banerjee, S.~Ghosh, and R.~Gonzo, ``{BMS symmetry of celestial OPE},''
  \href{http://dx.doi.org/10.1007/JHEP04(2020)130}{{\em JHEP} {\bfseries 04}
  (2020) 130}, \href{http://arxiv.org/abs/2002.00975}{{\ttfamily
  arXiv:2002.00975 [hep-th]}}.

\bibitem{Jiang:2021csc}
H.~Jiang, ``{Celestial OPEs and $ w_{1+\infty}$ algebra from worldsheet in
  string theory},'' \href{http://arxiv.org/abs/2110.04255}{{\ttfamily
  arXiv:2110.04255 [hep-th]}}.

\bibitem{Adamo:2021zpw}
T.~Adamo, W.~Bu, E.~Casali, and A.~Sharma, ``{Celestial operator products from
  the worldsheet},'' \href{http://arxiv.org/abs/2111.02279}{{\ttfamily
  arXiv:2111.02279 [hep-th]}}.

\bibitem{Bittleston:2022jeq}
R.~Bittleston, ``{On the associativity of 1-loop corrections to the celestial
  operator product in gravity},''
  \href{http://arxiv.org/abs/2211.06417}{{\ttfamily arXiv:2211.06417
  [hep-th]}}.

\bibitem{Nandan:2019jas}
D.~Nandan, A.~Schreiber, A.~Volovich, and M.~Zlotnikov, ``{Celestial
  Amplitudes: Conformal Partial Waves and Soft Limits},''
  \href{http://dx.doi.org/10.1007/JHEP10(2019)018}{{\em JHEP} {\bfseries 10}
  (2019) 018},
\href{http://arxiv.org/abs/1904.10940}{{\ttfamily arXiv:1904.10940 [hep-th]}}.

\bibitem{Fan:2021isc}
W.~Fan, A.~Fotopoulos, S.~Stieberger, T.~R. Taylor, and B.~Zhu, ``{Conformal
  Blocks from Celestial Gluon Amplitudes},''
  \href{http://arxiv.org/abs/2103.04420}{{\ttfamily arXiv:2103.04420
  [hep-th]}}.

\bibitem{Atanasov:2021cje}
A.~Atanasov, W.~Melton, A.-M. Raclariu, and A.~Strominger, ``{Conformal Block
  Expansion in Celestial CFT},''
  \href{http://arxiv.org/abs/2104.13432}{{\ttfamily arXiv:2104.13432
  [hep-th]}}.

\bibitem{Fan:2021pbp}
W.~Fan, A.~Fotopoulos, S.~Stieberger, T.~R. Taylor, and B.~Zhu, ``{Conformal
  blocks from celestial gluon amplitudes. Part II. Single-valued
  correlators},'' \href{http://dx.doi.org/10.1007/JHEP11(2021)179}{{\em JHEP}
  {\bfseries 11} (2021) 179}, \href{http://arxiv.org/abs/2108.10337}{{\ttfamily
  arXiv:2108.10337 [hep-th]}}.

\bibitem{Guevara:2021tvr}
A.~Guevara, ``{Celestial OPE blocks},''
  \href{http://arxiv.org/abs/2108.12706}{{\ttfamily arXiv:2108.12706
  [hep-th]}}.

\bibitem{Hu:2022syq}
Y.~Hu, L.~Lippstreu, M.~Spradlin, A.~Y. Srikant, and A.~Volovich, ``{Four-point
  correlators of light-ray operators in CCFT},''
  \href{http://dx.doi.org/10.1007/JHEP07(2022)104}{{\em JHEP} {\bfseries 07}
  (2022) 104}, \href{http://arxiv.org/abs/2203.04255}{{\ttfamily
  arXiv:2203.04255 [hep-th]}}.

\bibitem{De:2022gjn}
S.~De, Y.~Hu, A.~Yelleshpur~Srikant, and A.~Volovich, ``{Correlators of four
  light-ray operators in CCFT},''
  \href{http://dx.doi.org/10.1007/JHEP10(2022)170}{{\em JHEP} {\bfseries 10}
  (2022) 170}, \href{http://arxiv.org/abs/2206.08875}{{\ttfamily
  arXiv:2206.08875 [hep-th]}}.

\bibitem{Fan:2022vbz}
W.~Fan, A.~Fotopoulos, S.~Stieberger, T.~R. Taylor, and B.~Zhu, ``{Elements of
  Celestial Conformal Field Theory},''
  \href{http://arxiv.org/abs/2202.08288}{{\ttfamily arXiv:2202.08288
  [hep-th]}}.

\bibitem{Fan:2022kpp}
W.~Fan, A.~Fotopoulos, S.~Stieberger, T.~R. Taylor, and B.~Zhu, ``{Celestial
  Yang-Mills amplitudes and D = 4 conformal blocks},''
  \href{http://dx.doi.org/10.1007/JHEP09(2022)182}{{\em JHEP} {\bfseries 09}
  (2022) 182}, \href{http://arxiv.org/abs/2206.08979}{{\ttfamily
  arXiv:2206.08979 [hep-th]}}.

\bibitem{Garcia-Sepulveda:2022lga}
D.~Garc\'\i{}a-Sep\'ulveda, A.~Guevara, J.~Kulp, and J.~Wu, ``{Notes on
  resonances and unitarity from celestial amplitudes},''
  \href{http://dx.doi.org/10.1007/JHEP09(2022)245}{{\em JHEP} {\bfseries 09}
  (2022) 245}, \href{http://arxiv.org/abs/2205.14633}{{\ttfamily
  arXiv:2205.14633 [hep-th]}}.

\bibitem{Jorge-Diaz:2022dmy}
C.~Jorge-Diaz, S.~Pasterski, and A.~Sharma, ``{Celestial amplitudes in an
  ambidextrous basis},'' \href{http://arxiv.org/abs/2212.00962}{{\ttfamily
  arXiv:2212.00962 [hep-th]}}.

\bibitem{Pasterski:2017kqt}
S.~Pasterski and S.-H. Shao, ``{Conformal basis for flat space amplitudes},''
  \href{http://dx.doi.org/10.1103/PhysRevD.96.065022}{{\em Phys. Rev.}
  {\bfseries D96} no.~6, (2017) 065022},
\href{http://arxiv.org/abs/1705.01027}{{\ttfamily arXiv:1705.01027 [hep-th]}}.

\bibitem{Crawley:2021ivb}
E.~Crawley, N.~Miller, S.~A. Narayanan, and A.~Strominger, ``{State-operator
  correspondence in celestial conformal field theory},''
  \href{http://dx.doi.org/10.1007/JHEP09(2021)132}{{\em JHEP} {\bfseries 09}
  (2021) 132}, \href{http://arxiv.org/abs/2105.00331}{{\ttfamily
  arXiv:2105.00331 [hep-th]}}.

\bibitem{Lam:2017ofc}
H.~T. Lam and S.-H. Shao, ``{Conformal Basis, Optical Theorem, and the Bulk
  Point Singularity},''
\href{http://arxiv.org/abs/1711.06138}{{\ttfamily arXiv:1711.06138 [hep-th]}}.

\bibitem{Ren:2022sws}
L.~Ren, M.~Spradlin, A.~Yelleshpur~Srikant, and A.~Volovich, ``{On effective
  field theories with celestial duals},''
  \href{http://dx.doi.org/10.1007/JHEP08(2022)251}{{\em JHEP} {\bfseries 08}
  (2022) 251}, \href{http://arxiv.org/abs/2206.08322}{{\ttfamily
  arXiv:2206.08322 [hep-th]}}.

\bibitem{Bhardwaj:2022anh}
R.~Bhardwaj, L.~Lippstreu, L.~Ren, M.~Spradlin, A.~Yelleshpur~Srikant, and
  A.~Volovich, ``{Loop-level gluon OPEs in celestial holography},''
  \href{http://arxiv.org/abs/2208.14416}{{\ttfamily arXiv:2208.14416
  [hep-th]}}.

\bibitem{Ball:2022bgg}
A.~Ball, ``{Celestial Locality and the Jacobi Identity},''
  \href{http://arxiv.org/abs/2211.09151}{{\ttfamily arXiv:2211.09151
  [hep-th]}}.

\bibitem{Fitzpatrick:2010zm}
A.~L. Fitzpatrick, E.~Katz, D.~Poland, and D.~Simmons-Duffin, ``{Effective
  Conformal Theory and the Flat-Space Limit of AdS},''
  \href{http://dx.doi.org/10.1007/JHEP07(2011)023}{{\em JHEP} {\bfseries 07}
  (2011) 023}, \href{http://arxiv.org/abs/1007.2412}{{\ttfamily arXiv:1007.2412
  [hep-th]}}.

\bibitem{Fitzpatrick:2011dm}
A.~L. Fitzpatrick and J.~Kaplan, ``{Unitarity and the Holographic S-Matrix},''
  \href{http://dx.doi.org/10.1007/JHEP10(2012)032}{{\em JHEP} {\bfseries 10}
  (2012) 032}, \href{http://arxiv.org/abs/1112.4845}{{\ttfamily arXiv:1112.4845
  [hep-th]}}.

\bibitem{SimmonsDuffin:2012uy}
D.~Simmons-Duffin, ``{Projectors, Shadows, and Conformal Blocks},''
  \href{http://dx.doi.org/10.1007/JHEP04(2014)146}{{\em JHEP} {\bfseries 04}
  (2014) 146},
\href{http://arxiv.org/abs/1204.3894}{{\ttfamily arXiv:1204.3894 [hep-th]}}.

\bibitem{Czech:2016xec}
B.~Czech, L.~Lamprou, S.~McCandlish, B.~Mosk, and J.~Sully, ``{A Stereoscopic
  Look into the Bulk},'' \href{http://dx.doi.org/10.1007/JHEP07(2016)129}{{\em
  JHEP} {\bfseries 07} (2016) 129},
  \href{http://arxiv.org/abs/1604.03110}{{\ttfamily arXiv:1604.03110
  [hep-th]}}.

\bibitem{Gadde:2017sjg}
A.~Gadde, ``{In search of conformal theories},''
  \href{http://arxiv.org/abs/1702.07362}{{\ttfamily arXiv:1702.07362
  [hep-th]}}.

\bibitem{McLoughlin:2016uwa}
T.~McLoughlin and D.~Nandan, ``{Multi-Soft gluon limits and extended current
  algebras at null-infinity},''
  \href{http://dx.doi.org/10.1007/JHEP08(2017)124}{{\em JHEP} {\bfseries 08}
  (2017) 124}, \href{http://arxiv.org/abs/1610.03841}{{\ttfamily
  arXiv:1610.03841 [hep-th]}}.

\bibitem{Nande:2017dba}
A.~Nande, M.~Pate, and A.~Strominger, ``{Soft Factorization in QED from 2D
  Kac-Moody Symmetry},'' \href{http://dx.doi.org/10.1007/JHEP02(2018)079}{{\em
  JHEP} {\bfseries 02} (2018) 079},
\href{http://arxiv.org/abs/1705.00608}{{\ttfamily arXiv:1705.00608 [hep-th]}}.

\bibitem{Fan:2020xjj}
W.~Fan, A.~Fotopoulos, S.~Stieberger, and T.~R. Taylor, ``{On Sugawara
  construction on Celestial Sphere},''
  \href{http://dx.doi.org/10.1007/JHEP09(2020)139}{{\em JHEP} {\bfseries 09}
  (2020) 139}, \href{http://arxiv.org/abs/2005.10666}{{\ttfamily
  arXiv:2005.10666 [hep-th]}}.

\bibitem{Ebert:2020nqf}
S.~Ebert, A.~Sharma, and D.~Wang, ``{Descendants in celestial CFT and emergent
  multi-collinear factorization},''
  \href{http://arxiv.org/abs/2009.07881}{{\ttfamily arXiv:2009.07881
  [hep-th]}}.

\bibitem{Ball:2023sdz}
A.~Ball, Y.~Hu, and S.~Pasterski, ``{Multicollinear singularities in celestial
  CFT},'' \href{http://dx.doi.org/10.1007/JHEP02(2024)219}{{\em JHEP}
  {\bfseries 02} (2024) 219}, \href{http://arxiv.org/abs/2309.16602}{{\ttfamily
  arXiv:2309.16602 [hep-th]}}.

\bibitem{Guevara:2024ixn}
A.~Guevara, Y.~Hu, and S.~Pasterski, ``{Multiparticle Contributions to the
  Celestial OPE},'' \href{http://arxiv.org/abs/2402.18798}{{\ttfamily
  arXiv:2402.18798 [hep-th]}}.

\bibitem{deGioia:2024yne}
L.~P. de~Gioia and A.-M. Raclariu, ``{Celestial amplitudes from conformal
  correlators with bulk-point kinematics},''
  \href{http://arxiv.org/abs/2405.07972}{{\ttfamily arXiv:2405.07972
  [hep-th]}}.

\bibitem{Alday:2024yyj}
L.~F. Alday, M.~Nocchi, R.~Ruzziconi, and A.~Yelleshpur~Srikant, ``{Carrollian
  Amplitudes from Holographic Correlators},''
  \href{http://arxiv.org/abs/2406.19343}{{\ttfamily arXiv:2406.19343
  [hep-th]}}.

\bibitem{Iacobacci:2024laa}
L.~Iacobacci and K.~Nguyen, ``{Celestial decomposition of Wigner's
  particles},'' \href{http://arxiv.org/abs/2411.19219}{{\ttfamily
  arXiv:2411.19219 [hep-th]}}.

\bibitem{Prabhu:2022zcr}
K.~Prabhu, G.~Satishchandran, and R.~M. Wald, ``{Infrared finite scattering
  theory in quantum field theory and quantum gravity},''
  \href{http://dx.doi.org/10.1103/PhysRevD.106.066005}{{\em Phys. Rev. D}
  {\bfseries 106} no.~6, (2022) 066005},
  \href{http://arxiv.org/abs/2203.14334}{{\ttfamily arXiv:2203.14334
  [hep-th]}}.

\bibitem{Caron-Huot:2023vxl}
S.~Caron-Huot, M.~Giroux, H.~S. Hannesdottir, and S.~Mizera, ``{What can be
  measured asymptotically?},''
  \href{http://arxiv.org/abs/2308.02125}{{\ttfamily arXiv:2308.02125
  [hep-th]}}.

\bibitem{Haag:1958vt}
R.~Haag, ``{Quantum field theories with composite particles and asymptotic
  conditions},'' \href{http://dx.doi.org/10.1103/PhysRev.112.669}{{\em Phys.
  Rev.} {\bfseries 112} (1958) 669--673}.

\bibitem{Araki:1962zhd}
H.~Araki, K.~Hepp, and D.~Ruelle, ``{On the asymptotic behaviour of Wightman
  functions in space-like directions},''
  \href{http://dx.doi.org/10.5169/seals-113273}{{\em Helv. Phys. Acta}
  {\bfseries 35} no.~III, (1962) 164--174}.

\bibitem{Collins:2019ozc}
J.~Collins, ``{A new approach to the LSZ reduction formula},''
  \href{http://arxiv.org/abs/1904.10923}{{\ttfamily arXiv:1904.10923
  [hep-ph]}}.

\bibitem{bogolubov1990haag}
N.~Bogolubov, A.~Logunov, A.~Oksak, I.~Todorov, and G.~Gould, ``{Haag-Ruelle
  Scattering Theory},'' in {\em General Principles of Quantum Field Theory},
  pp.~486--502.
\newblock Springer, 1990.

\bibitem{Banks:1998dd}
T.~Banks, M.~R. Douglas, G.~T. Horowitz, and E.~J. Martinec, ``{AdS dynamics
  from conformal field theory},''
  \href{http://arxiv.org/abs/hep-th/9808016}{{\ttfamily arXiv:hep-th/9808016}}.

\bibitem{Harlow:2011ke}
D.~Harlow and D.~Stanford, ``{Operator Dictionaries and Wave Functions in
  AdS/CFT and dS/CFT},'' \href{http://arxiv.org/abs/1104.2621}{{\ttfamily
  arXiv:1104.2621 [hep-th]}}.

\bibitem{He:2020ifr}
T.~He and P.~Mitra, ``{Covariant Phase Space and Soft Factorization in
  Non-Abelian Gauge Theories},''
  \href{http://dx.doi.org/10.1007/JHEP03(2021)015}{{\em JHEP} {\bfseries 03}
  (2021) 015}, \href{http://arxiv.org/abs/2009.14334}{{\ttfamily
  arXiv:2009.14334 [hep-th]}}.

\bibitem{Kim:2023qbl}
S.~Kim, P.~Kraus, R.~Monten, and R.~M. Myers, ``{S-matrix path integral
  approach to symmetries and soft theorems},''
  \href{http://dx.doi.org/10.1007/JHEP10(2023)036}{{\em JHEP} {\bfseries 10}
  (2023) 036}, \href{http://arxiv.org/abs/2307.12368}{{\ttfamily
  arXiv:2307.12368 [hep-th]}}.

\bibitem{Jain:2023fxc}
D.~Jain, S.~Kundu, S.~Minwalla, O.~Parrikar, S.~G. Prabhu, and P.~Shrivastava,
  ``{The S-matrix and boundary correlators in flat space},''
  \href{http://arxiv.org/abs/2311.03443}{{\ttfamily arXiv:2311.03443
  [hep-th]}}.

\bibitem{Kraus:2024gso}
P.~Kraus and R.~M. Myers, ``{Carrollian Partition Functions and the Flat Limit
  of AdS},'' \href{http://arxiv.org/abs/2407.13668}{{\ttfamily arXiv:2407.13668
  [hep-th]}}.

\bibitem{vanRees:2022itk}
B.~C. van Rees and X.~Zhao, ``{Quantum Field Theory in AdS Space instead of
  Lehmann-Symanzik-Zimmerman Axioms},''
  \href{http://dx.doi.org/10.1103/PhysRevLett.130.191601}{{\em Phys. Rev.
  Lett.} {\bfseries 130} no.~19, (2023) 191601},
  \href{http://arxiv.org/abs/2210.15683}{{\ttfamily arXiv:2210.15683
  [hep-th]}}.

\bibitem{Donnay:2022sdg}
L.~Donnay, S.~Pasterski, and A.~Puhm, ``{Goldilocks Modes and the Three
  Scattering Bases},'' \href{http://arxiv.org/abs/2202.11127}{{\ttfamily
  arXiv:2202.11127 [hep-th]}}.

\bibitem{Jorstad:2024yzm}
E.~J\o{}rstad and S.~Pasterski, ``{A Comment on Boundary Correlators: Soft
  Omissions and the Massless S-Matrix},''
  \href{http://arxiv.org/abs/2410.20296}{{\ttfamily arXiv:2410.20296
  [hep-th]}}.

\bibitem{Jorstad:2023ajr}
E.~J\o{}rstad, S.~Pasterski, and A.~Sharma, ``{Equating extrapolate
  dictionaries for massless scattering},''
  \href{http://dx.doi.org/10.1007/JHEP02(2024)228}{{\em JHEP} {\bfseries 02}
  (2024) 228}, \href{http://arxiv.org/abs/2310.02186}{{\ttfamily
  arXiv:2310.02186 [hep-th]}}.

\bibitem{Geroch:1978us}
R.~P. Geroch, ``{Null infinity is not a good initial data surface},''
  \href{http://dx.doi.org/10.1063/1.523827}{{\em J. Math. Phys.} {\bfseries 19}
  (1978) 1300--1303}.

\bibitem{Tjoa:2022mly}
E.~Tjoa and F.~Gray, ``{Modest holography and bulk reconstruction in
  asymptotically flat spacetimes},''
  \href{http://dx.doi.org/10.1103/PhysRevD.106.025021}{{\em Phys. Rev. D}
  {\bfseries 106} no.~2, (2022) 025021},
  \href{http://arxiv.org/abs/2204.13133}{{\ttfamily arXiv:2204.13133 [gr-qc]}}.

\bibitem{Ashtekar:1987tt}
A.~Ashtekar, {\em {Asymptotic Quantization: Based on 1984 Naples Lectures}}.
\newblock Monographs and Textbooks in Physical Science, 2,
1987.
\newblock

\bibitem{Ashtekar:1981sf}
A.~Ashtekar, ``{Asymptotic Quantization of the Gravitational Field},''
\href{http://dx.doi.org/10.1103/PhysRevLett.46.573}{{\em Phys. Rev. Lett.}
  {\bfseries 46} (1981) 573--576}.

\bibitem{He:2014laa}
T.~He, V.~Lysov, P.~Mitra, and A.~Strominger, ``{BMS supertranslations and
  Weinberg's soft graviton theorem},''
  \href{http://dx.doi.org/10.1007/JHEP05(2015)151}{{\em JHEP} {\bfseries 05}
  (2015) 151},
\href{http://arxiv.org/abs/1401.7026}{{\ttfamily arXiv:1401.7026 [hep-th]}}.

\bibitem{Kapec:2014opa}
D.~Kapec, V.~Lysov, S.~Pasterski, and A.~Strominger, ``{Semiclassical Virasoro
  symmetry of the quantum gravity $ \mathcal{S}$-matrix},''
  \href{http://dx.doi.org/10.1007/JHEP08(2014)058}{{\em JHEP} {\bfseries 08}
  (2014) 058},
\href{http://arxiv.org/abs/1406.3312}{{\ttfamily arXiv:1406.3312 [hep-th]}}.

\bibitem{Heemskerk:2009pn}
I.~Heemskerk, J.~Penedones, J.~Polchinski, and J.~Sully, ``{Holography from
  Conformal Field Theory},''
  \href{http://dx.doi.org/10.1088/1126-6708/2009/10/079}{{\em JHEP} {\bfseries
  10} (2009) 079}, \href{http://arxiv.org/abs/0907.0151}{{\ttfamily
  arXiv:0907.0151 [hep-th]}}.

\bibitem{Fitzpatrick:2011jn}
A.~L. Fitzpatrick and J.~Kaplan, ``{Scattering States in AdS/CFT},''
  \href{http://arxiv.org/abs/1104.2597}{{\ttfamily arXiv:1104.2597 [hep-th]}}.

\bibitem{Fitzpatrick:2011hu}
A.~L. Fitzpatrick and J.~Kaplan, ``{Analyticity and the Holographic
  S-Matrix},'' \href{http://dx.doi.org/10.1007/JHEP10(2012)127}{{\em JHEP}
  {\bfseries 10} (2012) 127}, \href{http://arxiv.org/abs/1111.6972}{{\ttfamily
  arXiv:1111.6972 [hep-th]}}.

\bibitem{Karateev:2018oml}
D.~Karateev, P.~Kravchuk, and D.~Simmons-Duffin, ``{Harmonic Analysis and Mean
  Field Theory},'' \href{http://dx.doi.org/10.1007/JHEP10(2019)217}{{\em JHEP}
  {\bfseries 10} (2019) 217}, \href{http://arxiv.org/abs/1809.05111}{{\ttfamily
  arXiv:1809.05111 [hep-th]}}.

\bibitem{Chang:2023ttm}
C.-M. Chang, R.~Liu, and W.-J. Ma, ``{Split representation in celestial
  holography},'' \href{http://arxiv.org/abs/2311.08736}{{\ttfamily
  arXiv:2311.08736 [hep-th]}}.

\bibitem{Salzer:2023jqv}
J.~Salzer, ``{An embedding space approach to Carrollian CFT correlators for
  flat space holography},''
  \href{http://dx.doi.org/10.1007/JHEP10(2023)084}{{\em JHEP} {\bfseries 10}
  (2023) 084}, \href{http://arxiv.org/abs/2304.08292}{{\ttfamily
  arXiv:2304.08292 [hep-th]}}.

\bibitem{Nguyen:2023miw}
K.~Nguyen, ``{Carrollian conformal correlators and massless scattering
  amplitudes},'' \href{http://arxiv.org/abs/2311.09869}{{\ttfamily
  arXiv:2311.09869 [hep-th]}}.

\bibitem{Dumitrescu:2015fej}
T.~T. Dumitrescu, T.~He, P.~Mitra, and A.~Strominger, ``{Infinite-Dimensional
  Fermionic Symmetry in Supersymmetric Gauge Theories},''
  \href{http://arxiv.org/abs/1511.07429}{{\ttfamily arXiv:1511.07429
  [hep-th]}}.

\bibitem{Bekaert:2006py}
X.~Bekaert and N.~Boulanger, ``{The unitary representations of the Poincar{\'e}
  group in any spacetime dimension},''
  \href{http://dx.doi.org/10.21468/SciPostPhysLectNotes.30}{{\em SciPost Phys.
  Lect. Notes} {\bfseries 30} (2021) 1},
  \href{http://arxiv.org/abs/hep-th/0611263}{{\ttfamily arXiv:hep-th/0611263}}.

\bibitem{inonu1953contraction}
E.~Inonu and E.~P. Wigner, ``On the contraction of groups and their
  representations,'' {\em Proceedings of the National Academy of Sciences}
  {\bfseries 39} no.~6, (1953) 510--524.

\bibitem{mickelsson1972contractions}
J.~Mickelsson and J.~Niederle, ``{Contractions of representations of de Sitter
  groups},'' {\em Communications in Mathematical Physics} {\bfseries 27} (1972)
  167--180.

\bibitem{dooley1983contractions}
A.~H. Dooley and J.~Rice, ``Contractions of rotation groups and their
  representations,'' in {\em Mathematical Proceedings of the Cambridge
  Philosophical Society}, vol.~94-3, pp.~509--517, Cambridge University Press.
\newblock 1983.

\bibitem{dooley1985contractions}
A.~H. Dooley and J.~Rice, ``{On contractions of semisimple Lie groups},'' {\em
  Transactions of the American Mathematical Society} {\bfseries 289} no.~1,
  (1985) 185--202.

\bibitem{cishahayo1994inonu}
C.~Cishahayo, ``{In{\"o}n{\"u}-Wigner Contraction of Kinematical Group
  Representations},'' {\em Quantization and Infinite-Dimensional Systems}
  (1994) 165--173.

\bibitem{de1992quantum}
S.~De~Bi{\`e}vre and A.~M. El~Gradechi, ``{Quantum mechanics and coherent
  states on the anti-de Sitter spacetime and their Poincar{\'e} contraction},''
  in {\em Annales de l'IHP Physique th{\'e}orique}, vol.~57-4, pp.~403--428.
\newblock 1992.

\bibitem{el1993phase}
A.~M. El~Gradechi, ``{Phase space localization for anti-de Sitter quantum
  mechanics and its zero curvature limit},'' in {\em NASA CONFERENCE
  PUBLICATION}, pp.~235--235, NASA.
\newblock 1993.

\bibitem{ElGradechi:1992te}
A.~M. El~Gradechi and S.~De~Bievre, ``{Phase space quantum mechanics on the
  anti-De Sitter space-time and its Poincare contraction},''
  \href{http://dx.doi.org/10.1006/aphy.1994.1089}{{\em Annals Phys.} {\bfseries
  235} (1994) 1--34}, \href{http://arxiv.org/abs/hep-th/9210133}{{\ttfamily
  arXiv:hep-th/9210133}}.

\bibitem{cmp25}
J.~Caminiti, R.~C. Myers, and S.~Pasterski, ``{Swing Surfaces in AdS/CFT},''
  {\em to appear} .

\bibitem{Banerjee:2024yir}
S.~Banerjee, ``{Boundary operators in asymptotically flat space-time},''
  \href{http://arxiv.org/abs/2406.06690}{{\ttfamily arXiv:2406.06690
  [hep-th]}}.

\bibitem{Banerjee:2024hvb}
S.~Banerjee, R.~Basu, and S.~Atul~Bhatkar, ``{Light transformation: A Celestial
  and Carrollian perspective},''
  \href{http://arxiv.org/abs/2407.08379}{{\ttfamily arXiv:2407.08379
  [hep-th]}}.

\bibitem{Pasterski:2017ylz}
S.~Pasterski, S.-H. Shao, and A.~Strominger, ``{Gluon Amplitudes as 2d
  Conformal Correlators},''
  \href{http://dx.doi.org/10.1103/PhysRevD.96.085006}{{\em Phys. Rev.}
  {\bfseries D96} no.~8, (2017) 085006},
\href{http://arxiv.org/abs/1706.03917}{{\ttfamily arXiv:1706.03917 [hep-th]}}.

\bibitem{Weinberg:1995mt}
S.~Weinberg, {\em {The Quantum theory of fields. Vol. 1: Foundations}}.
\newblock Cambridge University Press, 6, 2005.

\bibitem{rideau1966reduction}
G.~Rideau, ``{On the reduction of the regular representation of the
  Poincar{\'e} group},''.

\bibitem{hai1969harmonic}
N.~X. Hai, ``{Harmonic analysis on the Poincar{\'e} group: I. Generalized
  matrix elements},'' {\em Communications in Mathematical Physics} {\bfseries
  12} (1969) 331--350.

\bibitem{hai1971harmonic}
N.~X. Hai, ``{Harmonic analysis on the Poincar{\'e} group: II. The Fourier
  transform},'' {\em Communications in Mathematical Physics} {\bfseries 22}
  (1971) 301--320.

\bibitem{lomont1960decomposition}
J.~Lomont, ``Decomposition of direct products of representations of the
  inhomogeneous lorentz group,'' {\em Journal of Mathematical Physics}
  {\bfseries 1} no.~3, (1960) 237--243.

\bibitem{lomont1967reduction}
J.~Lomont and H.~Moses, ``Reduction of reducible representations of the
  infinitesimal generators of the proper, orthochronous, inhomogeneous lorentz
  group,'' {\em Journal of Mathematical Physics} {\bfseries 8} no.~4, (1967)
  837--850.

\bibitem{raczka1986theory}
R.~Raczka and A.~O. Barut, {\em Theory of group representations and
  applications}.
\newblock World Scientific Publishing Company, 1986.

\bibitem{mackey1951induced}
G.~W. Mackey, ``On induced representations of groups,'' {\em American Journal
  of Mathematics} {\bfseries 73} no.~3, (1951) 576--592.

\bibitem{mackey1952induced}
G.~W. Mackey, ``{Induced representations of locally compact groups I},'' {\em
  Annals of Mathematics} (1952) 101--139.

\bibitem{mackey1953induced}
G.~W. Mackey, ``{Induced representations of locally compact groups II. The
  Frobenius reciprocity theorem},'' {\em Annals of Mathematics} (1953)
  193--221.

\bibitem{Mackey1968InducedRO}
G.~W. Mackey, {\em Induced representations of groups and quantum mechanics}.
\newblock W.A. Benjamin, Inc. and Editore Boringhieri, 1968.

\bibitem{Farrill_Lectures}
J.~Figueroa-O'Farrill, ``{The Theory of Induced Representations in Field
  Theory},''.
  \href{https://www.maths.ed.ac.uk/~jmf/Teaching/Projects/Poincare/IndReps.pdf}{https://www.maths.ed.ac.uk/~jmf/Teaching/Projects/Poincare/IndReps.pdf}.

\bibitem{Oblak:2016eij}
B.~Oblak, \href{http://dx.doi.org/10.1007/978-3-319-61878-4}{{\em {BMS
  Particles in Three Dimensions}}}.
\newblock PhD thesis, U. Brussels, Brussels U., 2016.
\newblock \href{http://arxiv.org/abs/1610.08526}{{\ttfamily arXiv:1610.08526
  [hep-th]}}.

\bibitem{shapiro1956expansion}
I.~Shapiro, ``{Expansion of a Wave Function in Irreducible Representation of
  the Lorentz Group},'' in {\em Soviet Physics Doklady}, vol.~1, p.~91.
\newblock 1956.

\bibitem{zastavenko1960integral}
L.~Zastavenko and K.~Chou, ``Integral transfomations of the {I.S.} {S}hapiro
  type for zero mass particles,'' {\em Zhur. Eksptl'. i Teoret. Fiz.}
  {\bfseries 38} (1960) 77--80.

\bibitem{Joos:1962qq}
H.~Joos, ``{Zur Darstellungstheorie der inhomogenen Lorentzgruppe als Grundlage
  quantenmechanischer Kinematik},''
  \href{http://dx.doi.org/10.1002/prop.2180100302}{{\em Fortsch. Phys.}
  {\bfseries 10} (1962) 65--146}.

\bibitem{Banerjee:2018gce}
S.~Banerjee, ``{Null Infinity and Unitary Representation of The Poincare
  Group},'' \href{http://dx.doi.org/10.1007/JHEP01(2019)205}{{\em JHEP}
  {\bfseries 01} (2019) 205}, \href{http://arxiv.org/abs/1801.10171}{{\ttfamily
  arXiv:1801.10171 [hep-th]}}.

\bibitem{mukunda1968zero}
N.~Mukunda, ``{Zero-Mass Representations of the Poincar{\'e} Group in an O(3,
  1) Basis},'' {\em Journal of Mathematical Physics} {\bfseries 9} no.~4,
  (1968) 532--536.

\bibitem{chakrabarti1968lorentz}
A.~Chakrabarti, M.~Levy-Nahas, and R.~Seneor, ``{``Lorentz Basis''of the
  Poincar{\'e} Group},'' {\em Journal of Mathematical Physics} {\bfseries 9}
  no.~8, (1968) 1274--1283.

\bibitem{ruhl1969convolution}
W.~R{\"u}hl, ``{The convolution of Fourier transforms and its application to
  the decomposition of the momentum operator on the homogeneous Lorentz
  group},'' {\em Nuovo Cimento} {\bfseries 63} no.~CERN-TH-1016, (1969)
  1131--1162.

\bibitem{chakrabarti1971lorentz}
A.~Chakrabarti, ``{Lorentz basis of the Poincar{\'e} group. II}'' {\em Journal
  of Mathematical Physics} {\bfseries 12} no.~9, (1971) 1822--1840.

\bibitem{macdowell1972reduction}
W.~Macdowell and R.~Roskies, ``{Reduction of the Poincar{\'e} group with
  respect to the Lorentz group},'' {\em Journal of Mathematical Physics}
  {\bfseries 13} no.~10, (1972) 1585--1592.

\bibitem{cantoni1975structure}
V.~Cantoni, ``{On the structure of the representations of the Poincar{\'e}
  group with zero mass and discrete helicity},'' {\em Rendiconti del Circolo
  Matematico di Palermo} {\bfseries 24} (1975) 35--50.

\bibitem{Mizera:2022sln}
S.~Mizera and S.~Pasterski, ``{Celestial geometry},''
  \href{http://dx.doi.org/10.1007/JHEP09(2022)045}{{\em JHEP} {\bfseries 09}
  (2022) 045}, \href{http://arxiv.org/abs/2204.02505}{{\ttfamily
  arXiv:2204.02505 [hep-th]}}.

\bibitem{Conrady:2010sx}
F.~Conrady and J.~Hnybida, ``{Unitary irreducible representations of SL(2,C) in
  discrete and continuous SU(1,1) bases},''
  \href{http://dx.doi.org/10.1063/1.3533393}{{\em J. Math. Phys.} {\bfseries
  52} (2011) 012501}, \href{http://arxiv.org/abs/1007.0937}{{\ttfamily
  arXiv:1007.0937 [gr-qc]}}.

\bibitem{novozhilov1969representations}
Y.~V. Novozhilov and E.~V. Prokhvatilov, ``{Representations of the Poincar\'e
  group in E(2) bases},'' {\em Teoreticheskaya i Matematicheskaya Fizika}
  {\bfseries 1} no.~1, (1969) 101--119.

\bibitem{itzykson1969group}
C.~Itzykson, ``Group representation in a continuous basis: An example,'' {\em
  Journal of Mathematical Physics} {\bfseries 10} no.~6, (1969) 1109--1114.

\bibitem{flajolet1995mellin}
P.~Flajolet, X.~Gourdon, and P.~Dumas, ``Mellin transforms and asymptotics:
  Harmonic sums,'' {\em Theoretical computer science} {\bfseries 144} no.~1-2,
  (1995) 3--58.

\bibitem{bertrand1995mellin}
J.~Bertrand, P.~Bertrand, and J.-P. Ovarlez, ``{The Mellin transform},'' {\em
  The transforms and applications handbook} (1995) Ch.11.

\bibitem{Guevara:2019ypd}
A.~Guevara, ``{Notes on Conformal Soft Theorems and Recursion Relations in
  Gravity},''
\href{http://arxiv.org/abs/1906.07810}{{\ttfamily arXiv:1906.07810 [hep-th]}}.

\bibitem{dobrev1986lecture}
V.~Dobrev, V.~Petkova, S.~Petrova, and I.~Todorov, ``{Harmonic analysis on the
  n-dimensional Lorentz group and its application to conformal quantum field
  theory},'' {\em Lecture Notes in Physics} {\bfseries 63} (1986) 281.

\bibitem{Law:2019glh}
Y.~T.~A. Law and M.~Zlotnikov, ``{Poincar\'e Constraints on Celestial
  Amplitudes},''
\href{http://arxiv.org/abs/1910.04356}{{\ttfamily arXiv:1910.04356 [hep-th]}}.

\bibitem{Chang:2022seh}
C.-M. Chang and W.-J. Ma, ``{Missing corner in the sky: massless three-point
  celestial amplitudes},''
  \href{http://dx.doi.org/10.1007/JHEP04(2023)051}{{\em JHEP} {\bfseries 04}
  (2023) 051}, \href{http://arxiv.org/abs/2212.07025}{{\ttfamily
  arXiv:2212.07025 [hep-th]}}.

\bibitem{Simmons-Duffin:2017nub}
D.~Simmons-Duffin, D.~Stanford, and E.~Witten, ``{A spacetime derivation of the
  Lorentzian OPE inversion formula},''
  \href{http://dx.doi.org/10.1007/JHEP07(2018)085}{{\em JHEP} {\bfseries 07}
  (2018) 085}, \href{http://arxiv.org/abs/1711.03816}{{\ttfamily
  arXiv:1711.03816 [hep-th]}}.

\bibitem{Mazac:2018qmi}
D.~Maz\'a\v{c}, ``{A Crossing-Symmetric OPE Inversion Formula},''
  \href{http://dx.doi.org/10.1007/JHEP06(2019)082}{{\em JHEP} {\bfseries 06}
  (2019) 082}, \href{http://arxiv.org/abs/1812.02254}{{\ttfamily
  arXiv:1812.02254 [hep-th]}}.

\bibitem{ShuHengTalk}
S.-H. Shao, ``{Celestial Free Fields},''. Corfu Summer Workshop on Celestial
  Amplitudes and Flat Space Holography, 2021,
  \href{https://www.youtube.com/watch?v=IlmiP8rsS2A&t=750s}{https://www.youtube.com/watch?v=IlmiP8rsS2A}.

\bibitem{Fan:2023lky}
W.~Fan, ``{Celestial conformal blocks of massless scalars and analytic
  continuation of the Appell function F$_{1}$},''
  \href{http://dx.doi.org/10.1007/JHEP01(2024)145}{{\em JHEP} {\bfseries 01}
  (2024) 145}, \href{http://arxiv.org/abs/2311.11345}{{\ttfamily
  arXiv:2311.11345 [hep-th]}}.

\bibitem{Law:2020xcf}
Y.~T.~A. Law and M.~Zlotnikov, ``{Relativistic partial waves for celestial
  amplitudes},'' \href{http://dx.doi.org/10.1007/JHEP11(2020)149}{{\em JHEP}
  {\bfseries 11} (2020) 149}, \href{http://arxiv.org/abs/2008.02331}{{\ttfamily
  arXiv:2008.02331 [hep-th]}}.

\bibitem{Law:2020tsg}
Y.~A. Law and M.~Zlotnikov, ``{Massive Spinning Bosons on the Celestial
  Sphere},'' \href{http://dx.doi.org/10.1007/JHEP06(2020)079}{{\em JHEP}
  {\bfseries 06} (2020) 079}, \href{http://arxiv.org/abs/2004.04309}{{\ttfamily
  arXiv:2004.04309 [hep-th]}}.

\bibitem{Schreiber:2017jsr}
A.~Schreiber, A.~Volovich, and M.~Zlotnikov, ``{Tree-level gluon amplitudes on
  the celestial sphere},''
  \href{http://dx.doi.org/10.1016/j.physletb.2018.04.010}{{\em Phys. Lett.}
  {\bfseries B781} (2018) 349--357},
\href{http://arxiv.org/abs/1711.08435}{{\ttfamily arXiv:1711.08435 [hep-th]}}.

\bibitem{Muck:2020wtx}
L.~Iacobacci and W.~M\"uck, ``{Conformal Primary Basis for Dirac Spinors},''
  \href{http://dx.doi.org/10.1103/PhysRevD.102.106025}{{\em Phys. Rev. D}
  {\bfseries 102} no.~10, (2020) 106025},
  \href{http://arxiv.org/abs/2009.02938}{{\ttfamily arXiv:2009.02938
  [hep-th]}}.

\bibitem{Narayanan:2020amh}
S.~A. Narayanan, ``{Massive Celestial Fermions},''
  \href{http://arxiv.org/abs/2009.03883}{{\ttfamily arXiv:2009.03883
  [hep-th]}}.

\bibitem{Costa:2011mg}
M.~S. Costa, J.~Penedones, D.~Poland, and S.~Rychkov, ``{Spinning Conformal
  Correlators},'' \href{http://dx.doi.org/10.1007/JHEP11(2011)071}{{\em JHEP}
  {\bfseries 11} (2011) 071},
\href{http://arxiv.org/abs/1107.3554}{{\ttfamily arXiv:1107.3554 [hep-th]}}.

\bibitem{Chen:2023naw}
B.~Chen and Z.~Hu, ``{Bulk reconstruction in flat holography},''
  \href{http://arxiv.org/abs/2312.13574}{{\ttfamily arXiv:2312.13574
  [hep-th]}}.

\bibitem{Have:2024dff}
E.~Have, K.~Nguyen, S.~Prohazka, and J.~Salzer, ``{Massive carrollian fields at
  timelike infinity},'' \href{http://dx.doi.org/10.1007/JHEP07(2024)054}{{\em
  JHEP} {\bfseries 07} (2024) 054},
  \href{http://arxiv.org/abs/2402.05190}{{\ttfamily arXiv:2402.05190
  [hep-th]}}.

\bibitem{Jain:2014nza}
S.~Jain, M.~Mandlik, S.~Minwalla, T.~Takimi, S.~R. Wadia, and S.~Yokoyama,
  ``{Unitarity, Crossing Symmetry and Duality of the S-matrix in large N
  Chern-Simons theories with fundamental matter},''
  \href{http://dx.doi.org/10.1007/JHEP04(2015)129}{{\em JHEP} {\bfseries 04}
  (2015) 129}, \href{http://arxiv.org/abs/1404.6373}{{\ttfamily arXiv:1404.6373
  [hep-th]}}.

\bibitem{Mehta:2022lgq}
U.~Mehta, S.~Minwalla, C.~Patel, S.~Prakash, and K.~Sharma, ``{Crossing
  Symmetry in Matter Chern-Simons Theories at finite $N$ and $k$},''
  \href{http://dx.doi.org/10.4310/ATMP.2023.v27.n1.a5}{{\em Adv. Theor. Math.
  Phys.} {\bfseries 27} (2023) 193--310},
  \href{http://arxiv.org/abs/2210.07272}{{\ttfamily arXiv:2210.07272
  [hep-th]}}.

\bibitem{Minwalla:2022sef}
S.~Minwalla, A.~Mishra, N.~Prabhakar, and T.~Sharma, ``{The Hilbert space of
  large N Chern-Simons matter theories},''
  \href{http://dx.doi.org/10.1007/JHEP07(2022)025}{{\em JHEP} {\bfseries 07}
  (2022) 025}, \href{http://arxiv.org/abs/2201.08410}{{\ttfamily
  arXiv:2201.08410 [hep-th]}}.

\bibitem{Gabai:2022snc}
B.~Gabai, J.~Sandor, and X.~Yin, ``{Anyon scattering from lightcone
  Hamiltonian: the singlet channel},''
  \href{http://dx.doi.org/10.1007/JHEP09(2022)145}{{\em JHEP} {\bfseries 09}
  (2022) 145}, \href{http://arxiv.org/abs/2205.09144}{{\ttfamily
  arXiv:2205.09144 [hep-th]}}.

\bibitem{Hawking:2015qqa}
S.~Hawking, ``{The Information Paradox for Black Holes},''
  \href{http://arxiv.org/abs/1509.01147}{{\ttfamily arXiv:1509.01147
  [hep-th]}}.

\bibitem{Hawking:2016msc}
S.~W. Hawking, M.~J. Perry, and A.~Strominger, ``{Soft Hair on Black Holes},''
  \href{http://dx.doi.org/10.1103/PhysRevLett.116.231301}{{\em Phys. Rev.
  Lett.} {\bfseries 116} no.~23, (2016) 231301},
\href{http://arxiv.org/abs/1601.00921}{{\ttfamily arXiv:1601.00921 [hep-th]}}.

\bibitem{Hawking:2016sgy}
S.~W. Hawking, M.~J. Perry, and A.~Strominger, ``{Superrotation Charge and
  Supertranslation Hair on Black Holes},''
  \href{http://dx.doi.org/10.1007/JHEP05(2017)161}{{\em JHEP} {\bfseries 05}
  (2017) 161}, \href{http://arxiv.org/abs/1611.09175}{{\ttfamily
  arXiv:1611.09175 [hep-th]}}.

\bibitem{Satishchandran:2019pyc}
G.~Satishchandran and R.~M. Wald, ``{Asymptotic behavior of massless fields and
  the memory effect},''
  \href{http://dx.doi.org/10.1103/PhysRevD.99.084007}{{\em Phys. Rev. D}
  {\bfseries 99} no.~8, (2019) 084007},
  \href{http://arxiv.org/abs/1901.05942}{{\ttfamily arXiv:1901.05942 [gr-qc]}}.

\bibitem{Kulish:1970ut}
P.~P. Kulish and L.~D. Faddeev, ``{Asymptotic conditions and infrared
  divergences in quantum electrodynamics},''
  \href{http://dx.doi.org/10.1007/BF01066485}{{\em Theor. Math. Phys.}
  {\bfseries 4} (1970) 745}.

\bibitem{Hannesdottir:2019opa}
H.~Hannesdottir and M.~D. Schwartz, ``{$S$-Matrix for massless particles},''
  \href{http://dx.doi.org/10.1103/PhysRevD.101.105001}{{\em Phys. Rev. D}
  {\bfseries 101} no.~10, (2020) 105001},
  \href{http://arxiv.org/abs/1911.06821}{{\ttfamily arXiv:1911.06821
  [hep-th]}}.

\bibitem{Hannesdottir:2019umk}
H.~Hannesdottir and M.~D. Schwartz, ``{Finite $S$ matrix},''
  \href{http://dx.doi.org/10.1103/PhysRevD.107.L021701}{{\em Phys. Rev. D}
  {\bfseries 107} no.~2, (2023) L021701},
  \href{http://arxiv.org/abs/1906.03271}{{\ttfamily arXiv:1906.03271
  [hep-th]}}.

\bibitem{Zamolodchikov:1992zr}
A.~B. Zamolodchikov and A.~B. Zamolodchikov, ``{Massless factorized scattering
  and sigma models with topological terms},''
  \href{http://dx.doi.org/10.1016/0550-3213(92)90136-Y}{{\em Nucl. Phys. B}
  {\bfseries 379} (1992) 602--623}.

\bibitem{Fendley:1993xa}
P.~Fendley, H.~Saleur, and A.~B. Zamolodchikov, ``{Massless flows, 2. The Exact
  S matrix approach},'' \href{http://dx.doi.org/10.1142/S0217751X93002277}{{\em
  Int. J. Mod. Phys. A} {\bfseries 8} (1993) 5751--5778},
  \href{http://arxiv.org/abs/hep-th/9304051}{{\ttfamily arXiv:hep-th/9304051}}.

\bibitem{Hubner:2019sly}
M.~H\"ubner, C.~Talbot, P.~D. Lasky, and E.~Thrane, ``{Measuring
  gravitational-wave memory in the first LIGO/Virgo gravitational-wave
  transient catalog},''
  \href{http://dx.doi.org/10.1103/PhysRevD.101.023011}{{\em Phys. Rev. D}
  {\bfseries 101} no.~2, (2020) 023011},
  \href{http://arxiv.org/abs/1911.12496}{{\ttfamily arXiv:1911.12496
  [astro-ph.HE]}}.

\bibitem{Mitman:2020pbt}
K.~Mitman, J.~Moxon, M.~A. Scheel, S.~A. Teukolsky, M.~Boyle, N.~Deppe, L.~E.
  Kidder, and W.~Throwe, ``{Computation of displacement and spin gravitational
  memory in numerical relativity},''
  \href{http://dx.doi.org/10.1103/PhysRevD.102.104007}{{\em Phys. Rev. D}
  {\bfseries 102} no.~10, (2020) 104007},
  \href{http://arxiv.org/abs/2007.11562}{{\ttfamily arXiv:2007.11562 [gr-qc]}}.

\bibitem{kibble1968coherent1}
T.~Kibble, ``{Coherent Soft-Photon States and Infrared Divergences. I.
  Classical Currents},'' {\em Journal of Mathematical Physics} {\bfseries 9}
  no.~2, (1968) 315--324.

\bibitem{kibble1968coherent2}
T.~Kibble, ``{Coherent soft-photon states and infrared divergences. II.
  Mass-shell singularities of Green's functions},'' {\em Physical Review}
  {\bfseries 173} no.~5, (1968) 1527.

\bibitem{kibble1968coherent3}
T.~Kibble, ``{Coherent soft-photon states and infrared divergences. III.
  Asymptotic states and reduction formulas},'' {\em Physical Review} {\bfseries
  174} no.~5, (1968) 1882.

\bibitem{kibble1968coherent4}
T.~Kibble, ``{Coherent soft-photon states and infrared divergences. IV. The
  scattering operator},'' {\em Physical Review} {\bfseries 175} no.~5, (1968)
  1624.

\bibitem{Prabhu:2018gzs}
K.~Prabhu, ``{Conservation of asymptotic charges from past to future null
  infinity: Maxwell fields},''
  \href{http://dx.doi.org/10.1007/JHEP10(2018)113}{{\em JHEP} {\bfseries 10}
  (2018) 113}, \href{http://arxiv.org/abs/1808.07863}{{\ttfamily
  arXiv:1808.07863 [gr-qc]}}.

\bibitem{Bekaert:2024jxs}
X.~Bekaert, L.~Donnay, and Y.~Herfray, ``{BMS particles},''
  \href{http://arxiv.org/abs/2412.06002}{{\ttfamily arXiv:2412.06002
  [hep-th]}}.

\bibitem{Bondi:1962px}
H.~Bondi, M.~G.~J. van~der Burg, and A.~W.~K. Metzner, ``{Gravitational waves
  in general relativity. 7. Waves from axisymmetric isolated systems},''
\href{http://dx.doi.org/10.1098/rspa.1962.0161}{{\em Proc. Roy. Soc. Lond.}
  {\bfseries A269} (1962) 21--52}.

\bibitem{Barnich:2011mi}
G.~Barnich and C.~Troessaert, ``{BMS charge algebra},''
  \href{http://dx.doi.org/10.1007/JHEP12(2011)105}{{\em JHEP} {\bfseries 12}
  (2011) 105},
\href{http://arxiv.org/abs/1106.0213}{{\ttfamily arXiv:1106.0213 [hep-th]}}.

\bibitem{Campiglia:2020qvc}
M.~Campiglia and J.~Peraza, ``{Generalized BMS charge algebra},''
  \href{http://arxiv.org/abs/2002.06691}{{\ttfamily arXiv:2002.06691 [gr-qc]}}.

\bibitem{Duval:2014uoa}
C.~Duval, G.~W. Gibbons, P.~A. Horvathy, and P.~M. Zhang, ``{Carroll versus
  Newton and Galilei: two dual non-Einsteinian concepts of time},''
  \href{http://dx.doi.org/10.1088/0264-9381/31/8/085016}{{\em Class. Quant.
  Grav.} {\bfseries 31} (2014) 085016},
  \href{http://arxiv.org/abs/1402.0657}{{\ttfamily arXiv:1402.0657 [gr-qc]}}.

\bibitem{Duval:2014uva}
C.~Duval, G.~W. Gibbons, and P.~A. Horvathy, ``{Conformal Carroll groups and
  BMS symmetry},'' \href{http://dx.doi.org/10.1088/0264-9381/31/9/092001}{{\em
  Class. Quant. Grav.} {\bfseries 31} (2014) 092001},
  \href{http://arxiv.org/abs/1402.5894}{{\ttfamily arXiv:1402.5894 [gr-qc]}}.

\bibitem{Duval:2014lpa}
C.~Duval, G.~W. Gibbons, and P.~A. Horvathy, ``{Conformal Carroll groups},''
  \href{http://dx.doi.org/10.1088/1751-8113/47/33/335204}{{\em J. Phys. A}
  {\bfseries 47} no.~33, (2014) 335204},
  \href{http://arxiv.org/abs/1403.4213}{{\ttfamily arXiv:1403.4213 [hep-th]}}.

\bibitem{harish1966discrete}
B.~Harish-Chandra, ``{Discrete series for semisimple Lie groups II},'' {\em
  Acta Math} {\bfseries 116} (1966) 1--111.

\bibitem{Sun:2021thf}
Z.~Sun, ``{A note on the representations of $\text{SO}(1,d+1)$},''
  \href{http://arxiv.org/abs/2111.04591}{{\ttfamily arXiv:2111.04591
  [hep-th]}}.

\bibitem{Penedones:2023uqc}
J.~Penedones, K.~Salehi~Vaziri, and Z.~Sun, ``{Hilbert space of Quantum Field
  Theory in de Sitter spacetime},''
  \href{http://arxiv.org/abs/2301.04146}{{\ttfamily arXiv:2301.04146
  [hep-th]}}.

\bibitem{Gaiotto:2023hda}
D.~Gaiotto, ``{Sphere quantization of Higgs and Coulomb branches and Analytic
  Symplectic Duality},'' \href{http://arxiv.org/abs/2307.12396}{{\ttfamily
  arXiv:2307.12396 [hep-th]}}.

\bibitem{knapp2001representation}
A.~W. Knapp, ``Representation theory of semisimple groups: an overview based on
  examples,''.

\bibitem{bars1972operator}
I.~Bars and F.~G{\"u}rsey, ``{Operator Treatment of the Gel'fand-Naimark Basis
  for SL(2,C)},'' {\em Journal of Mathematical Physics} {\bfseries 13} no.~2,
  (1972) 131--143.

\bibitem{Osborn:2012vt}
H.~Osborn, ``{Conformal Blocks for Arbitrary Spins in Two Dimensions},''
  \href{http://dx.doi.org/10.1016/j.physletb.2012.09.045}{{\em Phys. Lett.}
  {\bfseries B718} (2012) 169--172},
\href{http://arxiv.org/abs/1205.1941}{{\ttfamily arXiv:1205.1941 [hep-th]}}.

\end{thebibliography}\endgroup

\end{document}